\documentclass[twocolumn]{aastex631}
\usepackage{newtxtext,newtxmath}
\usepackage{xcolor}
\usepackage{threeparttable}
\usepackage{verbatim}
\usepackage{hyperref}
\usepackage{tabularx}
\usepackage{makecell}

\DeclareMathOperator{\sinc}{sinc}

\begin{document}

\title{RV$\times$TESS I: Modeling Asteroseismic Signals with Simultaneous Photometry and RVs\footnote{This paper includes data gathered with the 6.5 meter Magellan Telescopes located at Las Campanas Observatory, Chile.}}

\correspondingauthor{Jiaxin Tang, Sharon X.~Wang}
\email{tangjx22@mails.tsinghua.edu.cn, sharonw@tsinghua.edu.cn}

\author[0000-0002-7563-7618]{Jiaxin Tang}
\affiliation{Department of Astronomy, Tsinghua University, Beijing 100084, People's Republic of China}

\author[0000-0002-6937-9034]{Sharon X.~Wang}
\affiliation{Department of Astronomy, Tsinghua University, Beijing 100084, People's Republic of China}

\author[0000-0003-3020-4437]{Yaguang Li} 
\affiliation{Institute for Astronomy, University of Hawai`i, 2680 Woodlawn Drive, Honolulu, HI 96822, USA}

\author[0000-0001-5222-4661]{Timothy R. Bedding} 
\affiliation{Sydney Institute for Astronomy, School of Physics, University of Sydney NSW 2006, Australia}

\author[0000-0001-6753-4611]{Guang-Yao Xiao} 
\author[0000-0001-6039-0555]{Fabo Feng}
\affiliation{Tsung-Dao Lee Institute, Shanghai Jiao Tong University, 1 Lisuo Road, Shanghai, 201210, People’s Republic Of China}

\author[0000-0002-0007-6211]{Jie Yu} 
\affiliation{School of Computing, Australian National University, Acton, ACT 2601, Australia}
\affiliation{Research School of Astronomy \& Astrophysics, Australian National University, Cotter Rd., Weston, ACT 2611, Australia.}

\author[0000-0001-9109-437X
]{Zun Wang} 
\affiliation{Theoretical Particle Physics and Cosmology, King’s College London, UK}

\author[0000-0002-0040-6815]{Jennifer~A.~Burt}
\affiliation{Jet Propulsion Laboratory, California Institute of Technology, 4800 Oak Grove Drive, Pasadena, CA 91109, USA}
\author[0000-0003-1305-3761]{R. Paul Butler}
\affiliation{Earth and Planets Laboratory, Carnegie Institution for Science, 5241 Broad Branch Road, NW, Washington, DC 20015, USA}
\author[0000-0003-0035-8769]{Brad Carter}
\affiliation{Centre for Astrophysics, University of Southern Queensland, Toowoomba, QLD 4350, Australia}
\author[0000-0002-5226-787X]{Jeffrey D. Crane}
\affil{Observatories of the Carnegie Institution for Science, 813 Santa Barbara Street, Pasadena, CA 91101, USA}
\author[0000-0002-2100-3257]{Mat\'ias R. D\'iaz}
\affiliation{Las Campanas Observatory, Carnegie Institution for Science, Raul Bitr\'an 1200, La Serena, Chile}
\author[0000-0003-4976-9980]{Samuel K. Grunblatt}
\affiliation{Department of Physics and Astronomy, University of Alabama, 514 University Avenue, Tuscaloosa, AL 35487-0324, USA}
\author[0000-0001-8832-4488]{Daniel Huber}
\affiliation{Institute for Astronomy, University of Hawai`i, 2680 Woodlawn Drive, Honolulu, HI 96822, USA}
\author[0000-0003-0433-3665]{Hugh R.A. Jones}
\affiliation{Center for Astronomy Research, University of Hertfordshire, Hatfield, UK}
\author[0000-0002-7084-0529]{Stephen R. Kane}
\affiliation{Department of Earth and Planetary Sciences, University of California, Riverside, CA 92521, USA}
\author[0000-0002-4927-9925]{Jacob K. Luhn}
\altaffiliation{NASA Postdoctoral Fellow}
\affiliation{Jet Propulsion Laboratory, California Institute of Technology, 4800 Oak Grove Drive, Pasadena, CA 91109, USA}
\author[0000-0002-8681-6136]{Stephen A. Shectman}
\affil{Observatories of the Carnegie Institution for Science, 813 Santa Barbara Street, Pasadena, CA 91101, USA}
\author[0009-0008-2801-5040]{Johanna Teske} 
\affiliation{Earth and Planets Laboratory, Carnegie Institution for Science, 5241 Broad Branch Road, NW, Washington, DC 20015, USA}
\affil{Observatories of the Carnegie Institution for Science, 813 Santa Barbara Street, Pasadena, CA 91101, USA}

\newcommand{\PSUAA}{Department of Astronomy \& Astrophysics, 525 Davey Laboratory, 251 Pollock Road, Penn State, University Park, PA, 16802, USA}
\newcommand{\PSUCEHW}{Center for Exoplanets and Habitable Worlds, 525 Davey Laboratory, 251 Pollock Road, Penn State, University Park, PA, 16802, USA}
\newcommand{\PSETI}{Penn State Extraterrestrial Intelligence Center, 525 Davey Laboratory, 251 Pollock Road, Penn State, University Park, PA, 16802, USA}
\author[0000-0001-9957-9304]{Rob Wittenmyer}
\affiliation{Centre for Astrophysics, University of Southern Queensland, Toowoomba, QLD 4350, Australia}
\author[0000-0001-6160-5888]{Jason T.\ Wright}
\affiliation{\PSUAA}
\affiliation{\PSUCEHW}
\affiliation{\PSETI}
\email{astrowright@gmail.com}
\author{Jeremy Bailey}
\affiliation{School of Physics and Australian Centre for Astrobiology,University of New South Wales, Sydney 2052, Australia}
\author{Simon J. O'Toole}
\affiliation{Australian Astronomical Optics, Macquarie University, North
Ryde, NSW 1670, Australia}
\author{Chris G. Tinney}
\affiliation{School of Physics and Australian Centre for Astrobiology,University of New South Wales, Sydney 2052, Australia}

\begin{abstract}
Detecting small planets via the radial velocity method remains challenged by signals induced by stellar variability, versus the effects of the planet(s). Here, we explore using Gaussian Process (GP) regression with Transiting Exoplanet Survey Satellite (TESS) photometry in modeling radial velocities (RVs) to help to mitigate stellar jitter from oscillations and granulation for exoplanet detection. We applied GP regression to simultaneous TESS photometric and RV data of HD~5562, a G-type subgiant ($M_\star=1.09M_{\odot}$, $R_\star=1.88R_{\odot}$) with a V magnitude of 7.17, using photometry to inform the priors for RV fitting. The RV data is obtained by the Magellan Planet Finder Spectrograph (PFS). The photometry-informed GP regression reduced the RV scatter of HD~5562 from 2.03 to 0.51 m/s. We performed injection and recovery tests to evaluate the potential of GPs for discovering small exoplanets around evolved stars, which demonstrate that the GP provides comparable noise reduction to the binning method. We also found that the necessity of photometric data depends on the quality of the RV dataset. For long baseline and high-cadence RV observations, GP regression can effectively mitigate stellar jitter without photometric data. However, for intermittent RV observations, incorporating photometric data improves GP fitting and enhances detection capabilities.
\end{abstract}


\section{Introduction} \label{sec:intro}
In the past decade, precise radial velocity (RV) measurements have played a crucial role in detecting small exoplanets (non-gas-giant, smaller than Neptune), enabling the discrimination of the subtle gravitational perturbations exerted by orbiting planets on their host stars \citep[e.g.,][]{Hojjatpanah_2019,Brewer_2020,Gupta_2021}. With the RV technique being continually refined, enabling us to identify smaller and more distant planets, it brings us closer to the discovery of potentially habitable worlds \citep[e.g.,][]{Newman_2023,Hall_2018}. Ongoing efforts with extreme precision radial velocity (EPRV) instruments, for example, KPF \citep{Gibson_2016}, EXPRES \citep{Jurgenson_2016}, WIYN NEID \citep{Schwab_2016}, ESPRESSO \citep{Pepe_2021}, and so on, are currently targeting nearby stars to search for small planets or those within the habitable zone.

The bottleneck in the detection of Earth analogs through precise RV measurements is stellar jitter, an intrinsic stellar variability that can obscure the subtle signatures of Earth-mass planets \citep[e.g.,][]{astro_2020,Crass_2021}. Stellar jitter arises from both convective processes as well as magnetic activity within a star, leading to stochastic RV shifts. These features can change the depths and shapes of the stellar spectral lines, creating extra RV variations intrinsic to the stars. 

Magnetic jitter, being the most prominent bottleneck for detecting Earth analogs (e.g., \citealt{Luhn_2023, Gupta_2024}), has been extensively studied in the literature. Magnetic jitter manifests in multiple ways, typically on timescales of the stellar rotation, such as stellar inhomogeneities, including spots and faculae \citep[e.g., see review by][]{Berdyugina_2005}, as well as changes in the convective blueshift due to the inhibition of convection in magnetically active regions \citep[e.g.,][]{Bauer_2018, Meunier_2010, Lanza_2010, Dumusque_2014}. The complexity of the magnetic activity and how they interact with the star's convective motions make this type of jitter hard to model and remove from RV data, so multiple lines of works have been developed to target the magnetic jitter, for example with the help of spectral magnetic activity indicators such as Ca HK indices (e.g., \citealt{Rajpaul2015, Diaz_2018, Ma_2018, Burrows_2024}).

The convective processes, on the other hand, are caused by the stochastic excitation within the star related to the stellar convection, including stellar oscillation and granulation. Oscillation refers to quasi-periodic pulsations of stars, which are sound waves propagating through the star \citep[e.g., ][]{Aerts_2015}. Granulation jitter is caused by the motions of the stars' convective cells, the granules, which are large-scale, stochastic patterns of plasma movement \citep{Nordlund_2009}. Such phenomena are most extensively studied in cool stars, especially in Sun-like stars. From an asteroseismic perspective, cool stars exhibit solar-like p-mode oscillations that are stochastically excited by their outer convection zones, whereas hot stars lack such signals because their convective layers lie beneath the radiative envelope. From the exoplanet perspective, the detection of planetary companions is also more favorable around cool stars, due to relatively lower stellar masses yielding larger RV amplitudes. The typical RV amplitude of asteroseismic stellar jitter is on the order of m/s for Sun-like stars, which can be significant when compared to the RV signals produced by Earth-like planets around 10~cm/s (e.g. \citealt{Queloz_2001, Santos_2002, Dumusque_2011_HARPS}). Granulation is particularly challenging to mitigate (e.g., \citealt{Luhn_2023, Gupta_2024}), as it exhibits a more irregular pattern compared to the quasi-periodic stellar oscillations. 

One approach to mitigate asteroseismic stellar jitter is through averaging \citep{Dumusque_2011_limit}, especially efficient for stellar oscillations. Since oscillations usually occur over minutes to hours, whereas planetary signals have much longer periods (typically days), we can average the RVs within time bins of minutes to hours to smooth out the shorter-period oscillations. As noted by \cite{Chaplin_2019}, the optimal binning timescale should match the period of the stellar oscillation for maximum effectiveness, supported by analyzing the 8-hour RV data from UVES \citep{Dekker_2000} observations on $\alpha$ Cen A \citep{Butler_2004}. However, averaging means a loss of information, while the characteristics of oscillation and granulation signals carry additional information about the star itself. 

Alternatively, contemporaneous photometry has been known to be helpful. It has been widely used in the analysis of magnetic stellar jitter in RV studies (e.g. \citealt{Aigrain_2012, Oshagh_2017, Li&Basri_2024}). However, the application of photometry on RV time series to analyze asteroseismic signals is not as well-established \citep{Kjeldsen_2025}. Both asteroseismic and magnetic activity signals result in photometric and spectral variations due to changes in the star's photospheric temperature and bulk motions. Theoretically, photometry could assist in RV analysis by identifying and characterizing these variations. 

Now that the \textit{Kepler} \citep{Borucki_2010} and Transiting Exoplanet Survey Satellite (TESS) \citep{Ricker_2014} missions have provided more photometric measurements on numerous stars, using a photometry-trained Gaussian Process (GP) to describe the stellar activity has been increasingly adopted in different scenarios (e.g., \citealt{Grunblatt_2015,Robnik_2020, Pereira_2019, Beard_2024}). GP regression has proven to be a powerful tool in modeling time series, with advantages over frequency-domain analyses when dealing with irregularly sampled data \citealt{O'Sullivan_2024}. It already has numerous applications in asteroseismology \citep[e.g.][]{Brewer_2009,Grunblatt_2016,Hey_2024}. Recently, EPRV instruments have reached the precision and cadence to sample even the p-mode oscillations of K dwarfs \citep[e.g.,][]{Hon_2024,Campante_2024,Li_2025}, making time-series analyses of asteroseismic signals of small dwarf stars possible, motivating future work in the time domain. In this work,  we explore the effectiveness of GP in modeling the asteroseismic signals in precise RVs, especially evaluating the use of contemporaneous photometry from TESS.

In this paper, we present modeling of the simultaneous photometric and RV data of the G-type subgiant HD~5562, with a GP to measure its oscillation and granulation parameters. We validated the modeling method using injection recovery tests, and compared the efficiency of jitter reduction with our method to the averaging method. This work is part of the RV$\times$TESS program\footnote{\url{www.rvxtess.com}}. The paper is organized as follows: we first describe our observational data in Section~\ref{sec:observation}. We then include the stellar characterization results in Section~\ref{sec:stellar characterization}. Section~\ref{sec:result} presents the GP modeling on the observed data. In Section~\ref{sec:discussion}, we discuss our results and present a suite of planet injection-recovery tests and a comparison between GP modeling and the binning method in reducing stellar jitter.

\section{Observations} \label{sec:observation}

\subsection{TESS Photometry} \label{subsec:TESS}

\begin{figure}[ht]
    \centering
    \includegraphics[width=0.4\textwidth]{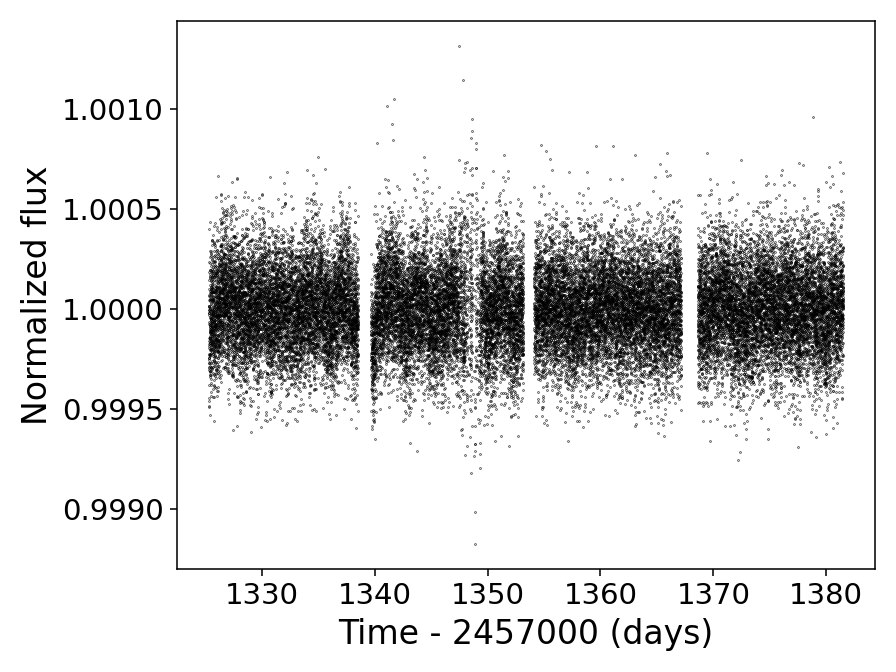}
    \caption{Normalized TESS 2-minute cadence PDCSAP light curve for HD~5562 from sectors 1 and 2. The large scatter during sector 1 near 1350 days originates from a misconfigured fine-pointing calibration. Note that the BJD time has an offset of 2457000 days. See Section~\ref{subsec:TESS} for more details.}
    \label{fig:lc}
\end{figure}

Our full asteroseismic analysis uses all available TESS data (see Sec.~\ref{subsec:asteroseismology}). And our analysis of the contemporaneous RVs incorporates the 2-minute cadence data from Sectors 1 and 2, which overlapped with the RV observations (Sec.~\ref{subsubsec:PFS}). 
During Sector 1, reaction wheel speeds were reset to low value every 2.5 days to improve pointing precision, and this "momentum dump" lasted about 5 minutes each. These maneuvers temporarily disrupt Fine Pointing mode, resulting in degraded pointing stability, which recovers to nominal levels after approximately 10 minutes. Thus the enhanced scatter observed near JD-1350 of sector 1 results from a misconfigured fine-pointing calibration, according to the TESS Data release note \footnote{\url{https://archive.stsci.edu/missions/tess/doc/tess_drn/tess_sector_01_drn01_v02.pdf}}. For the two 2-minute cadence sectors, we derived the Presearch Data Conditioning Simple Aperture Photometry (PDCSAP) light curves \citep{Smith_2012, Stumpe_2012, Stumpe_2014} using \texttt{lightkurve} package \citep{lightkurve}. We note that signals with periods at 1 day and 13.7 days due to the orbital period of Earth and TESS are not significant (illustrated as the grey vertical lines in the frequency domain, in Figure \ref{fig:lc_GP_psd}). We normalized the TESS light curves after subtracting an offset in absolute flux between sectors 1 and 2, and the result is illustrated in Figure \ref{fig:lc}.

\subsection{Precise Radial Velocities} \label{subsec:Precise RV}
\subsubsection{PFS} \label{subsubsec:PFS}
We obtained 188 RV data points of HD~5562 from the high-resolution spectroscopic observations using the Carnegie Planet Finder Spectrograph on the 6.5-meter Magellan II Clay Telescope located at Las Campanas Observatory in Chile (PFS; \citealt{Crane_2006,Crane_2008,Crane_2010}). PFS is a high-resolution iodine-free template optical \'echelle spectrograph, spanning a wavelength range of 391–734 nm with a resolving power of R $\sim$ 130,000, achieved through a slit with a dimension of 0.3$\arcsec \times$ 2.5$\arcsec$. The point-spread function (PSF) and wavelength calibration of the spectrograph are determined from the absorption lines imprinted by an iodine cell spanning 5000–6200\,\AA\ \citep{Valenti_1995,Butler_1996}. This procedure enables precise Doppler velocity estimates for each observation epoch, with an error budget that includes photon-counting noise, PSF residuals, and mismatch to the iodine-free template. 
Typically, PFS attains an RV precision of 0.5–1.0 m/s on nearby, bright, and photospherically quiet stars, within a distance of $\sim$100pc, V-mag around 8 to 10, and $\log R'_{\text{HK}}$ of about -5 \citep{Arriagada_2011}. The process of spectral data reduction and RV extraction was conducted through a customized pipeline \citep{Butler_1996}.
Our RV observations spanned 8 nights in 2018, comprising two distinct sections separated by a 23-day gap due to the scheduling of PFS runs on Magellan. We took 59 RV data points from UT August 2 to UT August 4 and 127 RV data points from UT August 24 to UT August 28, 2018. For the first and last night, the observations cover a $\sim$6-hour window continuously
, and for all other nights, the typical coverage is between 1300s and 1700s. The typical exposure time for each RV point is 160--420 seconds. The median RV uncertainty is 0.7~m/s, as derived from the data processing pipeline, and does not account for the additional uncertainties introduced by stellar jitter. Two additional data points were taken in September 2021 in order to better constrain the orbit of the stellar companion, which is described in Section \ref{subsec:companion}.

\subsubsection{AAT} \label{subsubsec:AAT}
HD~5562 was also observed several times from June 2002 to October 2014 by the UCLES \'echelle spectrograph \citep{Diego_1990} on the Anglo-Australian Telescope (AAT). These observations provided seven RV data points for HD5562, with a span of 4487 days (12 years), and we used them to examine the long-term trend seen in the PFS RV data caused by a stellar companion (see Section~\ref{subsec:companion}). We reduced the AAT data using a customized pipeline by \cite{Butler_1996}, similar to the one used for PFS data, ensuring consistency in the data processing approach across different telescopes and instruments. The mean RV uncertainty for the AAT data is 2.08 m/s.

\section{Stellar Characterization}\label{sec:stellar characterization}
We combined spectroscopic analysis, Spectral Energy Distribution (SED) fitting, isochrone fitting, and asteroseismic analyses to derive the stellar parameters for HD~5562, which are listed in Table~\ref{tab:stellar_param}. For $T_{\rm{eff}}$ and $\mathrm{[Fe/H]}$, we adopt the results from the \texttt{specmatch} spectral analysis. For the extinction A$_V$ and the luminosity, we adopted the results from SED fitting. For the mass, radius, $\log g$, age, $\nu_{\rm{max}}$, and $\delta \nu$, we adopted the results from asteroseismic analysis.

\begin{table*}[htbp]
    \caption{Basic information of HD~5562}
    \centering
    \begin{tabular}{lcccc}
        \hline\hline
        Parameter & Value & Description &\\
        \hline
        TIC ID$^1$ & 234516451 & TESS Input Catalog\\
        R.A. (J2000)$^2$ & 00:56:21.99 & Right Ascension \\
        Dec. (J2000)$^2$ & -63:57:27.84 & Declination \\
        $\mu_{\alpha}$ (mas/yr)$^2$ & 307.408 $\pm 0.080$ & RA Proper Motion \\
        $\mu_{\delta}$ (mas/yr)$^2$ & 141.567 $\pm 0.078$ & Dec Proper Motion  \\
        $\varpi$ (mas)$^2$ & 20.96 $\pm 0.067$ & Parallax distance \\
        d (pc)$^2$ & 47.20 $\pm 0.05$ & Stellar distance \\
        RV (km/s)$^2$ & -74.00 $\pm 0.44$ & Radial velocity \\
        $T^1$ & 6.451 $\pm$ 0.0061 & T band magnitude \\
        $V^1$ & 7.17 $\pm$ 0.03 & V band magnitude\\
        Spectral Type$^3$ & G9IV & Spectroscopy type from SED fitting\\
        \hline\hline
        Parameter & Description & Spectroscopy & SED & Asteroseimology \\  
        \hline
        $M_\star$ ($M_{\odot}$)$^3$ & Stellar mass & $1.080\pm0.080$ & 1.141 $^{+0.037}_{-0.064}$ & $\bf{1.089\pm0.067}$ \\
        $R_\star$ ($R_{\odot}$)$^3$ & Stellar radius & $2.045\pm0.421$ & 1.963 $^{+0.048}_{-0.054}$ & $\bf{1.881\pm0.043}$ \\
        Age (Gyr)$^3$ & Stellar age & 9.92$\pm 0.17$ & 7.89 $^{+0.59}_{-0.64}$ & $\bf{8.40\pm0.58}$\\
        A$_V^3$ & Extinction & -- & $\bf{0.033^{+0.026}_{-0.032}}$ & --\\
        $L_\star(L_{\odot})^3$ & Luminosity & -- & $\bf{2.92\pm0.22}$ & $2.76\pm 0.09$ \\
        $T_\mathrm{{eff}} (\rm K)^3$ & Effective temperature & $\bf{5319\pm110}$ & $5377^{+83}_{-60}$ & $5443\pm57$\\
        $\mathrm{[Fe/H]} (\rm{dex})^3$ & Stellar metalicity & $\bf{0.25\pm0.06}$ & $0.22^{+0.14}_{-0.13}$ & --\\
        $\log g^3$ & Surface gravity & $3.94\pm 0.16$ & $3.94\pm0.24$ & $\bf{3.92\pm 0.01}$\\
        $\nu_{\rm max} (\mu \mathrm{Hz})^3$  & Frequency of max oscillation power & -- & -- & $972.37\pm70.22$ \\
        $\Delta\nu (\mu \mathrm{Hz})^3$ & Frequency separation & -- & -- & $54.84\pm1.16$ \\
        \hline
        
    \end{tabular}
    \begin{tablenotes}
    \item[1] [1] \cite{Guerrero_2021}, [2] \cite{Gaia_2021}, [3] This work \\
    The final adopted stellar parameters are indicated in bold. 
    \end{tablenotes}
    \label{tab:stellar_param}
\end{table*}

\subsection{Spectroscopic parameters} \label{subsec:spectroscopic}
The spectrum we used to characterize HD~5562 comes from the Deconvolved Stellar Spectral Template (DSST), a product of the Magellan/PFS RV pipeline. The DSST was generated using a high-SNR spectrum stacked from a few spectral observations taken without the iodine cell. The stacked spectrum was deconvolved using the estimated instrumental point-spread function (PSF) of PFS based on iodine observations taken through calibration stars (fast-rotating hot stars, typical of spectral type O/B/A). In this procedure, the spectrum was divided into 1280 segments. Each segment has its own wavelength solution and best-fit PSF and was deconvolved independently. To create a contiguous spectrum for each spectral order, we stitched together the individual DSST segments, taking into account the overlaps between adjacent segments. We only used orders 25--59 among the 73 orders in our analyses, since an iodine absorption cell was used to provide wavelength solutions for these orders.

The DSST segments were normalized individually in the PFS pipeline, so we had to refine the normalization for the stitched spectra. We employed the Alpha-shape Fitting to Spectrum algorithm from \cite{Xu_2019} for normalization. This algorithm uses an alpha shape akin to a convex hull to estimate the high-level shape of the blaze function, to perform the normalization. 

We then passed the normalized spectrum to \texttt{SpecMatch-Emp} \citep{Yee_2017}, which compares the input spectrum with a library of high resolution ($R\sim55,000$) and high signal-to-noise ratio (SNR~\textgreater100) Keck/HIRES spectra taken by the California Planet Search. Briefly, it first shifts and matches the input spectrum to the rest frame of the library, and then it compares the input to the Keck/HIRES libraries to select five closest matches according to $\chi^2$ statistics. For each order of the input spectrum, \texttt{SpecMatch-Emp} uses a linear combination of parameters from the five best-match spectra to generate estimates for [Fe/H] and ${T_{\rm{eff}}}$. We calculated the average parameters from the 25 spectral orders and used these average values as our final spectroscopic parameters for the SED, isochrone, and asteroseismic analyses described in the following sections. To summarize, HD~5562 has an effective temperature of 5319$\pm 110$ K, a radius of 1.88 $\pm 0.04$ R$_\odot$, and metallicity of 0.25$\pm 0.06$. We take the larger value of the error reported package and the scatter, as the error. We used the $\log{g}$, [Fe/H] and $T_{\rm{eff}}$ as priors for the SED and isochrone fittings using \texttt{astroARIADNE} in the next subsection. 

\subsection{Spectral Energy Distribution and Isochrone Fitting} \label{subsec:sed}
\begin{figure}
    \centering
    \includegraphics[width=0.49   \textwidth]{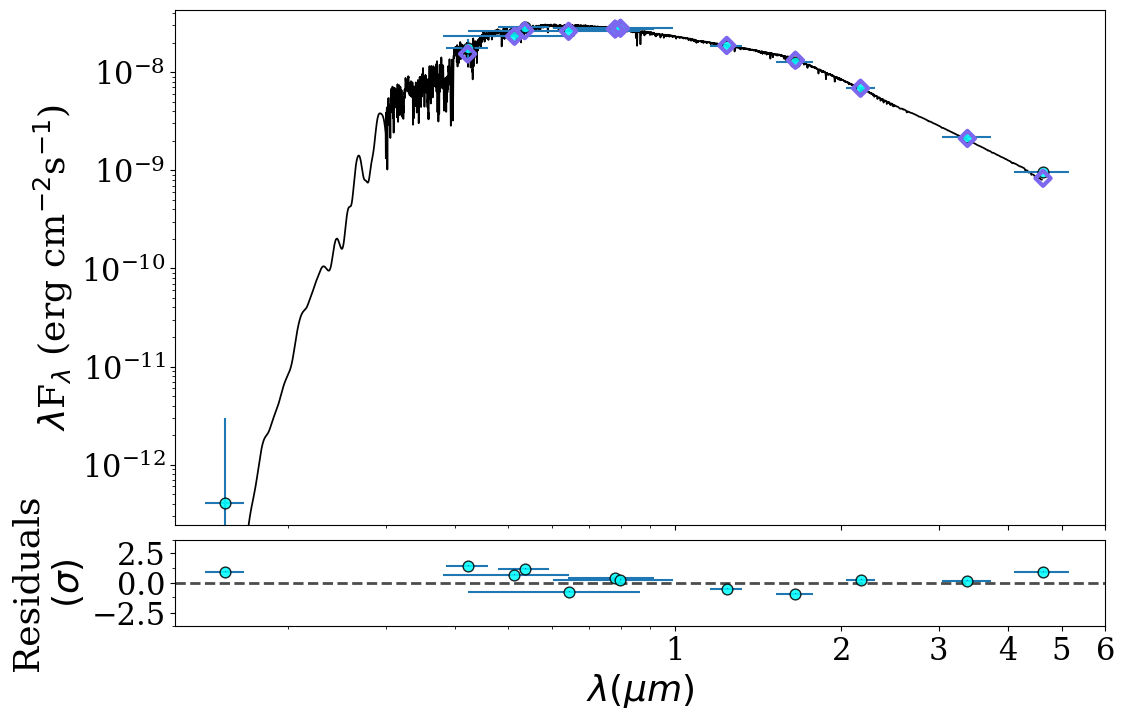}
    \caption{{\bf Top:} Stellar SED and the best-fit model. Blue points represent the photometry from the literature based on SIMBAD queries by \texttt{ARIADNE}, while purple diamonds denote synthetic photometry based on the best-fit stellar model (black line). Horizontal error bars indicate the filter bandpasses. {\bf Bottom:} Residuals of the fit in terms of ratios between residuals and uncertainties of the photometries. See Section~\ref{subsec:sed} for more details.}
    \label{fig:SED}
\end{figure}

Next, we analyzed the SED of HD~5562 with the \texttt{astroARIADNE} package \citep{Vines_2022}. We conducted SED fitting in conjunction with the Gaia DR2 parallax to estimate $T_{\rm eff}$ and radius. \footnote{We use the Gaia DR2 data in this analysis because the \texttt{astroARIADNE} package we are using defaults to DR2. We have carefully checked the Gaia data and found that the parallax differences of HD~5562 between DR2 and DR3 are within the 3-$\sigma$ range. This level of discrepancy is not significant enough to introduce substantial errors into our data. } \texttt{astroARIADNE} queries SIMBAD and gathers photometric measurements including FUV magnitudes from GALEX, BT and VT magnitudes from Tycho-2, JHKs magnitudes from 2MASS, W1–W2 magnitudes from WISE, TESS magnitude from TESS, and three Gaia magnitudes (G, GBP, GRP). Collectively, the available photometry covers a wavelength range from 0.14 to 5 $\mu$m (Figure~\ref{fig:SED}). \texttt{astroARIADNE} also queries Gaia DR2 for maximum line-of-sight extinction Av and incorporates this information into the prior.

With \texttt{astroARIADNE}, we performed the SED fitting with the Phoenix stellar atmosphere models \citep{Husser_2013}, with the priors for $T_{\rm eff}$, $\log{g}$, and [Fe/H] taken from the spectroscopic analysis (see below). We also used the Kurucz stellar atmosphere model embedded in \texttt{astroARIADNE} (ATLAS9; \citealt{Kurucz_1993}) for comparison. The results were consistent with the Phoenix model. We present the results using the Phoenix model here due to its smaller errors in stellar radius and age. Then \texttt{astroARIADNE} proceeds with the results from the SED fitting to fit for a set of $\log{g}$, radius, and mass by interpolating the MIST isochrones and performing Bayesian Model Averaging that accounts for model-specific systematic biases. The final results include the isochrone-derived mass, radius, age, and luminosity of the star HD5562, which is classified as G9IV (see Table \ref{tab:stellar_param} for detailed results).

\subsection{Asteroseismic analyses}\label{subsec:asteroseismology}
\begin{figure}
    \centering
    \includegraphics[width=0.49\textwidth]{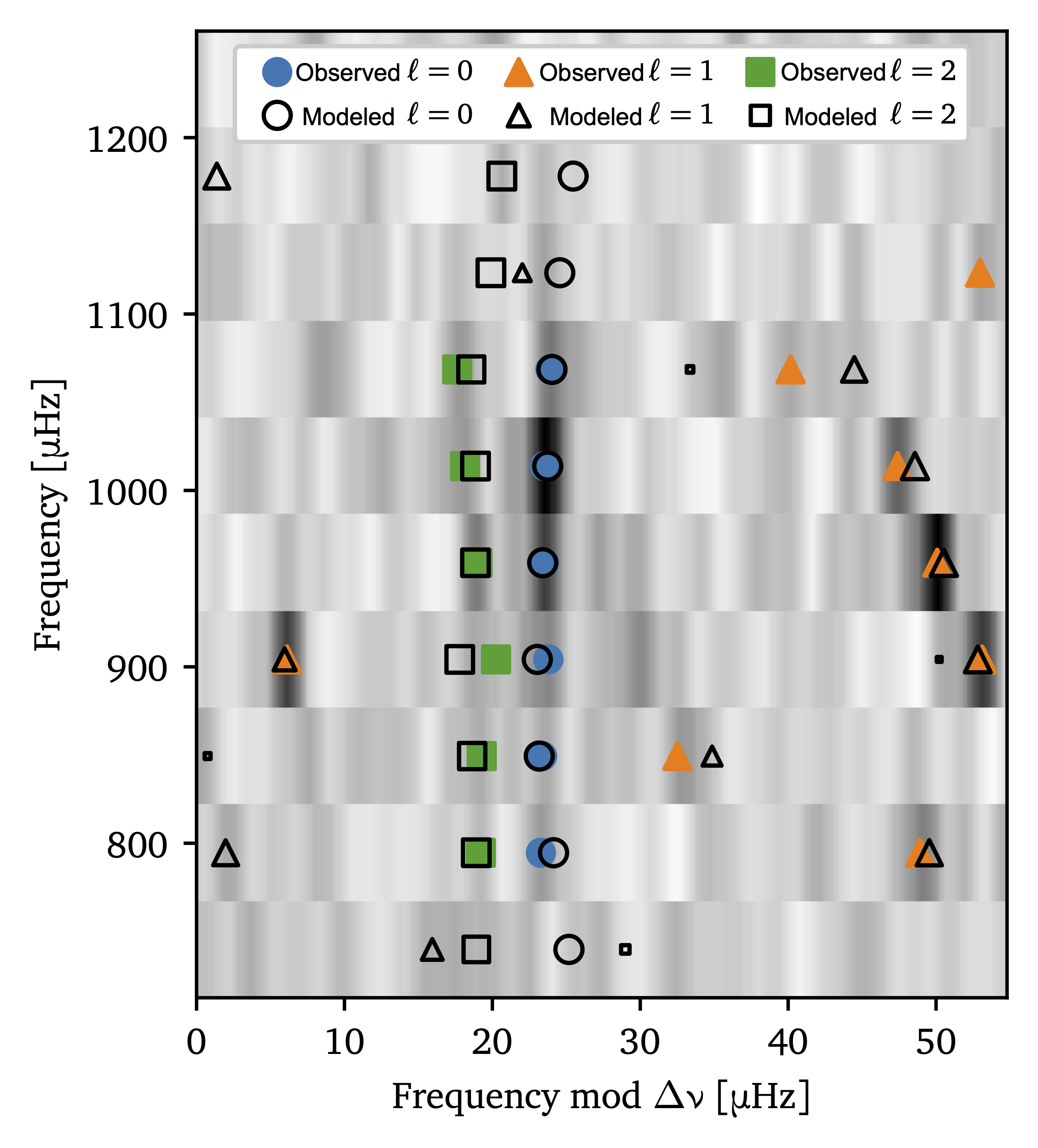}
    \caption{\'{E}chelle diagram from the asteroseismic analyses (Section~\ref{subsec:asteroseismology}) with the observed (colored symbols) and model (open symbols) frequencies highlighted. The frequency mod (x axis) shows the different modelled modes with the symbol sizes of $\ell=1$ and $2$ mofes inversely proportional to their mode inertia scaled to the closest $\ell=0$ modes, which indicates the mode amplitudes.}
    \label{fig:echelle}
\end{figure}

The pulsation spectrum reveals an oscillation pattern corresponding to an early subgiant, resembling the oscillations seen in $\beta$ Hyi \citep{Bedding_2007} and KIC 7747078 \citep{Liyg_2020}. The radial ($\ell=0$) and p-dominated quadrupole ($\ell=2$) modes were identified by their regularly spaced frequencies in $\Delta\nu$, forming two distinct ridges in an asteroseismic \'{e}chelle diagram (Figure~\ref{fig:echelle}). The dipole ($\ell=1$) modes exhibit a mixed character due to the coupling between p and g modes.
We identified 6 radial modes, 8 dipole modes, and 6 quadrupole modes. We extracted the mode frequencies by fitting the power spectrum using a sum of Lorentzian profiles, a process known as peak-bagging \citep{Handberg_2011,Davies_2016}. 
Each Lorentzian profile represents an oscillation mode that experiences damping over time, which is typical of solar-like oscillations. The fitting procedure followed that described in \citet{Liyg_2020}.

To perform asteroseismic model fitting, we used theoretical stellar models constructed with MESA \citep{paxton++2011-mesa, paxton++2013-mesa, paxton++2015-mesa, paxton++2018-mesa, paxton++2019-mesa, jermyn++2023-mesa} and GYRE \citep{townsend+2013-gyre}. The initial free parameters of the evolutionary tracks included stellar mass, helium abundance, metallicity, mixing length parameter, and convective core and envelope overshoot parameters. All stellar tracks evolved to the tip of the red giant branch (RGB). For each model, we corrected the surface effect of model oscillation modes following the procedures introduced in \citet{Ball_2014} combined with the amount of surface correction calibrated in \citet{Liyg_2023}. Although this method can produce erroneous results for mixed modes, \citet{Ong_2021} showed that this is less problematic for early subgiants with high coupling strength such as HD~5562.
We constructed a $\chi_i^2$ function for each model in the grid by incorporating $L$, $T_{\rm eff}$, [Fe/H], and oscillation frequencies as independent constraints. Each model $i$ was associated with a likelihood value $\mathcal{L}_i \propto \exp(-\chi_i^2/2)$. The derived stellar properties were taken as $\hat{\theta} = \sum_i \theta_i \mathcal{L}_i / \sum_i \mathcal{L}i$ and the uncertainties as $\sigma_\theta = \sqrt{\sum_i (\theta_i - \hat{\theta})^2 \mathcal{L}_i / \sum_i \mathcal{L}_i}$, where $\theta_i$ is each stellar property.

\subsection{Stellar Companion} \label{subsec:companion}
Upon examining the RV data obtained from PFS via the RV$\times$TESS program, a long-term trend was clearly visible, which appears to originate from a long-period stellar or sub-stellar companion. 
This is consistent with its entry in the Washington Double Star Catalog (link on VizieR \href{https://vizier.cds.unistra.fr/viz-bin/VizieR-5?-ref=VIZ682dfdd141a98&-out.add=.&-source=J/A%2bA/623/A72/hipgpma&recno=4350}{here}). Although speckle interferometry did not detect any companion \citep{Tokovinin2010}, \citet{Kervella_2019} analyzed proper-motion differences between Hipparcos and Gaia and specified this star in the WDS catalog. Notably, \cite{Hinkel2019} predicted that HD~5562 could be a host to giant planets based on stellar abundance patterns, and Gaia DR3 noted that HD~5562 is a spectroscopic binary \citep{GaiaCollaboration2023,GaiaBinary2022}. To further characterize this binary companion in order to better detrend our RV data, we collected two data points in addition to the RV$\times$TESS data and combined them with archival RVs from AAT. We employed the \texttt{Juliet} package \citep{Espinoza_2019} to perform a Keplerian fit and used \texttt{RadVel} \citep{Fulton_2018} to calculate the derived orbital parameters ($a$ and $M_b\sin{i}$). The best-fit orbital parameters and their prior settings are listed in Appendix Table~\ref{tab:rvorbitfit}. The RV fit indicates that the companion is an M dwarf with a minimal mass of 0.41 solar mass, an orbital period of 4066 days, and an eccentricity of 0.31. 

To further determine the nature of the companion, we performed a joint orbital fitting with the RVs and the Hipparcos-Gaia astrometry using the method developed by \citet{Feng2023MNRAS} (hereafter Feng23, a modified python version is available in \citealt{Xiao2024MNRAS})\footnote{\url{https://github.com/gyxiaotdli/mini_Agatha}}. In addition to Hipparcos intermediate astrometry data (IAD; new reduction of \citealt{vanLeeuwen2007}), Feng23 has also been optimized to use Gaia second and third data releases (GDR2 and GDR3; \citealt{GaiaCollaboration2018, GaiaCollaboration2023}) by simulating the Gaia epoch data with Gaia Observation Forecast Tool\footnote{\url{https://gaia.esac.esa.int/gost/index.jsp}} (GOST). We also considered the potential contamination to the system's photocenter from the companion using the mass-luminosity relation of \citet{Pecaut_Mamajek2013ApJS}. More details on the methodology can be found in Appendix B of \citet{Xiao2024MNRAS}. 

As shown in Table~\ref{tab:orbitfit}, the fitted parameters for our combined analysis of RVs (binned) and Hipparcos-Gaia astrometry include the orbital period $P$, RV semi-amplitude $K$, eccentricity $e$, argument of periastron $\omega$ of stellar reflex motion, orbital inclination $i$, longitude of ascending node $\Omega$, mean anomaly $M_{0}$ at the minimum epoch of RV data and five astrometric offsets ($\Delta \alpha*$, $\Delta \delta$, $\Delta \varpi$, $\Delta \mu_{\alpha*}$ and $\Delta \mu_\delta$) of barycenter relative to GDR3. Other parameters such as the semi-major axis $a$, the mass of companion $M_{\rm B}$, and the epoch of periastron passage $T_{\rm B}$ were derived from the above orbital elements. The priors for each parameter are listed in the last column. 
We finally derive the orbital solution by sampling the posterior via the parallel-tempering Markov Chain Monte Carlo (MCMC) sampler \texttt{ptemcee} \citep{Vousden2016}. We employ 30 temperatures, 100 walkers, and 50,000 steps per chain to generate posterior distributions for all the free parameters, with the first 25,000 steps being discarded as burn-in. We used tempered chains to allow the visit of broader regions of the parameter space.

\begin{figure*}
    \centering
	\includegraphics[width=\textwidth]{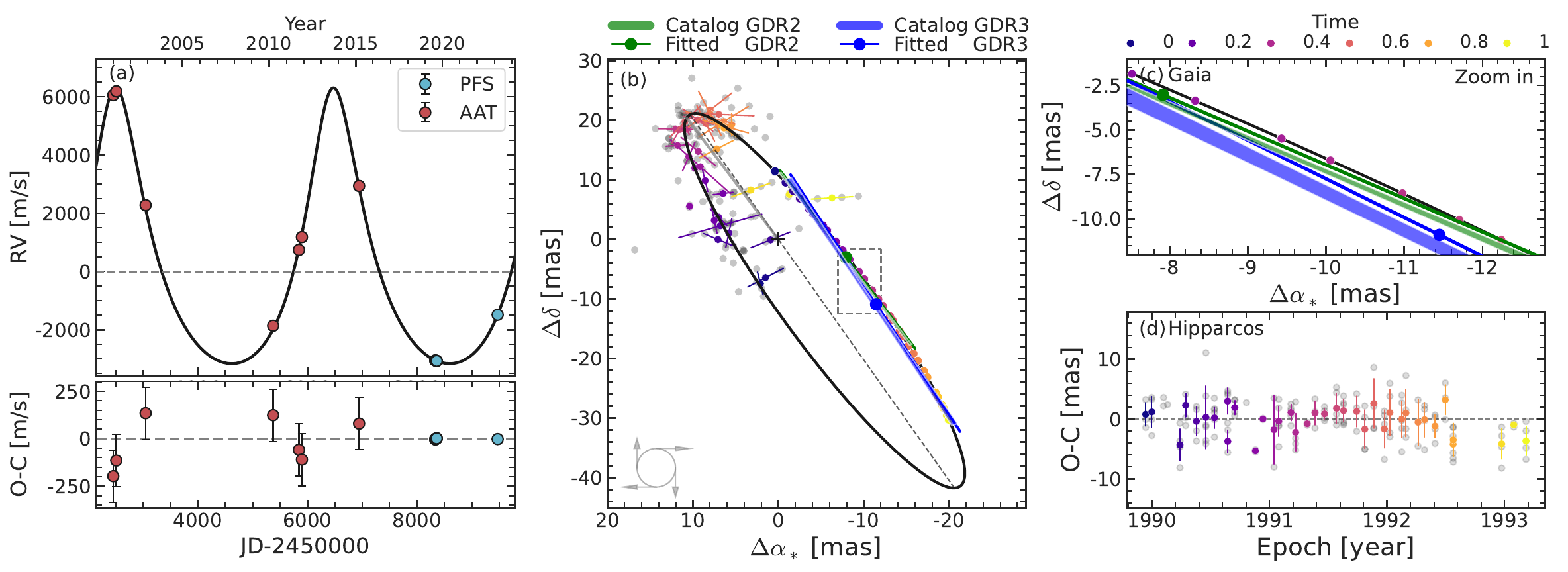}
    \caption{Joint fitting to the RVs, Hipparcos, and Gaia astrometry. (a) RV curve of HD\,5562\,B. The thick black line shows the best-fit Keplerian orbit. Residuals ($O-C$) between the observation and the model are plotted underneath.
    (b) The best-fitting astrometric orbit of HD\,5562. The black dashed line inside the orbit connects the ascending node and the descending node. The plus symbol denotes the system's barycenter, and the grey line connects it with the periapsis. The post-fit Hipparcos abscissa residuals are projected into the R.A. and decl. axes (grey dots) and have been binned into single points with colors. The brightness of these points gradually increases with observation time (the temporal baseline of each satellite is normalized to 1). The orientations of the error bars of each point denote the along-scan direction of Hipparcos. The curl at the lower left corner denotes the orientation of the orbital motion. (c) Zoom-in of the rectangular region of panel (b) which depicts the best fit to Gaia GOST data and the comparison between best-fit and catalog astrometry (positions and proper motions) at GDR2 and GDR3 reference epochs. The blue shaded regions represent the uncertainty of catalog positions and proper motions after removing the motion of the system's barycenter. The dot and slope of two lines (blue and green) indicate the best-fit position and proper motion offsets induced by the companion. (d) The residual ($O-C$) of Hipparcos abscissa.}
    \label{fig:rv_ast}
\end{figure*}

The optimal orbit of the companion has a period of ${10.885}_{-0.017}^{+0.016}$\,yr, an eccentricity of ${0.3351}_{-0.0063}^{+0.0070}$ and an inclination of ${99.83}_{-0.93}^{+0.92}\degr$, suggesting a retrograde orbital motion. Given the stellar mass of $M_{\star}={1.141}_{-0.064}^{+0.037}\,M_{\odot}$ from SED fitting, we derived a mass of ${488}_{-14}^{+10}\,M_{\rm Jup}$ (${0.4658}_{-0.0134}^{+0.0095}\,M_\odot$) and a semi-major axis of ${5.151}_{-0.082}^{+0.079}$ au for the companion. Therefore, HD\,5562\,B should be an M dwarf star. We compared our orbital solution for HD~5562B with that of \citet{Barbato_2023}. The two analyses yield consistent orbital parameters within their uncertainties, except for the argument of periastron ($\omega$), which differs at the $\sim$3.5$\sigma$ level. This discrepancy may be related to differences in the phase coverage of the RVs or in the astrometric fitting methods. More generally, comparing our results with other estimates available in the literature \citep[e.g.][]{Barbato_2023,Soubiran_2022,Gaia_2021}, all reported orbital parameters are mutually consistent within 3$\sigma$. Our estimate of the stellar luminosity agrees with those in all other studies, with the exception of that inferred from the CORALIE data \citep{Barbato_2023}, which may be attributable to a metallicity bias in the SED analysis adopted therein. The spectroscopic [Fe/H] values in our study and all others are in agreement well within 3$\sigma$. Fig.~\ref{fig:rv_ast} depicts the best-fit orbit to RVs, Hipparcos IAD, and Gaia GOST data, while the posterior distributions of selected orbital parameters are presented in Appendix Fig.~\ref{fig:rv_ast_corner}. 

\begin{table*}
\centering
\caption{The optimal orbital parameters and the prior settings for HD\,5562\,B. Log-$\mathcal{U}(a, b)$ is the logarithmic uniform distribution between $a$ and $b$, Cos$i$-$\mathcal{U}(a, b)$ is the cosine uniform distribution between $a$ and $b$, and $\mathcal{N}(a, b)$ is the Gaussian distribution with mean $a$ and standard deviation $b$.}\label{tab:orbitfit}
\begin{tabular*}{\textwidth}{@{}@{\extracolsep{\fill}}lcccc@{}}
\hline \hline
Parameter & Unit & Meaning & value & Prior \\
\hline
$P$&day&Orbital period&${3975.8}_{-6.3}^{+5.9}$ & Log-$\mathcal{U}(-1,16)$\\
$K$&m s$^{-1}$&RV semi-amplitude&${4737}_{-39}^{+43}$ & $\mathcal{U}(10^{-6},10^{6})$\\
$e$&---&Eccentricity&${0.3351}_{-0.0063}^{+0.0070}$ & $\mathcal{U}(0,1)$\\
$\omega$&deg&Argument of periapsis&${-8.72}_{-0.91}^{+0.87}$ & $\mathcal{U}(0,2\pi)$\\
$M_0$&deg&Mean anomaly at JD 2452456&${0.32}_{-0.24}^{+0.54}$& $\mathcal{U}(0,2\pi)$\\
$i$&deg&Inclination&${99.83}_{-0.93}^{+0.92}$& Cos$i$-$\mathcal{U}(-1,1)$\\
$\Omega$&deg&Longitude of ascending node&${26.35}_{-0.21}^{+0.22}$& $\mathcal{U}(0,2\pi)$\\\hline
$\Delta \alpha*$&mas&$\alpha*$ offset&${-10.93}_{-0.53}^{+0.52}$& $\mathcal{U}(-10^{6},10^{6})$\\
$\Delta \delta$&mas&$\delta$ offset&${-10.66}_{-0.70}^{+0.63}$& $\mathcal{U}(-10^{6},10^{6})$\\
$\Delta \mu_{\alpha*}$&mas\,yr$^{-1}$&$\mu_{\alpha*}$ offset &${-7.338}_{-0.032}^{+0.032}$& $\mathcal{U}(-10^{6},10^{6})$\\
$\Delta \mu_\delta$&mas\,yr$^{-1}$&$\mu_\delta$ offset&${-15.576}_{-0.031}^{+0.031}$& $\mathcal{U}(-10^{6},10^{6})$\\
$\Delta \varpi$&mas&$\varpi$ offset&${-0.218}_{-0.024}^{+0.024}$& $\mathcal{U}(-10^{6},10^{6})$\\\hline
$P$&yr&Orbital period &${10.885}_{-0.017}^{+0.016}$& ---\\
$a$&au&Semi-major axis &${5.151}_{-0.082}^{+0.079}$& ---\\
$M_{\rm B}$&$M_{\odot}$&Companion mass in solar mass &${0.4658}_{-0.0134}^{+0.0095}$& ---\\
$T_{\rm B}-2400000$&JD&Periapsis epoch&${52452.8}_{-6.0}^{+2.6}$& ---\\
\hline
$\gamma^{\rm AAT}$&m\,s$^{-1}$&RV offset for AAT&${-2300}_{-71}^{+66}$&$\mathcal{U}(-10^{-6},10^{6})$\\
$\gamma^{\rm PFS}$&m\,s$^{-1}$&RV offset for PFS&${3066}_{-19}^{+20}$&$\mathcal{U}(-10^{-6},10^{6})$\\
$J^{\rm AAT}$&m\,s$^{-1}$&RV jitter for AAT&${173}_{-58}^{+99}$&$\mathcal{U}(0,10^{6})$\\
$J^{\rm PFS}$&m\,s$^{-1}$&RV jitter for PFS&${3.00}_{-0.89}^{+1.4}$&$\mathcal{U}(0,10^{6})$\\
\hline 
\end{tabular*}
\end{table*}

\section{GP Analysis and Results} \label{sec:result}
We analyzed the RV time series of HD~5562 to model the stellar oscillation and granulation signals, and evaluate the efficacy of various choices of models and priors under the framework of GP. For all analyses in this section, we have removed the Keplerian signals caused by the stellar companion in the RVs. 

We first describe the overall GP model setup in Section~\ref{subsec:gp_general}. Then in Sections ~\ref{subsec:tradition} and~\ref{subsec:lcgp}, we introduce the informative priors for the GP regression from frequency analyses or GP fitting using the TESS light curves. We then describe our GP modeling with the high-cadence and intermittent RV observations on HD~5562 in Sections~\ref{subsec:rvgp} and~\ref{subsec:intermittent}, respectively.

\subsection{Gaussian Process Model Setup}\label{subsec:gp_general}
We briefly introduce our specific choice of model and kernel setup with GP in this section, while a more general and in-depth introduction to GP regression can be found in \cite{rasmussen2006gaussian}. GP regression provides a flexible, non-parametric approach to modeling time series data, treating each data point as a correlated random variable. With given datasets, a multivariate normal distribution is constructed with a mean vector and a covariance matrix, where each element encodes the covariance matrix between pairs of data points. This covariance matrix is described by a kernel function, which populates the matrix as a positive semi-definite matrix, describing the physical correlation between pairs of data points in the time series and can encode various properties such as smoothness, periodicity, and linearity. The covariance between any two data points $n$ and $m$ taken at times $t_n$ and $t_m$ is given by the function
\begin{equation}
    K_{\alpha} (\tau_{nm})= \sigma_{n}^2 \delta_{nm} + k_{\alpha}(\tau_{nm}),
\end{equation}
where $\sigma_{n}^2$ represents the measurement uncertainties, $\delta_{nm}$ is the Kronecker delta, $k_{\alpha}(\tau_{nm})$ is the kernel function parameterized by $\alpha$, and $\tau_{nm}$ is the absolute value of time difference between $t_n$ and $t_m$. With an appropriate selection of a kernel function, GP subsequently identifies the optimal set of kernel parameters that most accurately model the observed data in a probabilistic manner.

We modeled both light curves and RVs using the \texttt{celerite} package, a scalable library for one-dimension GP regression \citep{Foreman_Mackey_2017}. \texttt{celerite} offers a physically motivated kernel function for p-mode oscillations --- the stochastically driven simple harmonic oscillator (SHO). This kernel function is designed to capture the quasi-periodic nature of these oscillations and is particularly well-suited to asteroseismic signals. The kernel function \( k_{\alpha}(\tau_{nm}) \) in \texttt{celerite} is constructed using a mixture of exponential and periodic components, which can be expressed with only three hyper-parameters in the frequency domain as a power spectral density (PSD):
\begin{equation}
    S(\omega) = \sqrt{\frac{2}{\pi}} \frac{S_0 \omega_0^4}{(\omega-\omega_0^2)^2 +\omega_0^2 \omega^2/Q^2}. \label{eqn:SHO}
\end{equation}
Here, $S_0$ represents the amplitude, $\omega_0$ is the characteristic frequency (i.e., $\nu_{\rm max}$ for p-modes), and $Q$ is the quality factor. The value of $Q$ corresponds to different asteroseismic terms: as illustrated in Figure 1 of \cite{Foreman_Mackey_2017}, for $Q \leq 1/2$, the model exhibits no oscillatory behavior, indicating that the system's damping is sufficient to suppress periodic signals. For large $Q$, the shape of the PSD near the peak frequency approaches a Lorentzian. Specifically, low $Q \approx 1$ can capture granulation signal, while high $Q \gg 1$ is a good model for stellar oscillations. Generally, to reflect the critically damped nature of granulation, the granulation term is fixed at $1/ \sqrt{2}$ \citep{Michel_2009, Kallinger_2014, Foreman_Mackey_2017}. This PSD equation describing the stochastically driven SHO provides a flexible and accurate model for the observed asteroseismic signals.

\texttt{celerite} then allows the evaluation for maximum likelihood for a time series with $N$ measurements in $\mathcal{O}(N)$ operations. By maximizing the likelihood function, \texttt{celerite} determines the optimal hyper-parameters that best fit the given data with the model. 

\subsection{Preparing the GP Regression through Frequency Analyses} \label{subsec:tradition}

GP modeling using photometric data has shown to be helpful in disentangling stellar magnetic activity signals (e.g., spots) from planetary signals, either by training the GP model using photometry before applying it to the RV data, or performing joint fits to both photometry and RVs \citep[e.g.,][]{Tran_2024,Haywood_2014,Gan_2021}. In our study, we plan to follow the first approach to model the asteroseismic signals in RVs, with the first step being to establish a good GP model for the TESS photometry. For this work, we only used the TESS sectors that overlapped with the RV observations (Sectors 1 and 2), rather than using all available sectors, to better illustrate how the method might be applied to other stars for which only one or two contemporaneous sectors are available.

Our asteroseismic analysis of the full TESS light curve involved identifying and analyzing the individual oscillation modes (Sec.~\ref{subsec:asteroseismology}). However, for our GP modeling of the much shorter RV time series, we only need to capture the broad envelope of the oscillation signals and provide a good set of priors for the GP hyperparameters, i.e., $\nu_{\rm max}$, quality factor $Q$, and amplitude $S_0$. Although detailed fitting of individual oscillation modes is possible with tools like \texttt{celerite} \citep[e.g.][Section 6.4]{Foreman_Mackey_2017}, we chose not to pursue this approach because this would involve too many free parameters for fitting the RVs and would over-complicate the model. As shown below, the simple model adopted in this work seems to suffice in fitting the RV data, and our RV data do not show resolved individual modes. We thus focus on obtaining a good set of initial parameters and their priors for the GP hyperparameters describing the overall oscillation signals using the TESS light curve in this section. 

Initially, we attempted to perform GP fitting on the TESS light curve using uninformative priors. However, the high noise level in the light curve data made it difficult to capture the oscillation signal, leading to the SHO term converging toward an additional granulation-like signal. This issue highlights the need for a more robust approach to set informative priors for the GP hyper-parameters, especially for the oscillation quality factor $Q_{\rm osc}$ and oscillation frequency $\omega_{\rm osc}$. Therefore, to perform GP fitting on the light curve, we began with prior identification following the procedure described below. 

The power spectrum of the TESS 2-minute cadence SPOC light curves from Sectors 1 and 2, which overlap with the RV time series, shows an excess from the oscillations. Several peaks in the power spectrum near $1000 \mathrm{\mu Hz}$ are due to oscillations (see Section~\ref{sec:stellar characterization}), but it is hard to extract the broad envelope of the oscillation from these sporadic peaks. We thus extract the overall oscillation parameters through a Gaussian fit in frequency space following \cite{Kallinger_2014}, which tested different modeling approaches and determined the granulation and global oscillation parameters and relevant scaling relations for a large set of Kepler targets.

\begin{figure*}
    \centering
    \includegraphics[width=1\textwidth]{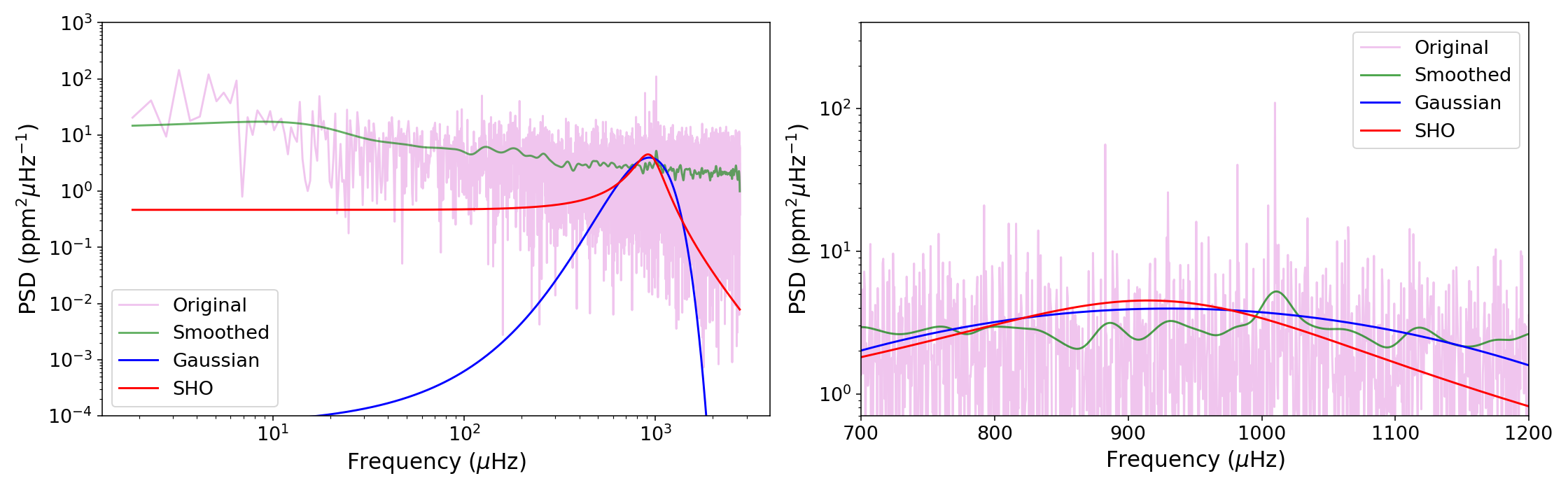}
    \caption{Power spectral density (PSD) of the light curve of HD~5562 from TESS Sectors 1 and 2, covering the whole frequency range (left) and a zoom-in view (right). The green line represents the smoothed spectrum for illustrative purposes, and the blue line is a Gaussian fit to the power spectrum between 700 and 1200 $\mu$Hz. The pink line represents the original, unsmoothed power spectrum. The red line is an initial guess for the SHO model in our GP regression to fit the light curve, which was derived based on the best-fit Gaussian model (blue) within 700-1200~$\mu$Hz. See Section~\ref{subsec:tradition} for more details.}
    \label{fig:Gaussian_SHO}
\end{figure*}

The frequency analysis follows a sequential process, as illustrated in Figure~\ref{fig:Gaussian_SHO}, consisting of the following steps: estimating $\nu_{\rm max}$, the power amplitude $P_g$, then the quality factor $Q$, providing both initial guess and priors for GP regression. 

The first step is to estimate $\nu_{\rm max}$. We first selected the region of the PSD where the oscillation hump is, specifically between 700$\mu$Hz and 1200$\mu$Hz, to fit a Gaussian model using the Levenberg-Marquardt Least Squares Fitter. The Gaussian model is defined by three parameters: amplitude, mean, and standard deviation. We then passed the best-fit value of $\nu_{\rm max}$ ($938\mu$Hz) in this step, which represents the frequency at the peak of the power hump, to the next step. 

The second step is to use the best-fit $\nu_{\rm max}$ to estimate the amplitude of the overall oscillation envelope, $P_g$. Since GP regression requires an initial guess and a prior for the quality factor $Q$, which is not easily determined from TESS data alone due to the relatively low SNR, we instead used the height of oscillation power excess hump $P_g$, to build a connection between the Gaussian form and SHO form of PSDs, which will be linked to the quality factor in the next step. Here, we fit the Gaussian shape based on \cite{Kallinger_2014} in which $P_g$ is the only free parameter, and the PSD in this Gaussian form is given by:
\begin{equation}
\begin{aligned}
    P(\nu)_{\rm Gaussian} &= A \exp\left(-\frac{(\nu- \nu_{\rm max})^2}{2 \sigma_{\nu_{\rm max}}^2}\right) \\
    &= \eta(\nu)^2 \left(P_g \exp\left(\frac{-(\nu - \nu_{\rm max})^2}{2\sigma^2}\right)\right), \label{eqn:traditional}
\end{aligned}
\end{equation}
where $P_g$, $\nu_{\rm max}$ and $\sigma$ are the height, central frequency and width of the hump, respectively. The term $P(\nu)$in this equation is equivalent to the oscillation PSD, $S(\omega)$, in the SHO model as presented in Equation~\ref{eqn:SHO}. From the previous step, we already have the value of $\nu_{\rm max}$, and $\sigma$ is derived from the full width at half maximum (FWHM) using the relation $FWHM / (2\sqrt{2\ln{2}})$, whereas FWHM is approximated as $\nu_{\rm max} /2$, based on the work of \cite{Campante_2016}. $\eta(\nu)^2$ is the attenuation factor due to the averaging over integrations, which equals $\sinc(\pi\nu_{\rm int})$, where $t_{\rm int}$ is the typical integration time of each exposure \citep[e.g.,][]{Barac_2022}. Finally, we determined the only free parameter $P_g$ to have a best-fit of 4.39~ppm$^2/\mu$Hz in this step.

The third step is to obtain a good constraint on $Q$ to be used in GP regression in Section~\ref{subsec:lcgp}. As $Q$ is related to the overall shape of the SHO function, we can obtain a rough estimate on Q by approximating the SHO function (Equation~\ref{eqn:SHO}) using the Gaussian function (Equation~\ref{eqn:traditional}), as both describe the shape of the PSD of the oscillation. Note that $\nu_{\rm max}$ in Equation~\ref{eqn:traditional} corresponds to $\omega_0$ in Equation~\ref{eqn:SHO}, which represents the central frequency of the oscillation hump. Using $P(\nu=\omega_0) = S(\omega = \omega_0)$ and $\omega_0 = \nu_{\rm max}$, we can obtain:
\begin{equation}
    S_0 = \frac{P_g}{Q^2} \sqrt{\frac{\pi}{2}}. 
    \label{eqn: peak relation}
\end{equation}
Among the three free parameters for the SHO function, $S_0$ and $Q$ are linked together via Equation~\ref{eqn: peak relation} above, and $\omega_0=\nu_{\rm max}$ has already been determined, so effectively, $Q$ is the only free parameter. The $Q$ value that would best match the SHO function to the shape of the Gaussian function is 3.08, with an uncertainty of $\sigma_Q = 0.15$ as estimated by the least-square optimizer. This $Q$ value and its uncertainty then provide a more accurate initial guess and tighter prior for the GP regression in the following sections.

\subsection{Light curve modeling with GP} \label{subsec:lcgp}
With relatively reliable initial guesses and priors given by the frequency-space analyses above, we then applied GP modeling on the TESS light curves using the SHO kernels.

The GP kernel function is based on Equation~\ref{eqn:SHO}, which is the power density for each component of the kernel, i.e., a granulation term to capture low-frequency stochastic variations, and an oscillation term to model high-frequency periodic variations. On top of those two components, our SHO kernel requires an extra diagonal jitter as the “white noise” term. The kernel function of the model is then the sum of these three terms. Granulation and oscillation terms have three hyperparameters each, being the amplitude $S_0$, quality factor $Q$, and frequency $\omega_0$. 

For the granulation term, the amplitude $S_{0, \rm gran}$ is set as a free parameter, and initial values are calculated from: 
\begin{equation}
    S_0 = \frac{1}{N} \frac{1}{\omega_0 Q}\sum_{i}^{N}\left(f_i-\bar{f}\right)^2,
    \label{eqn:S0}
\end{equation}
where $f$ stands for the flux for each light curve data point. The quality factor $Q_{\rm gran}$ is fixed at $1 /\sqrt{2}$ \citep{Foreman_Mackey_2017}. The frequency $\omega_{0, \rm gran}$ is set to have the initial guess value of 80$\mu$Hz, determined through visual inspection of the data since the characteristic frequency for granulation is about the turning point where the power falls blueward in frequency space, and the prior is an uninformative, wide log-uniform distribution (see Table~\ref{tab:GP_posterior}).

For the oscillation term, the amplitude $S_{0, \rm osc}$ also takes the initial guess from Equation~\ref{eqn:S0} and a wide, uninformative prior. We adopt an informative prior for the quality factor $Q_{\rm osc}$, a log-uniform prior covering the 3$\sigma$ range of the estimated $Q_{\rm osc}$ from the frequency analyses in Section~\ref{subsec:tradition}. We set the prior for the peak frequency $\omega_{0, \rm osc}$ in the same way as $Q_{\rm osc}$. The jitter term, $\sigma$, typically introduced to describe any additional white noise, is given a wide uninformative prior.

To summarize, the model has six free parameters, being: 1 granulation component with 2 hyperparameters -- the amplitude $S_{\rm 0,gran}$ and frequency $\omega_{\rm 0,gran}$; 1 oscillation component with 3 hyperparameters -- the amplitude $S_{\rm 0,osc}$, the frequency $\omega_{\rm 0,osc}$ and the quality factor $Q_{\rm osc}$; 1 jitter component with 1 hyperparameter -- a nuisance parameter, the jitter term $\sigma$. 

\begin{figure*}
    \centering
    \includegraphics[width=1\textwidth]{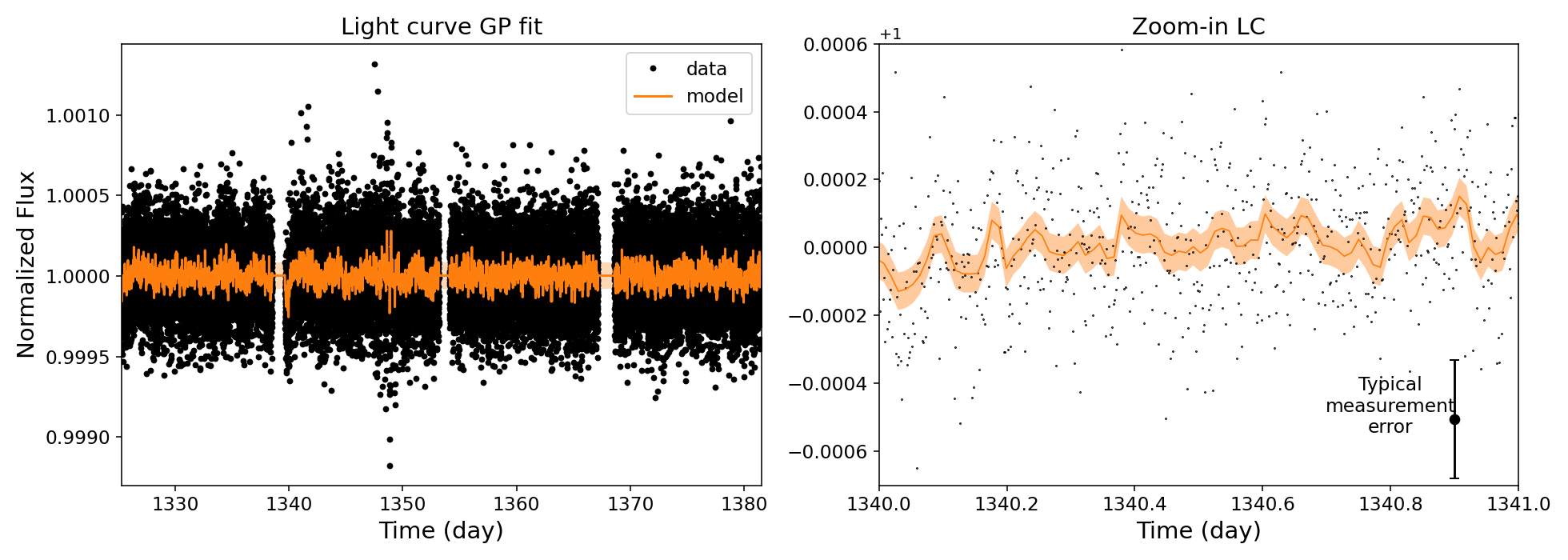}
    \caption{GP model in time series of both TESS Sectors 1 and 2 (left) and zoom-in plot of 1-day duration within Sector 1, better illustrating part of the GP model in the time domain (right). The orange line marks the best-fit GP model (with SHO kernels), and the black circles mark the TESS data points (same as Figure~\ref{fig:lc}). See Section~\ref{subsec:lcgp} for more details.}
    \label{fig:lc_GP_real}
\end{figure*}

\begin{figure}
    \centering
    \includegraphics [width=0.45\textwidth] {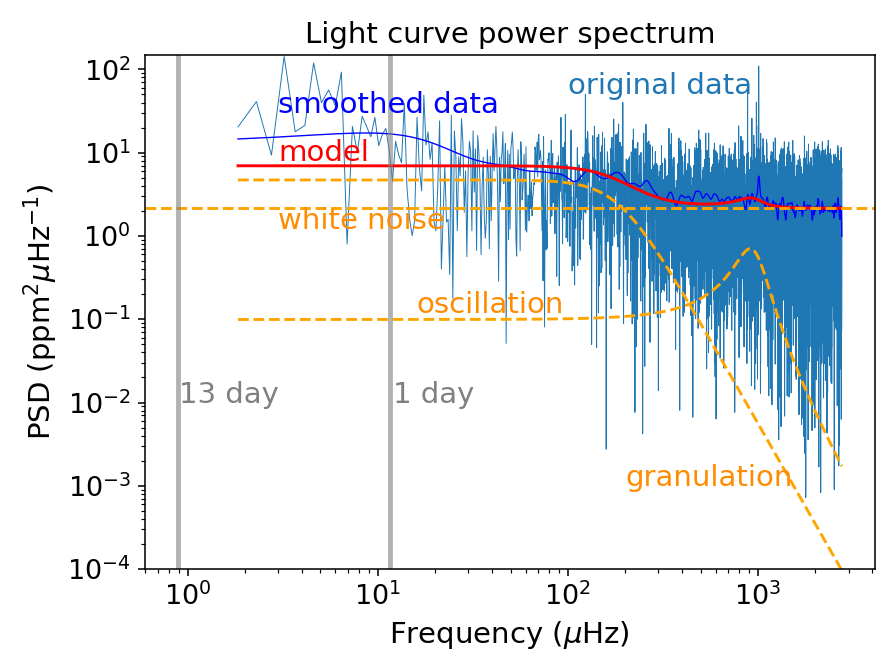}
    \caption{The TESS photometry and the best-fit GP model in Fourier space. The blue line represents the PSD of the TESS light curve using the Fast Chi-squared method of \cite{Palmer_2009}; the bright blue line is the smoothed PSD (labeled with "smoothed data"); the red line represents the best-fit GP model, while the dashed orange lines represent each kernel of the GP model with their names labeled. Grey vertical lines indicate two typical aliasing frequencies in TESS data as a result of the satellite's orbits and the rotation of the Earth.}
    \label{fig:lc_GP_psd}
\end{figure}

GP modeling operates directly in the time domain, as depicted in Figure \ref{fig:lc_GP_real}. To validate the fitting results, we can transform both the observed data and the model into the frequency domain using the Lomb-Scargle periodogram, as shown in Figure \ref{fig:lc_GP_psd}. We determined the uncertainties from MCMC using \texttt{emcee} \citep{emcee_Foreman-Mackey_2013} with posteriors illustrated in Figure~\ref{fig:lc_mcmc}, and the median value and uncertainties derived from the posterior distributions for the hyper-parameters are listed in Table~\ref{tab:GP_posterior}. The GP fitting on the TESS photometry gives a best-fit oscillation frequency $\omega_{\rm osc}$ of $952.8\mathrm{\mu Hz}$, and the best-fit characteristic frequency of granulation $\omega_{\rm gran}$ is $67.5\mathrm{\mu Hz}$.

\subsection{RV modeling with Gaussian Process} \label{subsec:rvgp}
We then applied GP modeling to the RV data, using the posterior parameters from the photometric GP analysis to inform the initial guesses and priors for RV modeling. The model parameters, priors, and fitting results are included in Table~\ref{tab:GP_posterior}. 

We constrain our model to four free hyperparameters plus a jitter term. For the granulation term, the quality factor $Q_{\rm gran}$ is kept fixed as $1/\sqrt{2}$, and the frequency $\omega_{\rm gran}$ has its prior informed by light curve GP fitting, being a wide log-uniform prior around the best-fit value (wider than $\pm6\sigma$). Similar to the previous subsection, the amplitude $S_{\rm gran}$ is a free parameter with its initial guess from Equation~\ref{eqn:S0} and an uninformative prior. The quality factor $Q_{\rm osc}$ gets its initial value from the light curve GP fitting result, and we adopt a very wide prior for $Q_{\rm osc}$ since the light curve GP modeling provided very little constraint on it. For the oscillation term, the frequency $\omega_{\rm osc}$ has a normal prior matching the LC's posterior, as this is the key prior constraint carried over from the LC. The initial value for oscillation amplitude $S_{\rm osc}$ is also calculated from Equation~\ref{eqn:S0} with an uninformative prior, although in principle this could be estimated from asteroseismic scaling relations (e.g., \citealt{Kjeldsen&Bedding_1995}).\footnote{We do not test the scaling relation between the light curve and RV oscillation amplitude in this paper as it could further complicate the parameterization. We leave it for future works.} 

With the set of priors provided by the light curve described above, we applied GP modeling to the entire RV dataset. However, due to a 21-day gap in the dataset, \texttt{celerite} encountered difficulties in numerical stability through such a discontinuous time series. To address this, we separated the data into two segments sharing the same set of hyperparameters and calculated the joint likelihood for GP regression. The first three nights were counted as the first segment while the other five nights were counted as the second. 

The RV data typically have weaker signals of granulation, since the spectroscopic data of this randomly distributed convection motion is averaged through the entire surface of the star, the microscopic changes of RVs caused by granules with smaller scales compared to oscillation will be averaged out during the observation, mainly due to the lack of long-baseline RV \citep{Harvey_1988, Grundahl_2007, Turck-Chieze_2008, Kjeldsen&Bedding_2011, Garcia_2013,Basu_2018B, Kjeldsen_2025}. Therefore we performed a model comparison to see whether granulation should be included. The model comparison showed that for RV data, having both granulation and oscillation could better fit the data than a model with the oscillation term alone. We run AIC and BIC tests with equation of $ \text{AIC} = 2k - 2\ln(\hat{L})$ and $\text{BIC} = k\ln(n) - 2\ln(\hat{L})$, where where $k$ is the number of parameters in the model, $n$ is the number of observations, and $\hat{L}$ is the maximum value of the likelihood function for the model. Specifically, the $\Delta$AIC between two-term case and the oscillation-only term is -12.9, while the $\Delta$BIC is -6.45. We therefore kept the granulation term.

\begin{figure*}
    \centering
    \includegraphics[width=1\textwidth]{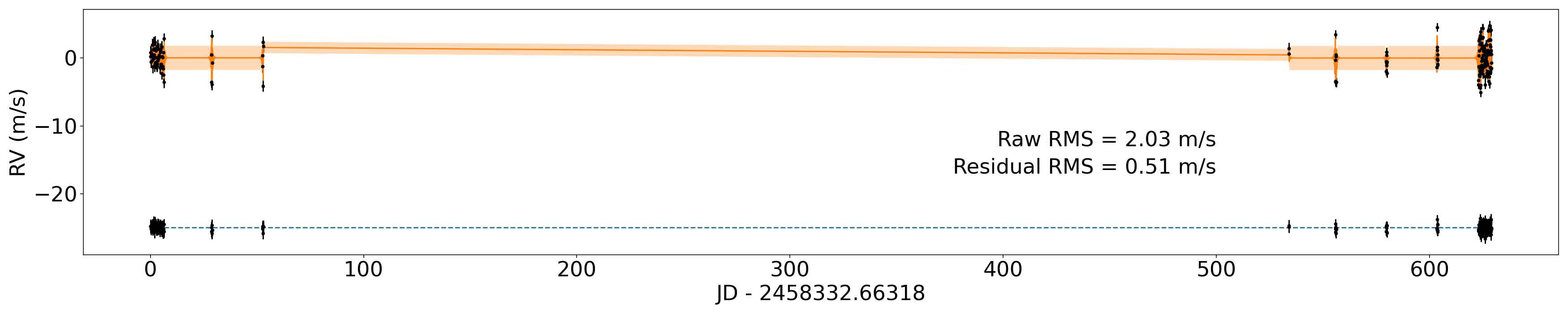}
    \caption{The PFS RVs taken for our program over the course of 27 nights (black circles along with the best-fit GP model (orange line with 1-$\sigma$ uncertainty in orange shades). The lower part of the figure shows the residuals to the fit with a constant offset of $-25$~m/s (blue dashed line) for clarity. See Section~\ref{subsec:rvgp} for more details.} 
    \label{fig:RV}
\end{figure*}

\begin{figure*}
    \centering
    \includegraphics[width=1\textwidth]{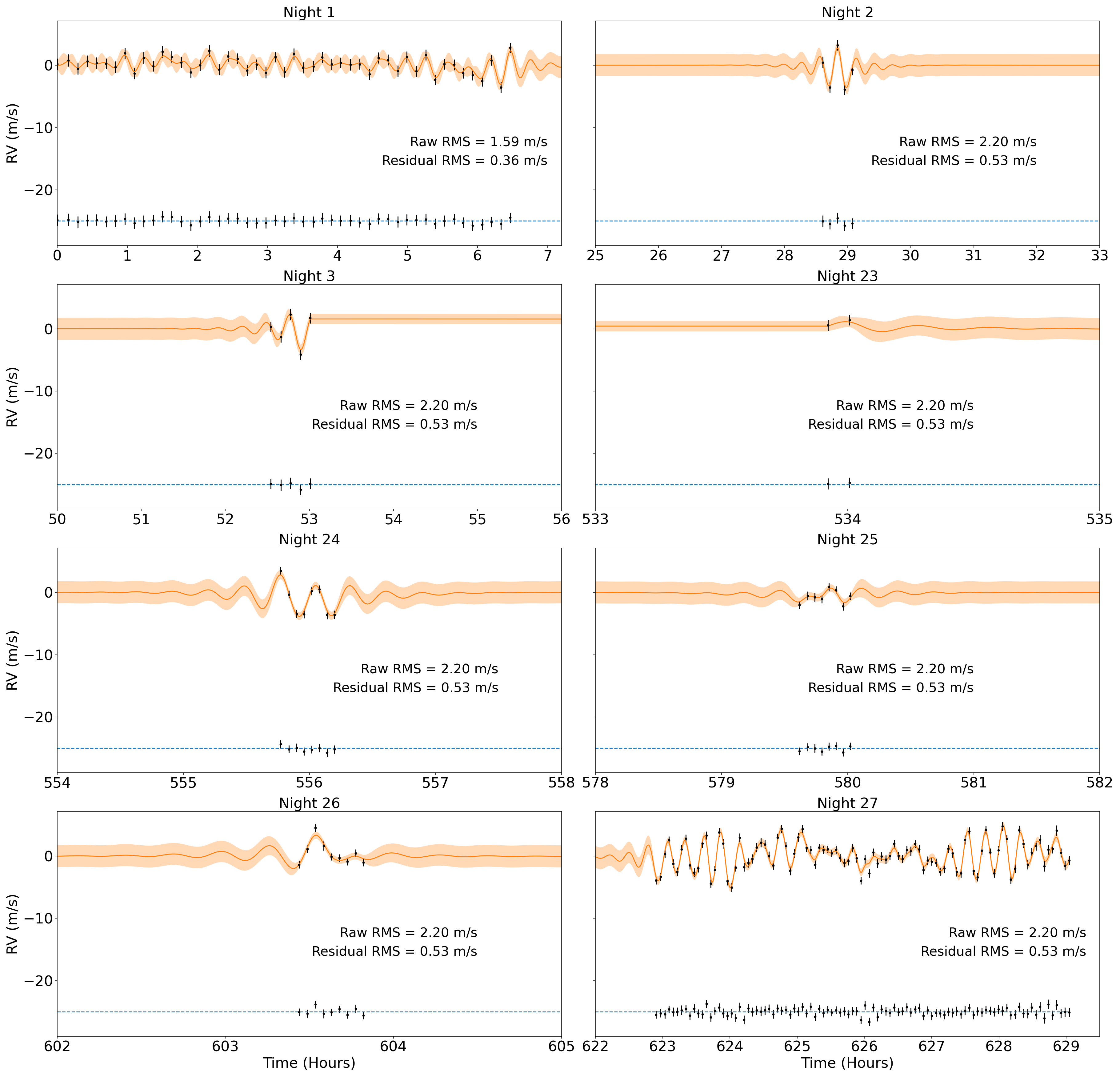}
    \caption{Zoom-in RV plots for Fig.~\ref{fig:RV} with the same legend. The RMS values are calculated for each observing night and displayed respectively in each subplot. Night 1--3 and Night 23--27 are each modeled with a GP kernel but fitted with a joint likelihood to share the same set of hyper-parameters for the GP, which is done to effectively accommodate the long gap in the baseline.}
    \label{fig:RV each}
\end{figure*}

The result for the RV GP modeling is listed in Table~\ref{tab:GP_posterior}. The best-fit values and their uncertainties are calculated from the median and standard deviation of the posteriors, plotted in Figure~\ref{fig:rv_mcmc}. The total RV scatter is reduced from 2.03~m/s to 0.51~m/s, close to the minimal reported RV uncertainty in our data set (0.58m/s), perhaps close to the RV systematic floor. Figure~\ref{fig:RV} shows the GP fit for the entire RV data set, and Figure~\ref{fig:RV each} shows the zoom-in plots for all nights. For the Sun, it is well established that the value of $\nu_{\max}$ derived from photometry is slightly lower than that from velocity measurements, with a typical difference of about 2\% in frequency \citep[e.g.][]{Howe_2020, Kjeldsen_2008, Andersen_2019}. This offset is thought to arise from differences in the atmospheric layers probed by intensity and velocity observations. In the case of HD 5562, the values of $\nu_{\max}$ obtained from the light curve and from the RV data are $\ln(\nu_{\max,\mathrm{LC}})=6.245\pm0.022$ and $\ln(\nu_{\max,\mathrm{RV}})=6.275\pm0.014$, respectively with unit of $\ln(\mathrm{cycles}/\mathrm{day})$. These are statistically consistent within their uncertainties, indicating that the TESS data do not have sufficient precision to detect a subtle solar-like offset between the photometric and RV determinations of $\nu_{\max}$. Therefore, we treat the two determinations as consistent. 

\begin{figure*}
    \centering
    \includegraphics[width=1\textwidth]{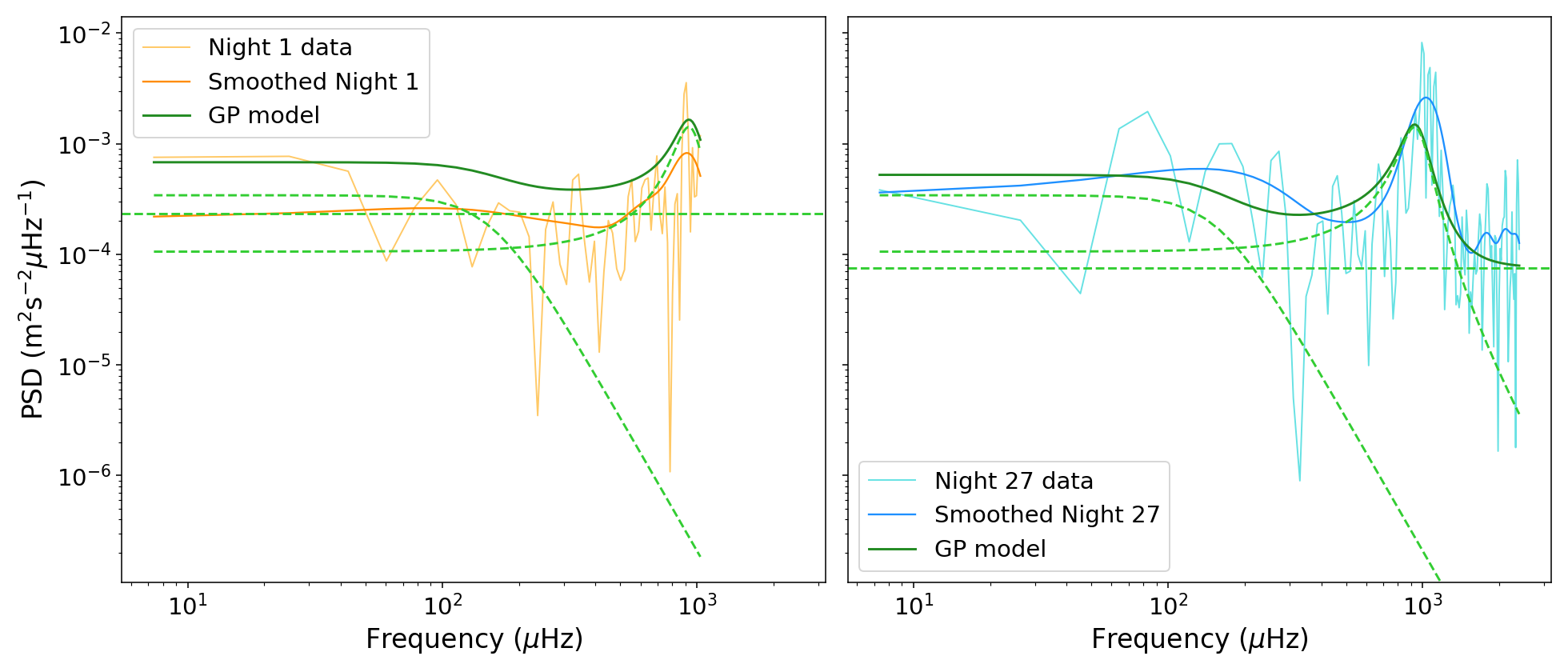}
    \caption{PSDs of the RVs taken on each of the two long-observation nights. The light orange and light blue lines represent the original PSDs, while the darker lines are the smoothed. The green solid lines represent the PSDs of the best-fit GP model (same model as in Figure~\ref{fig:RV}~and~\ref{fig:RV each}), while the dashed lines are for the model components (granulation, oscillation, and white noise). The PSDs validate the goodness-of-fit for the GP models in the frequency domain since the GP regression was only done in the time domain (Figure~\ref{fig:RV each}). See Section~\ref{subsec:rvgp} for more details.}
    \label{fig:RV power spec}
\end{figure*}

As the whole RV data set has very inhomogeneous sampling, we only show the power spectra for the first and the last nights with longer contiguous observations in Figure~\ref{fig:RV power spec}. 
We show the power spectra up to the effective Nyquist frequency which, given that the RV data for HD~5562 are not exactly regularly spaced, we calculated using the median sampling cadence: 
\begin{equation}
    f_{\rm Nyq} = 1/2\Delta t.
\end{equation}
Here, $\Delta t$ is the median cadence. For the first section (Nights 1--3) this is 481s, which results in a Nyquist frequency of 1040$\mu$Hz, while for the second section (Nights 23--27) it was 222s, resulting in a Nyquist frequency of 2252$\mu$Hz (with both times including the CCD readout time of $\sim$60s).\footnote{The exposure time was longer during the first night due to worse seeing, as we tried to maintain a similar SNR per exposure between observations, but back then we thought the $\nu_{max}$ of the star was much smaller ($\sim$40~min) as derived from the earlier stellar parameters and thought both exposure times were short enough.} Therefore, the upper limit of the frequency coverage is set at the larger value.  

The power spectrum shows that GP fits both granulation and oscillation features fairly well in the RV data. The power spectra shown in Figure~\ref{fig:RV power spec} illustrate that GP captures both terms in the RV data, although the granulation model does not look like a very good fit due to the limited RV coverage on longer timescales (at lower frequencies $\sim$days). 

Next, we test and show that the input priors from the photometric data are essential for RV GP modeling for our data set on HD~5562. When we modeled the RV data with only uninformative priors, the GP was able to capture the $\nu_{\rm max}$ value and reduce the RV jitter to a sub-meter level. However, the shape of the best-fit overall oscillation signal skewed heavily towards lower frequency signals, i.e., preferring a significantly smaller $Q_{\rm osc}$ value. This made the overall oscillation envelope wider than what seems reasonable, and the hyperparameters for the granulation term failed to converge. It was only when photometric data provided informative priors that the RV modeling could accurately represent the true power of both the granulation and oscillation terms.

Our RV cadence provides high sensitivity to oscillations but carries little leverage on longer-term granulation and activity signals, which therefore remain weakly constrained. Some of the weak residual structures (e.g., in the data of night 27) could plausibly originate from such incompletely modeled components. We note that the Generalized Lomb–Scargle power spectrum of the RV residuals shows no significant peaks that could be associated with residual asteroseismic oscillations or stellar rotation.

Several physical and methodological factors may also contribute to the low-level correlated structure remaining in the residuals. First, limitations in oscillation model complexity and frequency resolution may play a role. Our GP includes a single oscillation term to describe the solar-like envelope. We experimented with adding a second oscillation term, but AIC/BIC favored the one-term model for the present dataset. Given the limited frequency resolution and duty cycle, a weaker additional component could remain only partially captured and leak into the residuals. Second, window function and sampling effects can imprint additional structure. The irregular sampling and night-to-night cadence differences enhance spectral leakage and aliasing near the oscillation band, allowing small quasi-periodic features to appear in the time-domain residuals even when the overall RMS is low. Third, solar-like oscillations are stochastically excited and damped, with finite mode lifetimes (for subgiants, typically days to weeks). Because our first and last nights are separated by nearly a month, the oscillation realization (phases and amplitudes) need not be identical across nights. Although $\nu_\mathrm{max}$ remains unchanged, individual modes may be more or less damped at different epochs, naturally producing subtle differences in the residual patterns that are not individually significant but collectively visible.

We examined the $S_{\mathrm{HK}}$ and H$\alpha$ indices in both the time and frequency domains and found no significant periodicities or peaks indicative of magnetic activity. Therefore, stellar magnetic activity was not considered in this work.

To summarize, the RV modeling in this subsection demonstrates that GP could capture the oscillation and granulation signals in HD~5562 and help reduce the RV scatter. Although due to the limited data quality in the TESS light curve of HD~5562, we did not feel justified to fix any hyperparameters to help reduce the number of free parameters in the RV GP modeling, incorporating the priors from the light curve GP modeling provided considerable help to the RV GP regression, especially in constraining the granulation term, which is weak in the RV data.

\begin{table*}
\centering
\caption{The estimated hyperparameters and their prior settings for the GP fit on the light curve and RVs of HD~5562, respectively. Log-$\mathcal{U}(a, b)$ is the natural logarithmic uniform distribution between $a$ and $b$, and Log-$\mathcal{N}(a, b)$ is the logarithmic Gaussian distribution with mean $a$ and standard deviation $b$.}\label{tab:GP_posterior}
\begin{tabular*}{\textwidth}{@{}@{\extracolsep{\fill}}lcccc@{}}
\hline \hline
Parameter & Unit & Meaning & Value & Prior \\
\hline
\textit{TESS light curve} \\
$\ln(S_{\rm gran,lc})$& ppm$^{2}\mu$Hz$^{-1}$ &Amplitude of granulation in light curve& $4.219^{+0.064}_{-0.062}$ & Log-$\mathcal{U}(-15,15)$\\
$\ln(\omega_{\rm gran,lc})$&$\ln(\mathrm{cycles}/\mathrm{day})$&Frequency of granulation in light curve& $4.623^{+0.051}_{-0.052}$ & Log-$\mathcal{U}(0,8)$\\
$\ln(Q_{\rm gran,lc})$ & -- & Quality factor of granulation in light curve& $\ln(1/\sqrt{2})$ &Fixed \\
$\ln(S_{\rm osc,lc})$& ppm$^{2}\mu$Hz$^{-1}$ &Amplitude of oscillation in light curve& $0.280^{+0.111}_{-0.134}$ & Log-$\mathcal{U}(-15,15)$\\
$\ln(\omega_{\rm osc,lc})$ & $\ln(\mathrm{cycles}/\mathrm{day})$ & Frequency of oscillation in light curve & $6.245^{+0.022}_{-0.022}$ & Log-$\mathcal{U}(6.2,6.4)$ \\
$\ln(Q_{\rm osc,lc})$ & -- & Quality factor of oscillation in light curve & $0.997^{+0.060}_{-0.028}$ &Log-$\mathcal{U}(0.96,1.26)$\\
$\ln(\sigma_{\rm lc})$& ppm &Jitter term for white noise in light curve& $5.104^{+0.005}_{-0.005}$ & Log-$\mathcal{U}(-20,20)$\\
\hline
\textit{PFS RV} \\
$\ln(S_{\rm gran,rv})$& m$^{2}$s$^{-2}\mu$Hz$^{-1}$ &Amplitude of granulation in RV&${-5.240}_{-0.618}^{+0.621}$& Log-$\mathcal{U}(-9,15)$ \\
$\ln(\omega_{\rm gran,rv})$&$\ln(\mathrm{cycles}/\mathrm{day})$&Frequency of granulation in RV&$4.539_{-0.413}^{0.394}$& Log-$\mathcal{U}(3,6)$ \\
$\ln(Q_{\rm gran,rv})$ & -- & Quality factor of oscillation in RV & $\ln(1/\sqrt{2})$ & Fixed\\
$\ln(S_{\rm osc,rv})$&m$^{2}$s$^{-2}\mu$Hz$^{-1}$ &Amplitude of oscillation in RV&${-7.381}_{-0.318}^{+0.343}$& Log-$\mathcal{U}(-8,8)$\\
$\ln(\omega_{\rm osc,rv})$ & $\ln(\mathrm{cycles}/\mathrm{day})$ & Frequency of oscillation in RV & $6.275_{-0.014}^{+0.013}$ &Log-$\mathcal{N}(6.244, 0.023)$\\
$\ln(Q_{\rm osc,rv})$ & -- & Quality factor of oscillation in RV & $2.230_{0.329}^{0.353}$ & Log-$\mathcal{U}(0.34,3.00)$\\
$\ln(\sigma_{\rm rv})$ & m/s &Jitter term for white noise in RV& ${-0.508}_{-0.334}^{+0.226}$ & Log-$\mathcal{U}(-20,20)$\\
\hline 
\end{tabular*}
\end{table*}

\par

\subsection{Fitting the intermittent RVs}\label{subsec:intermittent}
Given that typical RV data sets for planet searches are sporadic or infrequent instead of having long-term, continuous observations like the ones on nights 1 and 27 presented here, we also tested fitting the intermittent RVs (i.e. Nights 2, 3, 23, 24, 25, 26) in our data of HD~5562 to assess the performance of GP regression. To be specific, we excluded the two 5-hour contiguous observations on nights 1 and 27, retaining only the RVs from the six nights with 30-minute continuous coverage each, typically consisting of a few data points (2--9 exposures; see Figure~\ref{fig:RV each}). 

We first tested whether granulation should be part of the GP model. The model comparison suggests that having only the oscillation kernel provides a better fit. Therefore granulation is excluded from all tests mentioned below. 

We then tested and summarized five different scenarios for modeling stellar jitter using GP using the HD~5562 data, varying the input data set and prior or model choices. We mostly focus on two sets of comparisons: using the full RV data set versus the intermittent one, and using the light curve fit to inform the priors versus uninformative priors. Table~\ref{tab:intermediate_RV} presents a description of these different scenarios, the model and prior choices, the corresponding RV RMS values, and comments on the results. The results suggest that, for sporadic RV measurements, GP regression can successfully fit the RV data and identify a suitable set of parameters, particularly the characteristic frequency of oscillation $\nu_{\rm max}$ can be identified even without the light curve providing the priors. However, the model exhibits overfitting, closely following most of the data points, especially in the cases with uninformative priors. The practical implications of this behavior will be further discussed in Section~\ref{sec:discussion}. 

\begin{table*}[htbp]
    \centering
    \caption{Summary of modeling results, varying the priors or the RV data set (see Section~\ref{sec:result})}
    \begin{tabularx}{\textwidth}{ccccp{0.35\textwidth}}
        \hline \hline
        Data & Model and Priors & Raw RMS & Residual RMS & Notes  \\
        \hline
        Full RV dataset& \parbox{4.5cm}{ \hspace{10pt} \\ model: granulation + oscillation;\\ input: scaling relations and RV PSD} & 2.03 m/s & 0.37 m/s & The best-fit model using least-square fitting is very sensitive to the initial guesses, but reasonable initial guesses on $\omega_{\rm osc}$ and $\omega_{\rm gran}$ are helpful, although the best-fit model has excessive power in granulation and a small $Q_{\rm osc}$. \\
        \hline
        Full RV dataset& \parbox{4.5cm}{ \hspace{5pt} \\ model: granulation + oscillation;\\ input: from LC fit} & 2.03 m/s & 0.51 m/s & A good balance between model accuracy and reduced jitter (Sec.~\ref{subsec:rvgp}).\\
         \hline
         Intermittent RV& \parbox{4.5cm}{ \hspace{10pt} \\ model: granulation + oscillation;\\ input: scaling relations and RV PSD} & 2.11 m/s & 0.36 m/s & The same as the result with the Full RV dataset with no LC input (i.e., excessive granulation power and too-small $Q_{\rm osc}$). \\
         \hline
         Intermittent RV & \parbox{4.5cm}{ \hspace{5pt} \\ model: granulation + oscillation;\\ input: from LC fit} & 2.11 m/s& 0.42 m/s& A good balance between model accuracy and reduced jitter. In particular, LC gives adequate information for a good guess and prior for granulation, which prevents the model from converging towards excessive power in granulation.\\
         \hline
         Intermittent RV & \parbox{4.5cm}{ \hspace{5pt} \\ model: oscillation;\\ input: from LC fit} & 2.11 m/s& 0.33 m/s& The goodness of fit is better than the granulation + oscillation model, but the best-fit model does not seem to capture the oscillation accurately (small $Q_{\rm osc}$). \\
         \hline\hline
    \end{tabularx} 
    \label{tab:intermediate_RV}
\end{table*}

\section{Discussion} \label{sec:discussion}

\subsection{Was Light Curve Helpful for Modeling the RVs?}
In general, precise light curves can be instrumental in RV fitting to guide the models on stellar jitter, particularly when RV data is limited in duration and cadence and thus insufficient in capturing the key timescales of stellar variability \citep[e.g.,][]{Haywood_2014, Gan_2021}. What we have seen with our results on modeling the asteroseismic signals in HD~5562 indicates that, if the RV data is of high cadence and spans a sufficiently long time, the light curve may not be absolutely necessary to provide additional constraints, though still helpful. However, when the RV data coverage is short and intermittent, the light curve can help restrict the RV model, improving its accuracy and reliability. 

To be specific, the first and last night of the HD~5562 RV dataset spans $\sim$6 hours with a high cadence of every $\sim$2 minutes, the GP modeling gets most of its constraints on the asteroseismic signals from these two nights, especially for the p-mode oscillation, which seems to be sufficiently constrained using these high-cadence RV data alone. However, without any prior information from the LC, the RV GP fitting is highly sensitive to the initial guesses and bounds given to the optimizer, and the way to mitigate this modeling uncertainty is to run the MCMC sampling to obtain the posteriors. The power spectrum of such high-quality RV data alone already provides sufficient information for a good initial guess, therefore, light curve is not required for a good model convergence, though it helps, especially for the granulation term to converge onto a more accurate one. 

On the other hand, intermittent RV alone has clear limitations in constraining the asteroseismic signals due to its short observing window per night, but such data are more realistic for exoplanet searches using RVs. Firstly, it is challenging to derive a good set of the initial guesses and bounds for GP fitting from intermittent RVs alone; secondly, we found that the oscillation term tends to converge towards having a lower quality factor $Q$ value, and the power spectrum of the best-fit model often shows almost no granulation feature. In fact, model comparison prefers a model without granulation. Therefore, using light curve GP fitting to inform RV GP fitting is crucial when dealing with only intermittent RVs.

\subsection{Injection-Recovery Tests for Planet Detections} \label{subsec:reduction efficiency}
We performed two sets of simple injection-recovery tests to assess the impact of stellar jitter on the detectability of exoplanets within the HD~5562 system, which currently does not have any known planets. We first tested the case with a simplistic approach where we subtracted the asteroseismic signals using a GP model a priori, and then the other case using a simultaneous Keplerian and GP fit. The GP model for the RVs is informed by the LC. 

\begin{figure*}
    \centering
    \includegraphics[width=1\textwidth]{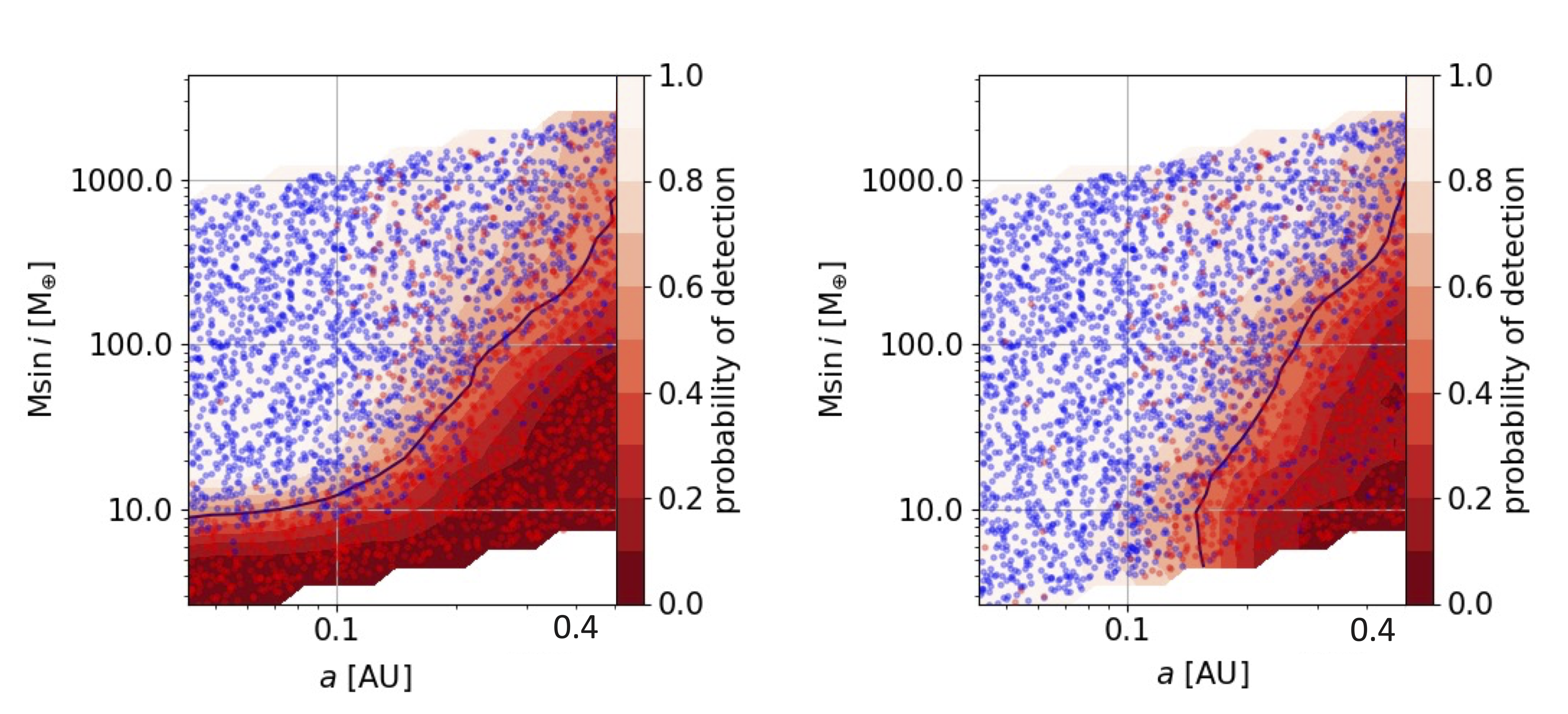}
    \caption{Completeness maps for planet searches using RVs of HD~5562 presented in this work before (left panel) and after (right panel) subtracting the best-fit GP model, which represents the stellar RV jitter from the asteroseismic signals. The comparison between the left and right panels illustrates the improved effectiveness of planet detection enabled under an idealized condition, where stellar jitter could be subtracted \textit{a priori} without affecting any underlying planetary signals, and thus it shows the best-case scenario. The circles are injected planet signals that can (blue) or cannot (red) be recovered. The red contours illustrate the probabilities of recovery, where the solid black line marks the 50\% detection probability. The maps were made using \texttt{RVsearch}. See Section~\ref{subsec:reduction efficiency} for more details.}
    \label{fig:inj_rec_5000}
\end{figure*}

We first performed injection-recovery tests on the RV data set to examine its sensitivity to planetary companions using just the original RVs, and then we subtracted the best-fit GP model (see Section~\ref{subsec:rvgp}) from the RVs and ran the test again for comparison. Using the \texttt{RVSearch} package \citep{Rosenthal_2021}, we first conducted an iterative periodogram search for existing periodic signals and subtracted them, and only the long-period stellar companion was subtracted in this first step. Subsequently, \texttt{RVSearch} injected a variety of synthetic planets and performed an iterative search again to test the recovery capabilities. No additional companions were detected either before or after the GP model subtraction. We conducted 5,000 injections and plotted the completeness contour maps based on the planet period and $M\sin i$ drawn from log-uniform distributions (Figure \ref{fig:inj_rec_5000}). Without GP model subtraction, planets with a mass less than 10$M_\oplus$ in the HD~5562 system were not recoverable. However, after applying GP to reduce the jitter scatter, the sensitivity to planetary signals improved significantly, with the minimum detectable $M\sin i$ decreasing from the $\sim 10M_\oplus$ level to around $1M_\oplus$.

Our injection-recovery test here on the RV dataset used in this work is meant for a quick demonstration only. As commented in previous sections, unlike the data here, in a typical exoplanet-monitoring cadence composed mainly of short, sporadic nights, the number of data points per night would only include a couple of oscillation cycles. A comprehensive evaluation of planet detection capabilities with realistic, intermittent RVs would require dedicated injection–recovery tests under different observing cadences, which represents an important but separate effort beyond the scope of this paper.

While this test provides an initial assessment of the gain in planet detection after removing the asteroseismic signals, it oversimplifies the treatment of stellar signals by directly subtracting the best-fit GP model \textit{a priori}. In practice, such prior knowledge of stellar signals is unrealistic when the planetary signals and stellar jitter signals co-exist in the RV data. To address this overestimation of detectability, we conducted another experiment, where we injected a single planetary signal into a set of simulated RV data and performed a planetary search using a simultaneous fitting approach. We chose to use simulated data instead of the real HD~5562 data because the high observing cadence of HD~5562 is not commonly available in real RV planet-search observations. 

We simulated a set of RV data based on the p-mode oscillation properties of HD~5562, without the granulation term in the SHO kernel, across 16 days of observations. Each night has only 15 minutes of observation (5 data points per night, 3-minute cadence on average, placed randomly at a time within the night), making the dataset a contiguous observation with half a month involving relatively high-cadence observations. We did not add granulation signals since the best-fit model for HD~5562 indicates that the power in granulation is significantly smaller than that in oscillations. In this scenario, we assume we only know some prior information on the p-mode oscillation signals (e.g., from scaling relations using stellar parameters), and the planet search process is a blind search. 

Based on the result on fitting the intermittent RVs in Section~\ref{subsec:intermittent}, for such a short coverage per night, it is most appropriate to apply GP with a kernel consisting of only one oscillation term for stellar jitter and white noise (a decision that perhaps one would make even without knowing that the input stellar jitter signals only contain oscillations). The \texttt{celerite} package allows an easy incorporation of Keplerian models, where the users can set the mean model of GP to be a customized model. This allowed us to account for planetary signals along with stellar oscillations simultaneously in the GP regression. With this Keplerian+GP model setup, we conducted four tests where a planet signal was injected with varying amplitudes, corresponding to different planet masses: 10m/s (which should be an easy task to detect), 3m/s (comparable in amplitude to stellar oscillations), 1m/s (close to the typical error bars of 0.7m/s for PFS RVs), and 0.5m/s (comparable to the lowest RMS achieved from the GP fitting in Section~\ref{sec:result}). All tests assumed an orbital period of 3.37 days for the injected planetary signal, chosen to maintain consistency across the tests and to represent a typical timescale for planetary signals that would be detectable for the selected baseline of 16 days. 

After fitting the Keplerian+GP model using maximum likelihood, we ran MCMC to estimate the significance of planet detection by evaluating the precision on period and RV semi-amplitude $K$ of the best-fit planet signal. We require the constraint on $K$ to be non-zero at least within 5$\sigma$ to count as a detection (recovery), with a 2--3$\sigma$ being a marginal detection. We found that our joint fit successfully recovered the planetary signals in the 10 m/s and 3 m/s cases with the correct period and RV amplitude, demonstrating its robustness in detecting strong and moderately strong signals. For the 1 m/s case, the model also recovered the planetary signal accurately without underestimating the planet's RV amplitude (a common concern when using GP), although it was a somewhat marginal detection. In the most extreme scenario of 0.5 m/s, the model failed to recover the correct planetary signal. This demonstrated that incorporating GP modeling on p-mode signals does not reduce the detectability of planetary signals or bias the RV amplitude, at least for scenarios close to the test case presented here (i.e., moderate cadence coverage close to 1/$\nu_{\rm max}$ nightly over a baseline of $\sim \times 4$ planetary period). The results match the expectations based on simple criteria of RV detection (e.g., \cite{Howard&Fulton_2016}), $K/\sigma_{\rm RV} \cdot \sqrt{N_{\rm obs}} > 6$, where if we assume that $\sigma_{RV} \sim 0.5$~m/s based on the residuals in the GP fits in Section~\ref{sec:result}, then detecting a $K=1$~m/s planet would be somewhat marginal while $K=0.5$~m/s would be not reachable, consistent with the results above.

Various mechanisms, such as magnetic activity and metallicity dependence, could suppress the oscillation amplitude, making the stellar jitter modeling more challenging \citep{Gupta_2022,Mathur_2019}. To fully answer the question of how GP modeling of asteroseismic signals affects or does not affect planet detections, one would need to run more extensive tests covering a larger parameter space in both the planet's and the stellar oscillation signals with a variety of RV observational cadences, which are beyond the scope of this study. Our simple tests highlight the potential of GP modeling in mitigating the asteroseismic jitter to boost planet detection sensitivities.

\subsection{Comparing GP Modeling with Binning} \label{subsec:bin}

Binning is another way to mitigate the stellar jitter originating from p-mode oscillations \citep[e.g.,][]{Dumusque_2011_limit}. We applied the binning method to the HD~5562 data to compare it with the GP modeling method. In Figure~3 of \cite{Chaplin_2019}, they analyzed the residual RV amplitude as a function of exposure duration (i.e., length of binning window) for the Sun. They suggested setting the exposure/binning window to be the inverse of the peak oscillation frequency, which appeared to be the most efficient timescale for binning. We tested a range of binning windows on HD~5562 RV data to determine how effectively binning reduces the asteroseismic signal. We applied binning windows of various lengths to the HD~5562 RV data and calculated the average RV within each bin. We then calculated the RMS of the new binned RV series and compared the results with \cite{Chaplin_2019}. 

\begin{figure*}
    \centering
    \includegraphics[width=1\textwidth]{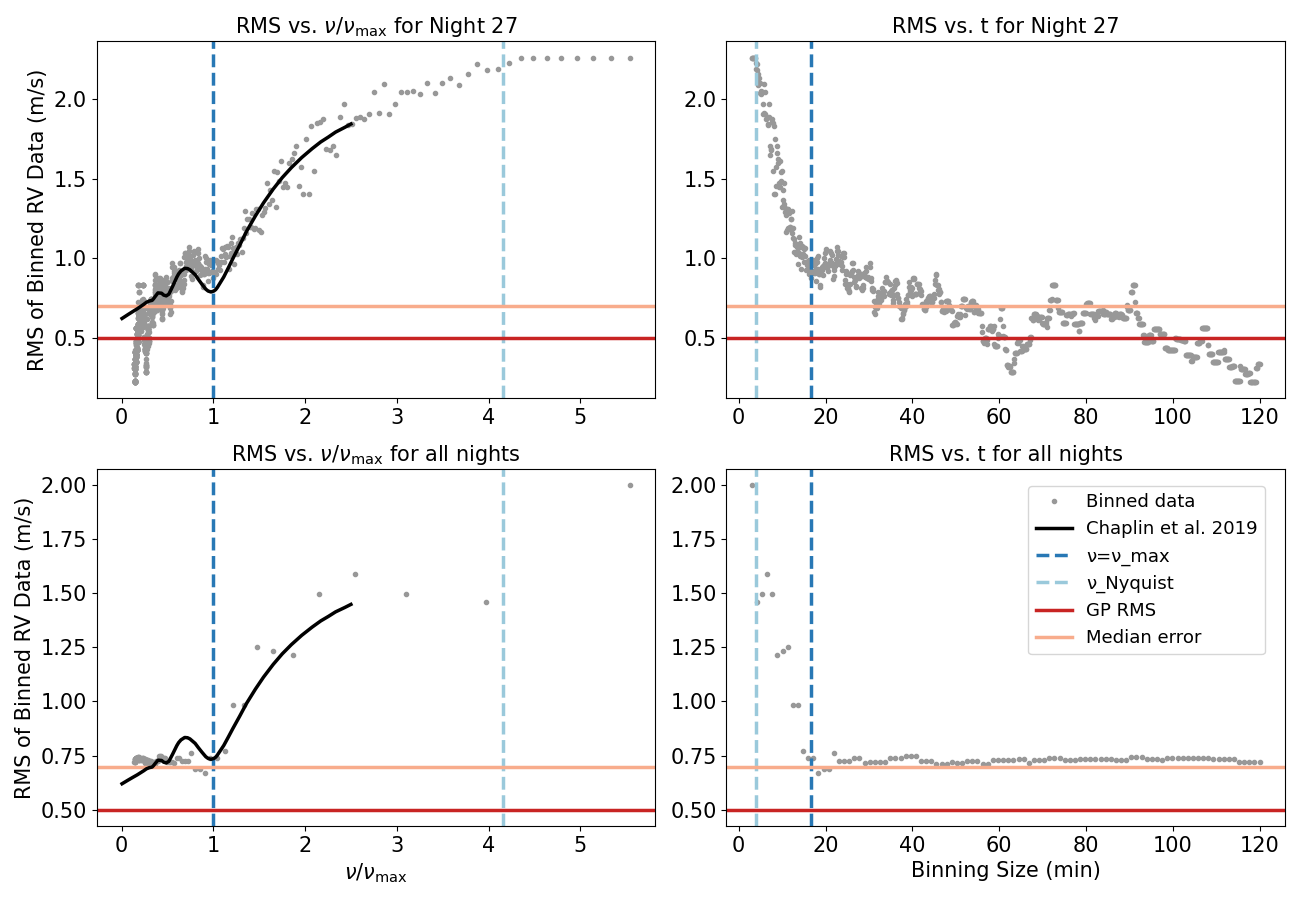}
    \caption{Illustration of how RV RMS decreases as we bin the RVs with changing bin sizes. The x-axes of the left panels are the lengths of the bin in units of $\nu_{\rm max}$, which is 17.14 minutes (Table~\ref{tab:stellar_param}), and the x-axes of the right panels are bin sizes in minutes. The top panels are for Night 27 data only, while the lower panels are for the entire RV dataset. The grey dots represent the RV RMS for each bin size. The lower panels have fewer bins due to the short baselines in Nights 2--26. The vertical dashed lines indicate the Nyquist frequency of our RV data (light blue) and $\nu_{\rm max}$ (dark blue). The horizontal lines mark the RMS level after subtracting the stellar jitter signal derived from GP modeling from the original data (red) and the median error of the RV dataset (orange). The black solid line denotes the predicted relation from Figure~3 of \cite{Chaplin_2019}, stretched and shifted to match our data for illustration purposes. See Section~\ref{subsec:bin} for more details. }
    \label{fig:bin_rms}
\end{figure*}

To better demonstrate the impacts of various binning windows, we tested the binning method on both the full RV data set (on eight nights) and also specifically on Night 27 (the night with the longest coverage and highest RV cadence), and the results are shown in Figure \ref{fig:bin_rms}. We overlaid the predicted relation between the binning window and the RMS shown in Figure~3 from \cite{Chaplin_2019} (black curve) with the results from our binning (gray dots). The horizontal or vertical lines mark the RMS from GP fitting (red), the median photon-limited RV error (orange), the Nyquist limit (light blue dashed line), and the oscillation peak (dark blue dashed line). The Night-27-only plots displayed more data points than the all-RV-used plots because for the intermittent nights (with nightly coverage $\sim$30~min),  binning windows considerably smaller than the nightly coverage length would yield similar RV RMS, leading to fewer distinct values on the plot. 

Our findings indicate that for our dataset on HD~5562, the binning method basically worked as expected, reaching the first minimum in RV RMS at the window length of $1/\nu_{\rm max}$, but this minimum RV RMS is larger than that of the GP modeling. Additionally, for relatively short total exposures (binning windows) not exceeding an hour per night, which are common in exoplanet detection, GP fitting is comparable or even outperforming the binning method in reducing the RV RMS, although it tends to overfit, as evidenced by the GP RMS being considerably lower than the photon limited error (but again, this does not seem to bias the planet detections, as described in the previous subsection). We speculate that this overfitting might improve if each mode of the stellar oscillation can be resolved and accounted for, leading to more precise and realistic GP fitting results. 

In conclusion, with sufficient sampling, GP modeling seems to be more advantageous than binning in reducing the RV RMS. In addition, GP modeling could also provide asteroseismic characterization of the star itself (even without the light curve), which is another advantage.

\section{Conclusion and Future Work}
In this work, we tested using GP regression to mitigate stellar jitter caused by stellar oscillation and granulation in RVs, with simultaneous photometry data from TESS providing informative priors. We conclude that: 
\begin{itemize}
    \item Similar to using GP to mitigate stellar magnetic activity, we found that when modeling asteroseismic signals, using GP to fit photometry data to give informative priors for RV GP fitting can help improve the RV precision for exoplanet detection. In our data of HD~5562, GP reduced the RV scatter from 2.03 to 0.51~m/s.
    \item With high cadence (sufficient to sample oscillations above Nyquist) and relatively long-baseline RV coverage ($\sim$1 night), GP can provide good constraints on the asteroseismic parameters and achieve effective stellar jitter mitigation using the RV data alone, i.e., without the help of photometric data.
    \item Our injection-recovery tests of planet detection indicate that GP modeling of stellar asteroseismic jitter has the potential to help discover smaller exoplanets around evolved stars without biasing the RV semi-amplitudes. 
    \item Compared with the binning method, GP can be just as effective or even outperform binning in terms of reducing the RV RMS. Although it might overfit the RVs, it could provide useful asteroseismic information on the star for stellar characterization.
\end{itemize}

Regarding future missions, an important question is that if we want to get by on intermittent RV data as part of a survey (e.g., HWO precursor survey), would these data benefit significantly if each star in the survey also had one night of high-quality RVs to characterize oscillations and granulation on a star-by-star basis? Our preliminary results suggest that a single night of high-cadence RVs or accurate stellar parameters may already suffice to constrain oscillation signals, whereas modeling granulation likely benefits from longer photometric baselines. 

The future goals of the RV$\times$TESS project include further testing the jitter reduction techniques for both evolved stars and dwarfs, as well as application to planet-hosting stars with significant asteroseismic signals that prevent the detection or confirmation of small planets, especially the transiting ones. Additionally, we aim to explore other correlations between the RV and photometric asteroseismic signals, including the amplitude scaling relation between RV and light curve variations \citep[e.g.,][]{Kjeldsen&Bedding_1995,Guo_2022}, or the phase effects between these two channels \citep[e.g.,][]{Jimenez_1999,Beck_2020}. By refining our understanding of these relationships and improving our methodologies of more accurate modeling to the stellar asteroseismic signals, we hope to enhance the sensitivity of RV surveys and, consequently, increase the chances of discovering new exoplanets, including those that may be habitable.

\section{Acknowledgments}
JT and SXW acknowledge support from the Tsinghua Dushi Fund (53121000123, 53121000124).
TRB acknowledges support from the Australian Research Council (FL220100117).
JAB acknowledges that part of this research was carried out at the Jet Propulsion Laboratory, California Institute of Technology, under a contract with the National Aeronautics and Space Administration (NASA).
We acknowledge the use of public TESS data from the pipelines at the TESS Science Office at the TESS Science Processing Operations Center. The computational resources supporting this research were provided by the NASA High-End Computing (HEC) Program through the NASA Advanced Supercomputing (NAS) Division at Ames Research Center, which facilitated the production of the SPOC data products. The data included in this paper were collected as part of the TESS mission and obtained from the MAST data archive at the Space Telescope Science Institute (STScI). The TESS mission is funded by the NASA Explorer Program. STScI is operated by the Association of Universities for Research in Astronomy, Inc., under NASA contract NAS 5–26555. The TESS data utilized in this paper are available in MAST: \href{https://archive.stsci.edu/doi/resolve/resolve.html?doi=10.17909/gvpc-0504}{10.17909/gvpc-0504}.

This research has made use of the SIMBAD database, CDS, Strasbourg Astronomical Observatory, France \citep{Wenger_2000_SIMBAD}.

We thank the anonymous referee for a thorough and constructive report that helped to improve the quality of this paper.

SIMBAD, VizieR, ADS

\vspace{5mm}

\appendix

\section{Plots of the Posterior Distributions and RV-only Orbital Fit Parameters}


\begin{figure*}
    \centering
	\includegraphics[width=\textwidth]{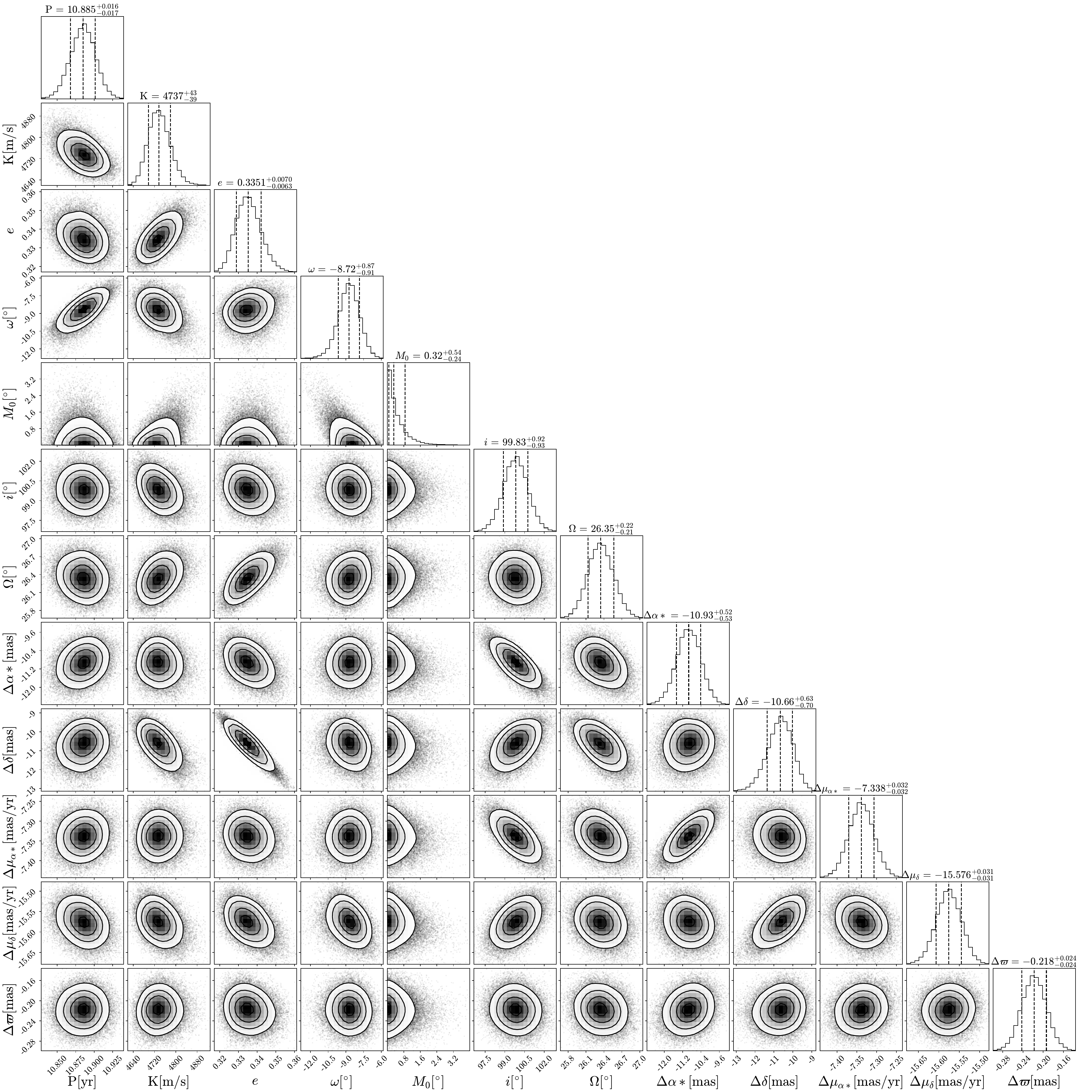}
    \caption{Posterior distributions for selected parameters from the combined analysis of RV and Hipparcos-Gaia astrometry. The median and the corresponding the $1\sigma$ confidence intervals are denoted by vertical dashed lines.}
    \label{fig:rv_ast_corner}
\end{figure*}

\begin{table*}[htbp]
    \centering
    \renewcommand{\arraystretch}{1.4}
    \caption{The prior settings, best-fit values, and the 68\% credible intervals of the companion of HD~5562. $\mathcal{N}$($\mu\ ,\ \sigma^{2}$) refers to normal priors with mean $\mu$ and standard deviation $\sigma$. $\mathcal{U}$(a\ ,\ b) refers to uniform priors in the range a--b. $\mathcal{LU}$(a\ ,\ b) signifies log uniform priors ranging from a - b.}
    \label{tab:rvorbitfit}
    \begin{tabular}{lccr}
        \hline\hline
        Parameter & Prior & Best fit & Description \\
        \hline
        \textit{Companion's parameters}\\
        $P$ (days) & $\mathcal{N}$ (4000\ ,\ 300) & $4066.39^{+18.59}_{-20.97}$ & Orbital period of the companion. \\
        $t_0$ (days) & $\mathcal{U}$ (2452000\ ,\ 2455000) & $2453109.72^{27.73}_{22.93}$ & Mid-transit time of the companion.\\
        $e$ & $\mathcal{U}$ (0.1\ ,\ 0.9) &  $0.3120^{+0.0113}_{-0.0099}$ & Eccentricity of the orbit.\\
        $\omega$ (deg) & $\mathcal{U}$ (5\ ,\ 175) & $6.4951^{+1.4388}_{-0.8850}$ & Argument of periastron passage. \\
        $M_b \sin i$ (M$_\odot$) & ... & $0.4108^{+0.0141}_{-0.0119}$ & Mass of the companion.\\
        $a$ (AU) & ...  & $1.4191^{+0.0356}_{-0.0342}$ & Semi-major axis of the companion. \\
        \hline
        \textit{RV parameters} \\
        $K$ (m~s$^{-1}$) & $\mathcal{U}$ (0\ ,\ 8000) & $4576.77^{+135.86}_{-116.23}$ & RV semi-amplitude of the orbit.\\
        \hline
        \textit{AAT and PFS parameters} \\
        $\mu_{AAT}$ (m~s$^{-1}$) & $\mathcal{U}$ (-4000\ ,\ 4000) & $-2318.17^{+130.39}_{-130.91}$ & Systemic offset for AAT.  \\
        $\mu_{PFS}$ (m~s$^{-1}$) & $\mathcal{U}$ (-4000\ ,\ 4000) & $3116.68^{+58.02}_{-66.64}$ & Systemic offset for PFS.\\
        $\sigma_{AAT}$ (m~s$^{-1}$) & $\mathcal{LU}$ (10$^{-2}$\ ,\  1000) & $428.52^{+141.18}_{-95.67}$ & Extra jitter term for AAT.\\
        $\sigma_{PFS}$ (m~s$^{-1}$) & $\mathcal{LU}$ (10$^{-2}$\ ,\  1000) & $1.22^{+0.65}_{-0.37}$ & Extra jitter term for PFS.\\
        \hline\hline
    \end{tabular}
\end{table*}

\begin{table}
    \centering
    \caption{RV data of HD~5562}
    \label{tab:RV_all}
    \begin{tabular}{lcccccc}
    \hline\hline
    JD 	&	RV (m/s) & $\sigma_{\rm RV}$ (m/s) & $S_{HK}$ & $\rm H\alpha$ & Photon counts & Exposure time (s)\\
    \hline 
    2458332.66318	&14.83	&0.92	&0.1576	&0.03303 &8043 &361 \\ 
    2458332.66975   &15.44  &0.99   &0.1497 &0.03295 &7169 &421 \\
    2458332.67551   &14.07  &0.91   &0.1384 &0.03289 &8679 &420 \\
    2458332.68104   &15.29  &0.93   &0.1422 &0.03288 &7079 &420 \\
    2458332.68664   &14.99  &0.92   &0.1384 &0.03290 &9020 &420 \\
    2458332.69234   &14.87  &0.86   &0.1381 &0.03304 &9724 &420 \\	 	
    ...&...&...&...&...\\
    \hline
    \end{tabular}
\tablecomments{This table is published in its entirety in the machine-readable format. A portion is shown here for guidance regarding its form and content.}
\end{table}

\begin{figure*}
    \centering
	\includegraphics[width=\textwidth]{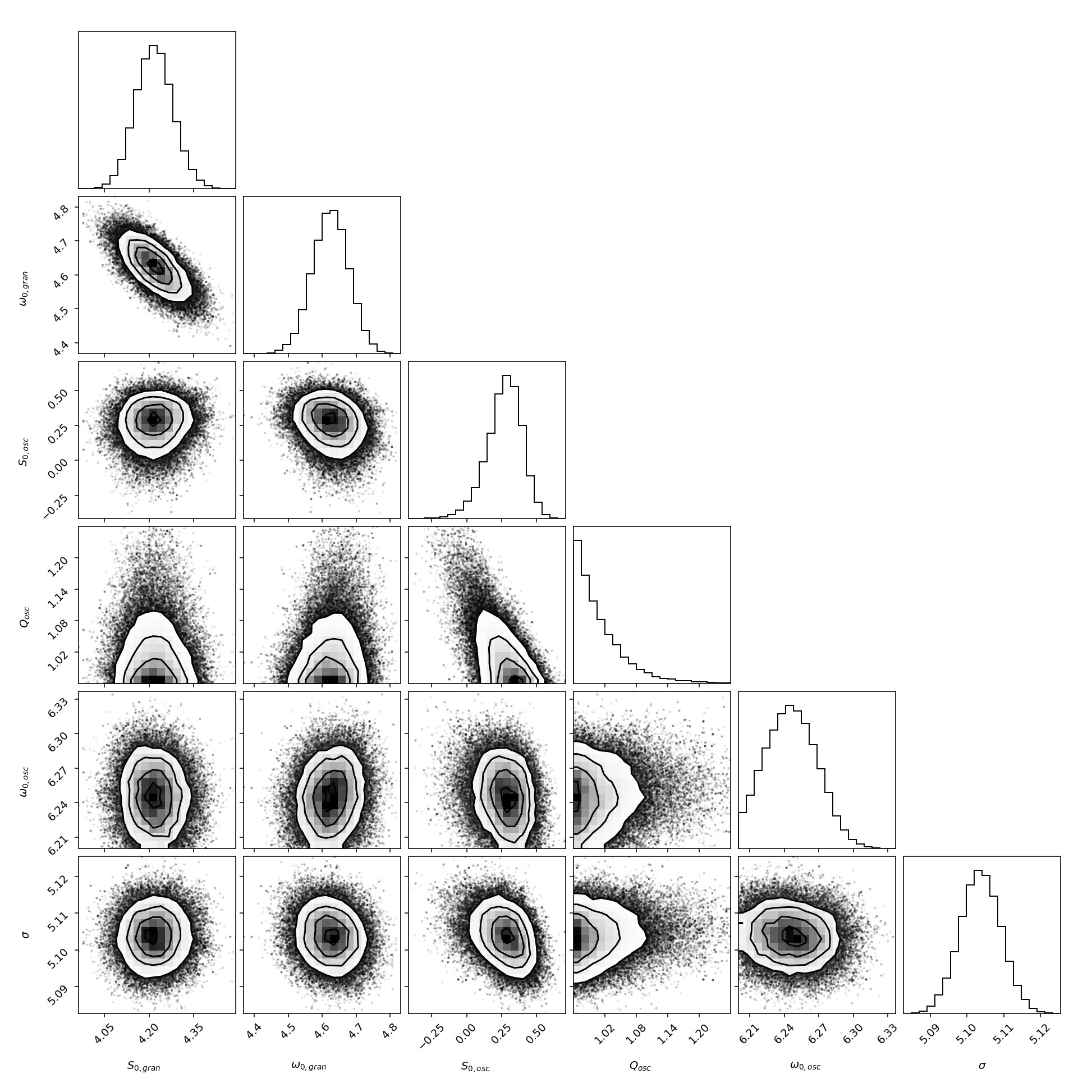}
    \caption{Posterior distributions for hyperparameters of light curve GP fit. All parameters are log values. $S_0$ is the amplitude, $Q$ is the quality factor, and $\sigma$ is the jitter term. Term belongings are indicated in the subscripts.}
    \label{fig:lc_mcmc}
\end{figure*}

\begin{figure*}
    \centering
	\includegraphics[width=\textwidth]{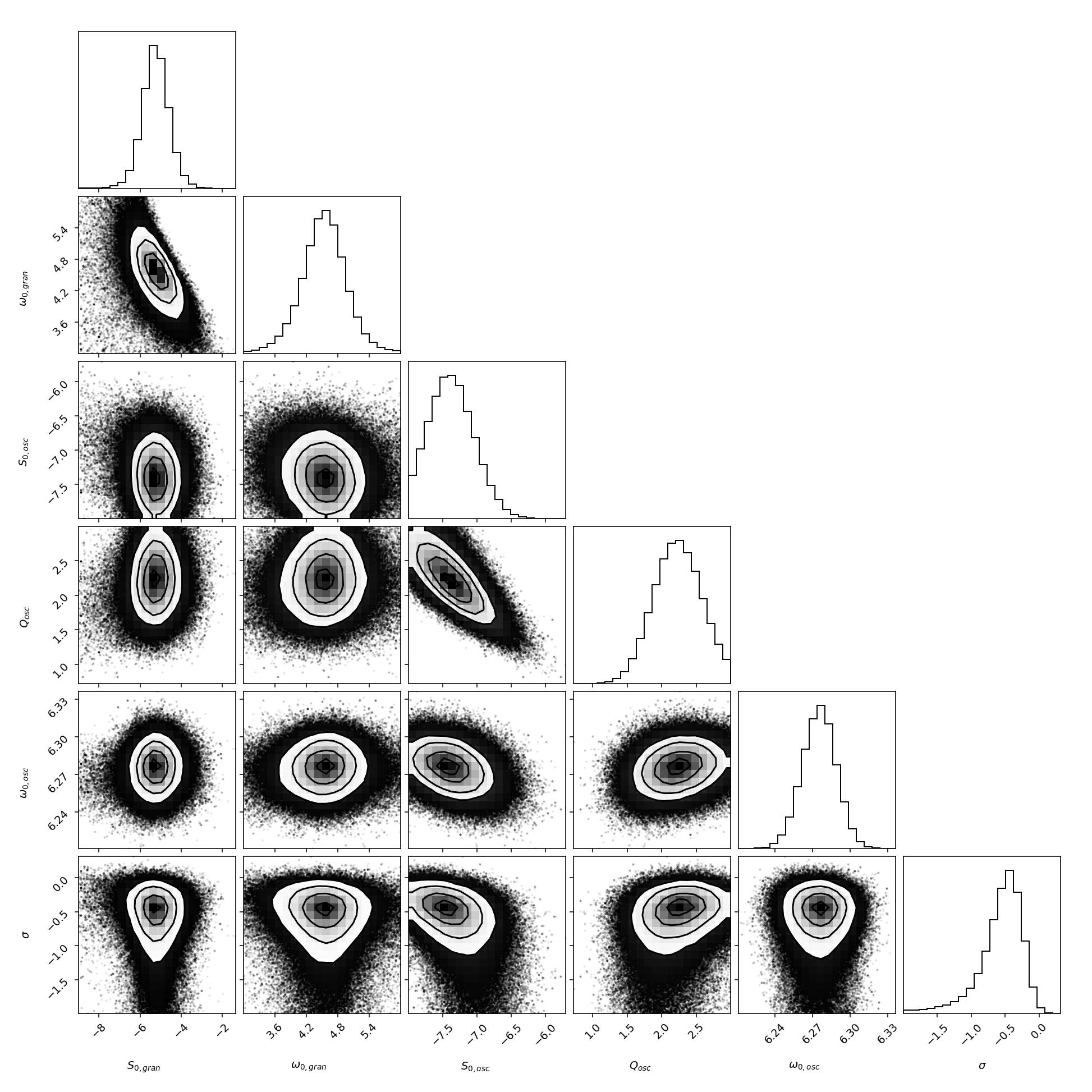}
    \caption{Posterior distributions for hyperparameters of RV GP fit. All parameters are log values. $S_0$ is the amplitude, $Q$ is the quality factor, and $\sigma$ is the jitter term. Term belongings are indicated in the subscripts.}
    \label{fig:rv_mcmc}
\end{figure*}


\bibliography{references}{}

@ARTICLE{Rajpaul2015,
       author = {{Rajpaul}, V. and {Aigrain}, S. and {Osborne}, M.~A. and {Reece}, S. and {Roberts}, S.},
        title = "{A Gaussian process framework for modelling stellar activity signals in radial velocity data}",
      journal = {\mnras},
         year = 2015,
        month = sep,
       volume = {452},
       number = {3},
        pages = {2269-2291},
          doi = {10.1093/mnras/stv1428},
archivePrefix = {arXiv},
       eprint = {1506.07304},
 primaryClass = {astro-ph.EP},
       adsurl = {https://ui.adsabs.harvard.edu/abs/2015MNRAS.452.2269R}
}

@ARTICLE{Pepe_2021,
       author = {{Pepe}, F. and {Cristiani}, S. and {Rebolo}, R. and {Santos}, N.~C. and {Dekker}, H. and {Cabral}, A. and {Di Marcantonio}, P. and {Figueira}, P. and {Lo Curto}, G. and {Lovis}, C. and {Mayor}, M. and {M{\'e}gevand}, D. and {Molaro}, P. and {Riva}, M. and {Zapatero Osorio}, M.~R. and {Amate}, M. and {Manescau}, A. and {Pasquini}, L. and {Zerbi}, F.~M. and {Adibekyan}, V. and {Abreu}, M. and {Affolter}, M. and {Alibert}, Y. and {Aliverti}, M. and {Allart}, R. and {Allende Prieto}, C. and {{\'A}lvarez}, D. and {Alves}, D. and {Avila}, G. and {Baldini}, V. and {Bandy}, T. and {Barros}, S.~C.~C. and {Benz}, W. and {Bianco}, A. and {Borsa}, F. and {Bourrier}, V. and {Bouchy}, F. and {Broeg}, C. and {Calderone}, G. and {Cirami}, R. and {Coelho}, J. and {Conconi}, P. and {Coretti}, I. and {Cumani}, C. and {Cupani}, G. and {D'Odorico}, V. and {Damasso}, M. and {Deiries}, S. and {Delabre}, B. and {Demangeon}, O.~D.~S. and {Dumusque}, X. and {Ehrenreich}, D. and {Faria}, J.~P. and {Fragoso}, A. and {Genolet}, L. and {Genoni}, M. and {G{\'e}nova Santos}, R. and {Gonz{\'a}lez Hern{\'a}ndez}, J.~I. and {Hughes}, I. and {Iwert}, O. and {Kerber}, F. and {Knudstrup}, J. and {Landoni}, M. and {Lavie}, B. and {Lillo-Box}, J. and {Lizon}, J. -L. and {Maire}, C. and {Martins}, C.~J.~A.~P. and {Mehner}, A. and {Micela}, G. and {Modigliani}, A. and {Monteiro}, M.~A. and {Monteiro}, M.~J.~P.~F.~G. and {Moschetti}, M. and {Murphy}, M.~T. and {Nunes}, N. and {Oggioni}, L. and {Oliveira}, A. and {Oshagh}, M. and {Pall{\'e}}, E. and {Pariani}, G. and {Poretti}, E. and {Rasilla}, J.~L. and {Rebord{\~a}o}, J. and {Redaelli}, E.~M. and {Santana Tschudi}, S. and {Santin}, P. and {Santos}, P. and {S{\'e}gransan}, D. and {Schmidt}, T.~M. and {Segovia}, A. and {Sosnowska}, D. and {Sozzetti}, A. and {Sousa}, S.~G. and {Span{\`o}}, P. and {Su{\'a}rez Mascare{\~n}o}, A. and {Tabernero}, H. and {Tenegi}, F. and {Udry}, S. and {Zanutta}, A.},
        title = "{ESPRESSO at VLT. On-sky performance and first results}",
      journal = {\aap},
         year = 2021,
        month = jan,
       volume = {645},
          eid = {A96},
        pages = {A96},
          doi = {10.1051/0004-6361/202038306},
archivePrefix = {arXiv},
       eprint = {2010.00316},
 primaryClass = {astro-ph.IM},
       adsurl = {https://ui.adsabs.harvard.edu/abs/2021A&A...645A..96P}
}

@INPROCEEDINGS{Jurgenson_2016,
       author = {{Jurgenson}, C. and {Fischer}, D. and {McCracken}, T. and {Sawyer}, D. and {Szymkowiak}, A. and {Davis}, A. and {Muller}, G. and {Santoro}, F.},
        title = "{EXPRES: a next generation RV spectrograph in the search for earth-like worlds}",
    booktitle = {Ground-based and Airborne Instrumentation for Astronomy VI},
         year = 2016,
       editor = {{Evans}, Christopher J. and {Simard}, Luc and {Takami}, Hideki},
       series = {Society of Photo-Optical Instrumentation Engineers (SPIE) Conference Series},
       volume = {9908},
        month = aug,
          eid = {99086T},
        pages = {99086T},
          doi = {10.1117/12.2233002},
archivePrefix = {arXiv},
       eprint = {1606.04413},
 primaryClass = {astro-ph.IM},
       adsurl = {https://ui.adsabs.harvard.edu/abs/2016SPIE.9908E..6TJ}
}

@INPROCEEDINGS{Schwab_2016,
       author = {{Schwab}, C. and {Rakich}, A. and {Gong}, Q. and {Mahadevan}, S. and {Halverson}, S.~P. and {Roy}, A. and {Terrien}, R.~C. and {Robertson}, P.~M. and {Hearty}, F.~R. and {Levi}, E.~I. and {Monson}, A.~J. and {Wright}, J.~T. and {McElwain}, M.~W. and {Bender}, C.~F. and {Blake}, C.~H. and {St{\"u}rmer}, J. and {Gurevich}, Y.~V. and {Chakraborty}, A. and {Ramsey}, L.~W.},
        title = "{Design of NEID, an extreme precision Doppler spectrograph for WIYN}",
    booktitle = {Ground-based and Airborne Instrumentation for Astronomy VI},
         year = 2016,
       editor = {{Evans}, Christopher J. and {Simard}, Luc and {Takami}, Hideki},
       series = {Society of Photo-Optical Instrumentation Engineers (SPIE) Conference Series},
       volume = {9908},
        month = aug,
          eid = {99087H},
        pages = {99087H},
          doi = {10.1117/12.2234411},
       adsurl = {https://ui.adsabs.harvard.edu/abs/2016SPIE.9908E..7HS}
}

@ARTICLE{Crass_2021,
       author = {{Crass}, Jonathan and {Gaudi}, B. Scott and {Leifer}, Stephanie and {Beichman}, Charles and {Bender}, Chad and {Blackwood}, Gary and {Burt}, Jennifer A. and {Callas}, John L. and {Cegla}, Heather M. and {Diddams}, Scott A. and {Dumusque}, Xavier and {Eastman}, Jason D. and {Ford}, Eric B. and {Fulton}, Benjamin and {Gibson}, Rose and {Halverson}, Samuel and {Haywood}, Rapha{\"e}lle D. and {Hearty}, Fred and {Howard}, Andrew W. and {Latham}, David W. and {L{\"o}hner-B{\"o}ttcher}, Johannes and {Mamajek}, Eric E. and {Mortier}, Annelies and {Newman}, Patrick and {Plavchan}, Peter and {Quirrenbach}, Andreas and {Reiners}, Ansgar and {Robertson}, Paul and {Roy}, Arpita and {Schwab}, Christian and {Seifahrt}, Andres and {Szentgyorgyi}, Andy and {Terrien}, Ryan and {Teske}, Johanna K. and {Thompson}, Samantha and {Vasisht}, Gautam},
        title = "{Extreme Precision Radial Velocity Working Group Final Report}",
      journal = {arXiv e-prints},
         year = 2021,
        month = jul,
          eid = {arXiv:2107.14291},
        pages = {arXiv:2107.14291},
          doi = {10.48550/arXiv.2107.14291},
archivePrefix = {arXiv},
       eprint = {2107.14291},
 primaryClass = {astro-ph.IM},
       adsurl = {https://ui.adsabs.harvard.edu/abs/2021arXiv210714291C}
}

@ARTICLE{Newman_2023,
       author = {{Newman}, Patrick D. and {Plavchan}, Peter and {Burt}, Jennifer A. and {Teske}, Johanna and {Mamajek}, Eric E. and {Leifer}, Stephanie and {Gaudi}, B. Scott and {Blackwood}, Gary and {Morgan}, Rhonda},
        title = "{Simulations for Planning Next-generation Exoplanet Radial Velocity Surveys}",
      journal = {\aj},
         year = 2023,
        month = apr,
       volume = {165},
       number = {4},
          eid = {151},
        pages = {151},
          doi = {10.3847/1538-3881/acad07},
archivePrefix = {arXiv},
       eprint = {2204.13968},
 primaryClass = {astro-ph.EP},
       adsurl = {https://ui.adsabs.harvard.edu/abs/2023AJ....165..151N}
}

@ARTICLE{Hall_2018,
       author = {{Hall}, Richard D. and {Thompson}, Samantha J. and {Handley}, Will and {Queloz}, Didier},
        title = "{On the Feasibility of Intense Radial Velocity Surveys for Earth-Twin Discoveries}",
      journal = {\mnras},
         year = 2018,
        month = sep,
       volume = {479},
       number = {3},
        pages = {2968-2987},
          doi = {10.1093/mnras/sty1464},
archivePrefix = {arXiv},
       eprint = {1806.00518},
 primaryClass = {astro-ph.EP},
       adsurl = {https://ui.adsabs.harvard.edu/abs/2018MNRAS.479.2968H}
}

@ARTICLE{Brewer_2020,
       author = {{Brewer}, John M. and {Fischer}, Debra A. and {Blackman}, Ryan T. and {Cabot}, Samuel H.~C. and {Davis}, Allen B. and {Laughlin}, Gregory and {Leet}, Christopher and {Ong}, J.~M. Joel and {Petersburg}, Ryan R. and {Szymkowiak}, Andrew E. and {Zhao}, Lily L. and {Henry}, Gregory W. and {Llama}, Joe},
        title = "{EXPRES. I. HD 3651 as an Ideal RV Benchmark}",
      journal = {\aj},
         year = 2020,
        month = aug,
       volume = {160},
       number = {2},
          eid = {67},
        pages = {67},
          doi = {10.3847/1538-3881/ab99c9},
archivePrefix = {arXiv},
       eprint = {2006.02303},
 primaryClass = {astro-ph.EP},
       adsurl = {https://ui.adsabs.harvard.edu/abs/2020AJ....160...67B}
}

@ARTICLE{Gupta_2021,
       author = {{Gupta}, Arvind F. and {Wright}, Jason T. and {Robertson}, Paul and {Halverson}, Samuel and {Luhn}, Jacob and {Roy}, Arpita and {Mahadevan}, Suvrath and {Ford}, Eric B. and {Bender}, Chad F. and {Blake}, Cullen H. and {Hearty}, Fred and {Kanodia}, Shubham and {Logsdon}, Sarah E. and {McElwain}, Michael W. and {Monson}, Andrew and {Ninan}, Joe P. and {Schwab}, Christian and {Stef{\'a}nsson}, Gu{\dh}mundur and {Terrien}, Ryan C.},
        title = "{Target Prioritization and Observing Strategies for the NEID Earth Twin Survey}",
      journal = {\aj},
         year = 2021,
        month = mar,
       volume = {161},
       number = {3},
          eid = {130},
        pages = {130},
          doi = {10.3847/1538-3881/abd79e},
archivePrefix = {arXiv},
       eprint = {2101.11689},
 primaryClass = {astro-ph.EP},
       adsurl = {https://ui.adsabs.harvard.edu/abs/2021AJ....161..130G}
}

@ARTICLE{Hojjatpanah_2019,
       author = {{Hojjatpanah}, S. and {Figueira}, P. and {Santos}, N.~C. and {Adibekyan}, V. and {Sousa}, S.~G. and {Delgado-Mena}, E. and {Alibert}, Y. and {Cristiani}, S. and {Gonz{\'a}lez Hern{\'a}ndez}, J.~I. and {Lanza}, A.~F. and {Di Marcantonio}, P. and {Martins}, J.~H.~C. and {Micela}, G. and {Molaro}, P. and {Neves}, V. and {Oshagh}, M. and {Pepe}, F. and {Poretti}, E. and {Rojas-Ayala}, B. and {Rebolo}, R. and {Su{\'a}rez Mascare{\~n}o}, A. and {Zapatero Osorio}, M.~R.},
        title = "{Catalog for the ESPRESSO blind radial velocity exoplanet survey}",
      journal = {\aap},
         year = 2019,
        month = sep,
       volume = {629},
          eid = {A80},
        pages = {A80},
          doi = {10.1051/0004-6361/201834729},
archivePrefix = {arXiv},
       eprint = {1908.04627},
 primaryClass = {astro-ph.EP},
       adsurl = {https://ui.adsabs.harvard.edu/abs/2019A&A...629A..80H}
}

@ARTICLE{Luhn_2023,
       author = {{Luhn}, Jacob K. and {Ford}, Eric B. and {Guo}, Zhao and {Gilbertson}, Christian and {Newman}, Patrick and {Plavchan}, Peter and {Burt}, Jennifer A. and {Teske}, Johanna and {Gupta}, Arvind F.},
        title = "{Impact of Correlated Noise on the Mass Precision of Earth-analog Planets in Radial Velocity Surveys}",
      journal = {\aj},
         year = 2023,
        month = mar,
       volume = {165},
       number = {3},
          eid = {98},
        pages = {98},
          doi = {10.3847/1538-3881/acad08},
archivePrefix = {arXiv},
       eprint = {2204.12512},
 primaryClass = {astro-ph.EP},
       adsurl = {https://ui.adsabs.harvard.edu/abs/2023AJ....165...98L}
}

@ARTICLE{Gupta_2024,
       author = {{Gupta}, Arvind F. and {Bedell}, Megan},
        title = "{Fishing for Planets: A Comparative Analysis of EPRV Survey Performance in the Presence of Correlated Noise}",
      journal = {\aj},
         year = 2024,
        month = jul,
       volume = {168},
       number = {1},
          eid = {29},
        pages = {29},
          doi = {10.3847/1538-3881/ad4ce6},
archivePrefix = {arXiv},
       eprint = {2303.14571},
 primaryClass = {astro-ph.EP},
       adsurl = {https://ui.adsabs.harvard.edu/abs/2024AJ....168...29G}
}

@ARTICLE{Gupta_2022,
       author = {{Gupta}, Arvind F. and {Luhn}, Jacob and {Wright}, Jason T. and {Mahadevan}, Suvrath and {Ford}, Eric B. and {Stef{\'a}nsson}, Gu{\dj}mundur and {Bender}, Chad F. and {Blake}, Cullen H. and {Halverson}, Samuel and {Hearty}, Fred and {Kanodia}, Shubham and {Logsdon}, Sarah E. and {McElwain}, Michael W. and {Ninan}, Joe P. and {Robertson}, Paul and {Roy}, Arpita and {Schwab}, Christian and {Terrien}, Ryan C.},
        title = "{Detection of p-mode Oscillations in HD 35833 with NEID and TESS}",
      journal = {\aj},
         year = 2022,
        month = dec,
       volume = {164},
       number = {6},
          eid = {254},
        pages = {254},
          doi = {10.3847/1538-3881/ac96f3},
archivePrefix = {arXiv},
       eprint = {2210.00544},
 primaryClass = {astro-ph.SR},
       adsurl = {https://ui.adsabs.harvard.edu/abs/2022AJ....164..254G}
}

@ARTICLE{Mathur_2019,
       author = {{Mathur}, Savita and {Garc{\'\i}a}, Rafael A. and {Bugnet}, Lisa and {Santos}, {\^A}ngela R.~G. and {Santiago}, Netsha and {Beck}, Paul G.},
        title = "{Revisiting the impact of stellar magnetic activity on the detection of solar-like oscillations by Kepler}",
      journal = {Frontiers in Astronomy and Space Sciences},
         year = 2019,
        month = jul,
       volume = {6},
          eid = {46},
        pages = {46},
          doi = {10.3389/fspas.2019.00046},
archivePrefix = {arXiv},
       eprint = {1907.01415},
 primaryClass = {astro-ph.SR},
       adsurl = {https://ui.adsabs.harvard.edu/abs/2019FrASS...6...46M}
}

@ARTICLE{Brewer_2009,
       author = {{Brewer}, B.~J. and {Stello}, D.},
        title = "{Gaussian process modelling of asteroseismic data}",
      journal = {\mnras},
         year = 2009,
        month = jun,
       volume = {395},
       number = {4},
        pages = {2226-2233},
          doi = {10.1111/j.1365-2966.2009.14679.x},
archivePrefix = {arXiv},
       eprint = {0902.3907},
 primaryClass = {astro-ph.SR},
       adsurl = {https://ui.adsabs.harvard.edu/abs/2009MNRAS.395.2226B}
}

@ARTICLE{Hey_2024,
       author = {{Hey}, Daniel and {Huber}, Daniel and {Ong}, Joel and {Stello}, Dennis and {Foreman-Mackey}, Daniel},
        title = "{Precise Time-Domain Asteroseismology and a Revised Target List for TESS Solar-Like Oscillators}",
      journal = {arXiv e-prints},
         year = 2024,
        month = mar,
          eid = {arXiv:2403.02489},
        pages = {arXiv:2403.02489},
          doi = {10.48550/arXiv.2403.02489},
archivePrefix = {arXiv},
       eprint = {2403.02489},
 primaryClass = {astro-ph.SR},
       adsurl = {https://ui.adsabs.harvard.edu/abs/2024arXiv240302489H}
}

@ARTICLE{Grunblatt_2015,
       author = {{Grunblatt}, Samuel K. and {Howard}, Andrew W. and {Haywood}, Rapha{\"e}lle D.},
        title = "{Determining the Mass of Kepler-78b with Nonparametric Gaussian Process Estimation}",
      journal = {\apj},
         year = 2015,
        month = aug,
       volume = {808},
       number = {2},
          eid = {127},
        pages = {127},
          doi = {10.1088/0004-637X/808/2/127},
archivePrefix = {arXiv},
       eprint = {1501.00369},
 primaryClass = {astro-ph.EP},
       adsurl = {https://ui.adsabs.harvard.edu/abs/2015ApJ...808..127G}
}

@ARTICLE{Grunblatt_2016,
       author = {{Grunblatt}, Samuel K. and {Huber}, Daniel and {Gaidos}, Eric J. and {Lopez}, Eric D. and {Fulton}, Benjamin J. and {Vanderburg}, Andrew and {Barclay}, Thomas and {Fortney}, Jonathan J. and {Howard}, Andrew W. and {Isaacson}, Howard T. and {Mann}, Andrew W. and {Petigura}, Erik and {Silva Aguirre}, Victor and {Sinukoff}, Evan J.},
        title = "{K2-97b: A (Re-?)Inflated Planet Orbiting a Red Giant Star}",
      journal = {\aj},
         year = 2016,
        month = dec,
       volume = {152},
       number = {6},
          eid = {185},
        pages = {185},
          doi = {10.3847/0004-6256/152/6/185},
archivePrefix = {arXiv},
       eprint = {1606.05818},
 primaryClass = {astro-ph.EP},
       adsurl = {https://ui.adsabs.harvard.edu/abs/2016AJ....152..185G}
}

@INPROCEEDINGS{Gibson_2016,
       author = {{Gibson}, Steven R. and {Howard}, Andrew W. and {Marcy}, Geoffrey W. and {Edelstein}, Jerry and {Wishnow}, Edward H. and {Poppett}, Claire L.},
        title = "{KPF: Keck Planet Finder}",
    booktitle = {Ground-based and Airborne Instrumentation for Astronomy VI},
         year = 2016,
       editor = {{Evans}, Christopher J. and {Simard}, Luc and {Takami}, Hideki},
       series = {Society of Photo-Optical Instrumentation Engineers (SPIE) Conference Series},
       volume = {9908},
        month = aug,
          eid = {990870},
        pages = {990870},
          doi = {10.1117/12.2233334},
       adsurl = {https://ui.adsabs.harvard.edu/abs/2016SPIE.9908E..70G}
}

@ARTICLE{Husser_2013,
       author = {{Husser}, T. -O. and {Wende-von Berg}, S. and {Dreizler}, S. and {Homeier}, D. and {Reiners}, A. and {Barman}, T. and {Hauschildt}, P.~H.},
        title = "{A new extensive library of PHOENIX stellar atmospheres and synthetic spectra}",
      journal = {\aap},
         year = 2013,
        month = may,
       volume = {553},
          eid = {A6},
        pages = {A6},
          doi = {10.1051/0004-6361/201219058},
archivePrefix = {arXiv},
       eprint = {1303.5632},
 primaryClass = {astro-ph.SR},
       adsurl = {https://ui.adsabs.harvard.edu/abs/2013A&A...553A...6H}
}

@ARTICLE{Kurucz_1993,
       author = {{Kurucz}, Robert},
        title = "{ATLAS9 Stellar Atmosphere Programs and 2 km/s grid.}",
      journal = {Robert Kurucz CD-ROM},
         year = 1993,
        month = jan,
       volume = {13},
       adsurl = {https://ui.adsabs.harvard.edu/abs/1993KurCD..13.....K}
}

@ARTICLE{Arriagada_2011,
       author = {{Arriagada}, Pamela},
        title = "{Chromospheric Activity of Southern Stars from the Magellan Planet Search Program}",
      journal = {\apj},
         year = 2011,
        month = jun,
       volume = {734},
       number = {1},
          eid = {70},
        pages = {70},
          doi = {10.1088/0004-637X/734/1/70},
archivePrefix = {arXiv},
       eprint = {1104.3186},
 primaryClass = {astro-ph.EP},
       adsurl = {https://ui.adsabs.harvard.edu/abs/2011ApJ...734...70A}
}

@ARTICLE{Palmer_2009,
       author = {{Palmer}, David M.},
        title = "{A Fast Chi-Squared Technique for Period Search of Irregularly Sampled Data}",
      journal = {\apj},
         year = 2009,
        month = apr,
       volume = {695},
       number = {1},
        pages = {496-502},
          doi = {10.1088/0004-637X/695/1/496},
archivePrefix = {arXiv},
       eprint = {0901.1913},
 primaryClass = {physics.data-an},
       adsurl = {https://ui.adsabs.harvard.edu/abs/2009ApJ...695..496P}
}

@ARTICLE{Diaz_2018,
       author = {{D{\'\i}az}, Mat{\'\i}as R. and {Jenkins}, James S. and {Tuomi}, Mikko and {Butler}, R. Paul and {Soto}, Maritza G. and {Teske}, Johanna K. and {Feng}, Fabo and {Shectman}, Stephen A. and {Arriagada}, Pamela and {Crane}, Jeffrey D. and {Thompson}, Ian B. and {Vogt}, Steven S.},
        title = "{The Test Case of HD 26965: Difficulties Disentangling Weak Doppler Signals from Stellar Activity}",
      journal = {\aj},
         year = 2018,
        month = mar,
       volume = {155},
       number = {3},
          eid = {126},
        pages = {126},
          doi = {10.3847/1538-3881/aaa896},
archivePrefix = {arXiv},
       eprint = {1801.03970},
 primaryClass = {astro-ph.EP},
       adsurl = {https://ui.adsabs.harvard.edu/abs/2018AJ....155..126D}
}

@ARTICLE{Bauer_2018,
       author = {{Bauer}, F.~F. and {Reiners}, A. and {Beeck}, B. and {Jeffers}, S.~V.},
        title = "{The influence of convective blueshift on radial velocities of F, G, and K stars}",
      journal = {\aap},
         year = 2018,
        month = feb,
       volume = {610},
          eid = {A52},
        pages = {A52},
          doi = {10.1051/0004-6361/201731227},
       adsurl = {https://ui.adsabs.harvard.edu/abs/2018A&A...610A..52B}
}

@ARTICLE{Meunier_2010,
       author = {{Meunier}, N. and {Desort}, M. and {Lagrange}, A. -M.},
        title = "{Using the Sun to estimate Earth-like planets detection capabilities . II. Impact of plages}",
      journal = {\aap},
         year = 2010,
        month = mar,
       volume = {512},
          eid = {A39},
        pages = {A39},
          doi = {10.1051/0004-6361/200913551},
archivePrefix = {arXiv},
       eprint = {1001.1638},
 primaryClass = {astro-ph.EP},
       adsurl = {https://ui.adsabs.harvard.edu/abs/2010A&A...512A..39M}
}

@ARTICLE{Lanza_2010,
       author = {{Lanza}, A.~F. and {Bonomo}, A.~S. and {Moutou}, C. and {Pagano}, I. and {Messina}, S. and {Leto}, G. and {Cutispoto}, G. and {Aigrain}, S. and {Alonso}, R. and {Barge}, P. and {Deleuil}, M. and {Auvergne}, M. and {Baglin}, A. and {Collier Cameron}, A.},
        title = "{Photospheric activity, rotation, and radial velocity variations of the planet-hosting star CoRoT-7}",
      journal = {\aap},
         year = 2010,
        month = sep,
       volume = {520},
          eid = {A53},
        pages = {A53},
          doi = {10.1051/0004-6361/201014403},
archivePrefix = {arXiv},
       eprint = {1005.3602},
 primaryClass = {astro-ph.SR},
       adsurl = {https://ui.adsabs.harvard.edu/abs/2010A&A...520A..53L}
}

@ARTICLE{Dumusque_2014,
       author = {{Dumusque}, X. and {Boisse}, I. and {Santos}, N.~C.},
        title = "{SOAP 2.0: A Tool to Estimate the Photometric and Radial Velocity Variations Induced by Stellar Spots and Plages}",
      journal = {\apj},
         year = 2014,
        month = dec,
       volume = {796},
       number = {2},
          eid = {132},
        pages = {132},
          doi = {10.1088/0004-637X/796/2/132},
archivePrefix = {arXiv},
       eprint = {1409.3594},
 primaryClass = {astro-ph.SR},
       adsurl = {https://ui.adsabs.harvard.edu/abs/2014ApJ...796..132D}
}

@ARTICLE{Li&Basri_2024,
       author = {{Li}, Canis and {Basri}, Gibor},
        title = "{Do Faculae Affect Autocorrelation Rotation Periods in Sun-like Stars?}",
      journal = {\apj},
         year = 2024,
        month = mar,
       volume = {963},
       number = {2},
          eid = {102},
        pages = {102},
          doi = {10.3847/1538-4357/ad1e59},
archivePrefix = {arXiv},
       eprint = {2401.13003},
 primaryClass = {astro-ph.SR},
       adsurl = {https://ui.adsabs.harvard.edu/abs/2024ApJ...963..102L}
}

@ARTICLE{Barbato_2023,
       author = {{Barbato}, D. and {S{\'e}gransan}, D. and {Udry}, S. and {Unger}, N. and {Bouchy}, F. and {Lovis}, C. and {Mayor}, M. and {Pepe}, F. and {Queloz}, D. and {Santos}, N.~C. and {Delisle}, J.~B. and {Figueira}, P. and {Marmier}, M. and {Matthews}, E.~C. and {Lo Curto}, G. and {Venturini}, J. and {Chaverot}, G. and {Cretignier}, M. and {Otegi}, J.~F. and {Stalport}, M.},
        title = "{The CORALIE survey for southern extrasolar planets. XIX. Brown dwarfs and stellar companions unveiled by radial velocity and astrometry}",
      journal = {\aap},
         year = 2023,
        month = jun,
       volume = {674},
          eid = {A114},
        pages = {A114},
          doi = {10.1051/0004-6361/202345874},
archivePrefix = {arXiv},
       eprint = {2303.16717},
 primaryClass = {astro-ph.EP},
       adsurl = {https://ui.adsabs.harvard.edu/abs/2023A&A...674A.114B}
}

@INPROCEEDINGS{Dekker_2000,
       author = {{Dekker}, Hans and {D'Odorico}, Sandro and {Kaufer}, Andreas and {Delabre}, Bernard and {Kotzlowski}, Heinz},
        title = "{Design, construction, and performance of UVES, the echelle spectrograph for the UT2 Kueyen Telescope at the ESO Paranal Observatory}",
    booktitle = {Optical and IR Telescope Instrumentation and Detectors},
         year = 2000,
       editor = {{Iye}, Masanori and {Moorwood}, Alan F.},
       series = {Society of Photo-Optical Instrumentation Engineers (SPIE) Conference Series},
       volume = {4008},
        month = aug,
        pages = {534-545},
          doi = {10.1117/12.395512},
       adsurl = {https://ui.adsabs.harvard.edu/abs/2000SPIE.4008..534D}
}

@ARTICLE{Borucki_2010,
       author = {{Borucki}, William J. and {Koch}, David and {Basri}, Gibor and {Batalha}, Natalie and {Brown}, Timothy and {Caldwell}, Douglas and {Caldwell}, John and {Christensen-Dalsgaard}, J{\o}rgen and {Cochran}, William D. and {DeVore}, Edna and {Dunham}, Edward W. and {Dupree}, Andrea K. and {Gautier}, Thomas N. and {Geary}, John C. and {Gilliland}, Ronald and {Gould}, Alan and {Howell}, Steve B. and {Jenkins}, Jon M. and {Kondo}, Yoji and {Latham}, David W. and {Marcy}, Geoffrey W. and {Meibom}, S{\o}ren and {Kjeldsen}, Hans and {Lissauer}, Jack J. and {Monet}, David G. and {Morrison}, David and {Sasselov}, Dimitar and {Tarter}, Jill and {Boss}, Alan and {Brownlee}, Don and {Owen}, Toby and {Buzasi}, Derek and {Charbonneau}, David and {Doyle}, Laurance and {Fortney}, Jonathan and {Ford}, Eric B. and {Holman}, Matthew J. and {Seager}, Sara and {Steffen}, Jason H. and {Welsh}, William F. and {Rowe}, Jason and {Anderson}, Howard and {Buchhave}, Lars and {Ciardi}, David and {Walkowicz}, Lucianne and {Sherry}, William and {Horch}, Elliott and {Isaacson}, Howard and {Everett}, Mark E. and {Fischer}, Debra and {Torres}, Guillermo and {Johnson}, John Asher and {Endl}, Michael and {MacQueen}, Phillip and {Bryson}, Stephen T. and {Dotson}, Jessie and {Haas}, Michael and {Kolodziejczak}, Jeffrey and {Van Cleve}, Jeffrey and {Chandrasekaran}, Hema and {Twicken}, Joseph D. and {Quintana}, Elisa V. and {Clarke}, Bruce D. and {Allen}, Christopher and {Li}, Jie and {Wu}, Haley and {Tenenbaum}, Peter and {Verner}, Ekaterina and {Bruhweiler}, Frederick and {Barnes}, Jason and {Prsa}, Andrej},
        title = "{Kepler Planet-Detection Mission: Introduction and First Results}",
      journal = {Science},
         year = 2010,
        month = feb,
       volume = {327},
       number = {5968},
        pages = {977},
          doi = {10.1126/science.1185402},
       adsurl = {https://ui.adsabs.harvard.edu/abs/2010Sci...327..977B}
}

@INPROCEEDINGS{Ricker_2014,
       author = {{Ricker}, George R. and {Winn}, Joshua N. and {Vanderspek}, Roland and {Latham}, David W. and {Bakos}, G{\'a}sp{\'a}r. {\'A}. and {Bean}, Jacob L. and {Berta-Thompson}, Zachory K. and {Brown}, Timothy M. and {Buchhave}, Lars and {Butler}, Nathaniel R. and {Butler}, R. Paul and {Chaplin}, William J. and {Charbonneau}, David and {Christensen-Dalsgaard}, J{\o}rgen and {Clampin}, Mark and {Deming}, Drake and {Doty}, John and {De Lee}, Nathan and {Dressing}, Courtney and {Dunham}, E.~W. and {Endl}, Michael and {Fressin}, Francois and {Ge}, Jian and {Henning}, Thomas and {Holman}, Matthew J. and {Howard}, Andrew W. and {Ida}, Shigeru and {Jenkins}, Jon and {Jernigan}, Garrett and {Johnson}, John A. and {Kaltenegger}, Lisa and {Kawai}, Nobuyuki and {Kjeldsen}, Hans and {Laughlin}, Gregory and {Levine}, Alan M. and {Lin}, Douglas and {Lissauer}, Jack J. and {MacQueen}, Phillip and {Marcy}, Geoffrey and {McCullough}, P.~R. and {Morton}, Timothy D. and {Narita}, Norio and {Paegert}, Martin and {Palle}, Enric and {Pepe}, Francesco and {Pepper}, Joshua and {Quirrenbach}, Andreas and {Rinehart}, S.~A. and {Sasselov}, Dimitar and {Sato}, Bun'ei and {Seager}, Sara and {Sozzetti}, Alessandro and {Stassun}, Keivan G. and {Sullivan}, Peter and {Szentgyorgyi}, Andrew and {Torres}, Guillermo and {Udry}, Stephane and {Villasenor}, Joel},
        title = "{Transiting Exoplanet Survey Satellite (TESS)}",
    booktitle = {Space Telescopes and Instrumentation 2014: Optical, Infrared, and Millimeter Wave},
         year = 2014,
       editor = {{Oschmann}, Jr., Jacobus M. and {Clampin}, Mark and {Fazio}, Giovanni G. and {MacEwen}, Howard A.},
       series = {Society of Photo-Optical Instrumentation Engineers (SPIE) Conference Series},
       volume = {9143},
        month = aug,
          eid = {914320},
        pages = {914320},
          doi = {10.1117/12.2063489},
archivePrefix = {arXiv},
       eprint = {1406.0151},
 primaryClass = {astro-ph.EP},
       adsurl = {https://ui.adsabs.harvard.edu/abs/2014SPIE.9143E..20R}
}

@ARTICLE{emcee_Foreman-Mackey_2013,
       author = {{Foreman-Mackey}, Daniel and {Hogg}, David W. and {Lang}, Dustin and {Goodman}, Jonathan},
        title = "{emcee: The MCMC Hammer}",
      journal = {\pasp},
         year = 2013,
        month = mar,
       volume = {125},
       number = {925},
        pages = {306},
          doi = {10.1086/670067},
archivePrefix = {arXiv},
       eprint = {1202.3665},
 primaryClass = {astro-ph.IM},
       adsurl = {https://ui.adsabs.harvard.edu/abs/2013PASP..125..306F}
}

@ARTICLE{Soubiran_2022,
       author = {{Soubiran}, C. and {Brouillet}, N. and {Casamiquela}, L.},
        title = "{Assessment of [Fe/H] determinations for FGK stars in spectroscopic surveys}",
      journal = {\aap},
         year = 2022,
        month = jul,
       volume = {663},
          eid = {A4},
        pages = {A4},
          doi = {10.1051/0004-6361/202142409},
archivePrefix = {arXiv},
       eprint = {2112.07545},
 primaryClass = {astro-ph.SR},
       adsurl = {https://ui.adsabs.harvard.edu/abs/2022A&A...663A...4S}
}

@ARTICLE{Ma_2018,
       author = {{Ma}, Bo and {Ge}, Jian and {Muterspaugh}, Matthew and {Singer}, Michael A. and {Henry}, Gregory W. and {Gonz{\'a}lez Hern{\'a}ndez}, Jonay I. and {Sithajan}, Sirinrat and {Jeram}, Sarik and {Williamson}, Michael and {Stassun}, Keivan and {Kimock}, Benjamin and {Varosi}, Frank and {Schofield}, Sidney and {Liu}, Jian and {Powell}, Scott and {Cassette}, Anthony and {Jakeman}, Hali and {Avner}, Louis and {Grieves}, Nolan and {Barnes}, Rory and {Zhao}, Bo and {Gilda}, Sankalp and {Grantham}, Jim and {Stafford}, Greg and {Savage}, David and {Bland}, Steve and {Ealey}, Brent},
        title = "{The first super-Earth detection from the high cadence and high radial velocity precision Dharma Planet Survey}",
      journal = {\mnras},
         year = 2018,
        month = oct,
       volume = {480},
       number = {2},
        pages = {2411-2422},
          doi = {10.1093/mnras/sty1933},
archivePrefix = {arXiv},
       eprint = {1807.07098},
 primaryClass = {astro-ph.EP},
       adsurl = {https://ui.adsabs.harvard.edu/abs/2018MNRAS.480.2411M}
}

@ARTICLE{Burrows_2024,
       author = {{Burrows}, Abigail and {Halverson}, Samuel and {Siegel}, Jared C. and {Gilbertson}, Christian and {Luhn}, Jacob and {Burt}, Jennifer and {Bender}, Chad F. and {Roy}, Arpita and {Terrien}, Ryan C. and {Vangstein}, Selma and {Mahadevan}, Suvrath and {Wright}, Jason T. and {Robertson}, Paul and {Ford}, Eric B. and {Stef{\'a}nsson}, Gumundur and {Ninan}, Joe P. and {Blake}, Cullen H. and {McElwain}, Michael W. and {Schwab}, Christian and {Zhao}, Jinglin},
        title = "{The Death of Vulcan: NEID Reveals That the Planet Candidate Orbiting HD 26965 Is Stellar Activity}",
      journal = {\aj},
         year = 2024,
        month = may,
       volume = {167},
       number = {5},
          eid = {243},
        pages = {243},
          doi = {10.3847/1538-3881/ad34d5},
archivePrefix = {arXiv},
       eprint = {2404.17494},
 primaryClass = {astro-ph.EP},
       adsurl = {https://ui.adsabs.harvard.edu/abs/2024AJ....167..243B}
}

@ARTICLE{Howe_2020,
       author = {{Howe}, Rachel and {Chaplin}, William J. and {Basu}, Sarbani and {Ball}, Warrick H. and {Davies}, Guy R. and {Elsworth}, Yvonne and {Hale}, Steven J. and {Miglio}, Andrea and {Nielsen}, Martin Bo and {Viani}, Lucas S.},
        title = "{Solar cycle variation of {\ensuremath{\nu}}$_{max}$ in helioseismic data and its implications for asteroseismology}",
      journal = {\mnras},
         year = 2020,
        month = mar,
       volume = {493},
       number = {1},
        pages = {L49-L53},
          doi = {10.1093/mnrasl/slaa006},
archivePrefix = {arXiv},
       eprint = {2001.10949},
 primaryClass = {astro-ph.SR},
       adsurl = {https://ui.adsabs.harvard.edu/abs/2020MNRAS.493L..49H}
}

@ARTICLE{Kjeldsen_2008,
       author = {{Kjeldsen}, Hans and {Bedding}, Timothy R. and {Arentoft}, Torben and {Butler}, R. Paul and {Dall}, Thomas H. and {Karoff}, Christoffer and {Kiss}, L{\'a}szl{\'o} L. and {Tinney}, C.~G. and {Chaplin}, William J.},
        title = "{The Amplitude of Solar Oscillations Using Stellar Techniques}",
      journal = {\apj},
         year = 2008,
        month = aug,
       volume = {682},
       number = {2},
        pages = {1370-1375},
          doi = {10.1086/589142},
archivePrefix = {arXiv},
       eprint = {0804.1182},
 primaryClass = {astro-ph},
       adsurl = {https://ui.adsabs.harvard.edu/abs/2008ApJ...682.1370K}
}

@ARTICLE{Andersen_2019,
       author = {{Andersen}, M.~F. and {Pall{\'e}}, P. and {Jessen-Hansen}, J. and {Wang}, K. and {Grundahl}, F. and {Bedding}, T.~R. and {Roca Cortes}, T. and {Yu}, J. and {Mathur}, S. and {Gacia}, R.~A. and {Arentoft}, T. and {R{\'e}gulo}, C. and {Tronsgaard}, R. and {Kjeldsen}, H. and {Christensen-Dalsgaard}, J.},
        title = "{Oscillations in the Sun with SONG: Setting the scale for asteroseismic investigations}",
      journal = {\aap},
         year = 2019,
        month = mar,
       volume = {623},
          eid = {L9},
        pages = {L9},
          doi = {10.1051/0004-6361/201935175},
archivePrefix = {arXiv},
       eprint = {1902.10717},
 primaryClass = {astro-ph.SR},
       adsurl = {https://ui.adsabs.harvard.edu/abs/2019A&A...623L...9A}
}

@ARTICLE{Butler_2004,
       author = {{Butler}, R. Paul and {Bedding}, Timothy R. and {Kjeldsen}, Hans and {McCarthy}, Chris and {O'Toole}, Simon J. and {Tinney}, Christopher G. and {Marcy}, Geoffrey W. and {Wright}, Jason T.},
        title = "{Ultra-High-Precision Velocity Measurements of Oscillations in {\ensuremath{\alpha}} Centauri A}",
      journal = {\apjl},
         year = 2004,
        month = jan,
       volume = {600},
       number = {1},
        pages = {L75-L78},
          doi = {10.1086/381434},
archivePrefix = {arXiv},
       eprint = {astro-ph/0311408},
 primaryClass = {astro-ph},
       adsurl = {https://ui.adsabs.harvard.edu/abs/2004ApJ...600L..75B}
}

@ARTICLE{Berdyugina_2005,
       author = {{Berdyugina}, Svetlana V.},
        title = "{Starspots: A Key to the Stellar Dynamo}",
      journal = {Living Reviews in Solar Physics},
         year = 2005,
        month = dec,
       volume = {2},
       number = {1},
          eid = {8},
        pages = {8},
          doi = {10.12942/lrsp-2005-8},
       adsurl = {https://ui.adsabs.harvard.edu/abs/2005LRSP....2....8B}
}

@ARTICLE{Aerts_2015,
       author = {{Aerts}, C.},
        title = "{The age and interior rotation of stars from asteroseismology}",
      journal = {Astronomische Nachrichten},
         year = 2015,
        month = jun,
       volume = {336},
       number = {5},
        pages = {477},
          doi = {10.1002/asna.201512177},
archivePrefix = {arXiv},
       eprint = {1503.06690},
 primaryClass = {astro-ph.SR},
       adsurl = {https://ui.adsabs.harvard.edu/abs/2015AN....336..477A}
}

@ARTICLE{Nordlund_2009,
       author = {{Nordlund}, {\r{A}}ke and {Stein}, Robert F. and {Asplund}, Martin},
        title = "{Solar Surface Convection}",
      journal = {Living Reviews in Solar Physics},
         year = 2009,
        month = dec,
       volume = {6},
       number = {1},
          eid = {2},
        pages = {2},
          doi = {10.12942/lrsp-2009-2},
       adsurl = {https://ui.adsabs.harvard.edu/abs/2009LRSP....6....2N}
}

@ARTICLE{Hon_2024,
       author = {{Hon}, Marc and {Huber}, Daniel and {Li}, Yaguang and {Metcalfe}, Travis S. and {Bedding}, Timothy R. and {Ong}, Joel and {Chontos}, Ashley and {Rubenzahl}, Ryan and {Halverson}, Samuel and {Garc{\'\i}a}, Rafael A. and {Kjeldsen}, Hans and {Stello}, Dennis and {Hey}, Daniel R. and {Campante}, Tiago and {Howard}, Andrew W. and {Gibson}, Steven R. and {Rider}, Kodi and {Roy}, Arpita and {Baker}, Ashley D. and {Edelstein}, Jerry and {Smith}, Chris and {Fulton}, Benjamin J. and {Walawender}, Josh and {Brodheim}, Max and {Brown}, Matt and {Chan}, Dwight and {Dai}, Fei and {Deich}, William and {Gottschalk}, Colby and {Grillo}, Jason and {Hale}, Dave and {Hill}, Grant M. and {Holden}, Bradford and {Householder}, Aaron and {Isaacson}, Howard and {Ishikawa}, Yuzo and {Jelinsky}, Sharon R. and {Kassis}, Marc and {Kaye}, Stephen and {Laher}, Russ and {Lanclos}, Kyle and {Lee}, Chien-Hsiu and {Lilley}, Scott and {McCarney}, Ben and {Miller}, Timothy N. and {Payne}, Joel and {Petigura}, Erik A. and {Poppett}, Claire and {Raffanti}, Michael and {Rockosi}, Constance and {Sanford}, Dale and {Schwab}, Christian and {Shaum}, Abby P. and {Sirk}, Martin M. and {Smith}, Roger and {Thorne}, Jim and {Valliant}, John and {Vandenberg}, Adam and {Wang}, Shin Ywan and {Wishnow}, Edward and {Wold}, Truman and {Yeh}, Sherry and {Baker}, Ashley and {Basu}, Sarbani and {Bedell}, Megan and {Cegla}, Heather M. and {Crossfield}, Ian and {Dressing}, Courtney and {Dumusque}, Xavier and {Knutson}, Heather and {Mawet}, Dimitri and {O'Meara}, John and {Stef{\'a}nsson}, Gu{\dj}mundur and {Teske}, Johanna and {Vasisht}, Gautam and {Wang}, Sharon Xuesong and {Weiss}, Lauren M. and {Winn}, Joshua N. and {Wright}, Jason T.},
        title = "{Asteroseismology of the Nearby K Dwarf {\ensuremath{\sigma}} Draconis Using the Keck Planet Finder and TESS}",
      journal = {\apj},
         year = 2024,
        month = nov,
       volume = {975},
       number = {1},
          eid = {147},
        pages = {147},
          doi = {10.3847/1538-4357/ad76a9},
archivePrefix = {arXiv},
       eprint = {2407.21234},
 primaryClass = {astro-ph.SR},
       adsurl = {https://ui.adsabs.harvard.edu/abs/2024ApJ...975..147H}
}

@ARTICLE{Campante_2024,
       author = {{Campante}, T.~L. and {Kjeldsen}, H. and {Li}, Y. and {Lund}, M.~N. and {Silva}, A.~M. and {Corsaro}, E. and {Gomes da Silva}, J. and {Martins}, J.~H.~C. and {Adibekyan}, V. and {Azevedo Silva}, T. and {Bedding}, T.~R. and {Bossini}, D. and {Buzasi}, D.~L. and {Chaplin}, W.~J. and {Costa}, R.~R. and {Cunha}, M.~S. and {Cristo}, E. and {Faria}, J.~P. and {Garc{\'\i}a}, R.~A. and {Huber}, D. and {Lundkvist}, M.~S. and {Metcalfe}, T.~S. and {Monteiro}, M.~J.~P.~F.~G. and {Neitzel}, A.~W. and {Nielsen}, M.~B. and {Poretti}, E. and {Santos}, N.~C. and {Sousa}, S.~G.},
        title = "{Expanding the frontiers of cool-dwarf asteroseismology with ESPRESSO. Detection of solar-like oscillations in the K5 dwarf ϵ Indi}",
      journal = {\aap},
         year = 2024,
        month = mar,
       volume = {683},
          eid = {L16},
        pages = {L16},
          doi = {10.1051/0004-6361/202449197},
archivePrefix = {arXiv},
       eprint = {2403.16333},
 primaryClass = {astro-ph.SR},
       adsurl = {https://ui.adsabs.harvard.edu/abs/2024A&A...683L..16C}
}

@ARTICLE{Li_2025,
       author = {{Li}, Yaguang and {Huber}, Daniel and {Ong}, J.~M. Joel and {van Saders}, Jennifer and {Costa}, R.~R. and {Reersted Larsen}, Jens and {Basu}, Sarbani and {Bedding}, Timothy R. and {Dai}, Fei and {Chontos}, Ashley and {Carmichael}, Theron W. and {Hey}, Daniel and {Kjeldsen}, Hans and {Hon}, Marc and {Campante}, Tiago L. and {Monteiro}, M\textbackslash'ario J.~P.~F.~G. and {Sloth Lundkvist}, Mia and {Saunders}, Nicholas and {Isaacson}, Howard and {Howard}, Andrew W. and {Gibson}, Steven R. and {Halverson}, Samuel and {Rider}, Kodi and {Roy}, Arpita and {Baker}, Ashley D. and {Edelstein}, Jerry and {Smith}, Chris and {Fulton}, Benjamin J. and {Walawender}, Josh},
        title = "{K-dwarf Radius Inflation and a 10-Gyr Spin-down Clock Unveiled through Asteroseismology of HD 219134 from the Keck Planet Finder}",
      journal = {arXiv e-prints},
         year = 2025,
        month = feb,
          eid = {arXiv:2502.00971},
        pages = {arXiv:2502.00971},
          doi = {10.48550/arXiv.2502.00971},
archivePrefix = {arXiv},
       eprint = {2502.00971},
 primaryClass = {astro-ph.SR},
       adsurl = {https://ui.adsabs.harvard.edu/abs/2025arXiv250200971L}
}

@BOOK{Basu_2018B,
       author = {{Basu}, Sarbani and {Chaplin}, William J.},
        title = "{Asteroseismic Data Analysis. Foundations and Techniques}",
         year = 2018,
       adsurl = {https://ui.adsabs.harvard.edu/abs/2018adaf.book.....B}
}

@BOOK{astro_2020,
  author  = {{National Academies of Sciences, Engineering, and Medicine}},
  title   = "{Pathways to Discovery in Astronomy and Astrophysics for the 2020s}",
  year    = {2021},
  doi     = {10.17226/26141},
  adsurl  = {https://ui.adsabs.harvard.edu/abs/2021pdaa.book.....N}
}

@ARTICLE{Dumusque_2011_HARPS,
       author = {{Dumusque}, X. and {Lovis}, C. and {S{\'e}gransan}, D. and {Mayor}, M. and {Udry}, S. and {Benz}, W. and {Bouchy}, F. and {Lo Curto}, G. and {Mordasini}, C. and {Pepe}, F. and {Queloz}, D. and {Santos}, N.~C. and {Naef}, D.},
        title = "{The HARPS search for southern extra-solar planets. XXX. Planetary systems around stars with solar-like magnetic cycles and short-term activity variation}",
      journal = {\aap},
         year = 2011,
        month = nov,
       volume = {535},
          eid = {A55},
        pages = {A55},
          doi = {10.1051/0004-6361/201117148},
archivePrefix = {arXiv},
       eprint = {1107.1748},
 primaryClass = {astro-ph.EP},
       adsurl = {https://ui.adsabs.harvard.edu/abs/2011A&A...535A..55D}
}

@ARTICLE{Dumusque_2011_limit,
       author = {{Dumusque}, X. and {Udry}, S. and {Lovis}, C. and {Santos}, N.~C. and {Monteiro}, M.~J.~P.~F.~G.},
        title = "{Planetary detection limits taking into account stellar noise. I. Observational strategies to reduce stellar oscillation and granulation effects}",
      journal = {\aap},
         year = 2011,
        month = jan,
       volume = {525},
          eid = {A140},
        pages = {A140},
          doi = {10.1051/0004-6361/201014097},
archivePrefix = {arXiv},
       eprint = {1010.2616},
 primaryClass = {astro-ph.EP},
       adsurl = {https://ui.adsabs.harvard.edu/abs/2011A&A...525A.140D}
}

@ARTICLE{Aigrain_2012,
       author = {{Aigrain}, S. and {Pont}, F. and {Zucker}, S.},
        title = "{A simple method to estimate radial velocity variations due to stellar activity using photometry}",
      journal = {\mnras},
         year = 2012,
        month = feb,
       volume = {419},
       number = {4},
        pages = {3147-3158},
          doi = {10.1111/j.1365-2966.2011.19960.x},
archivePrefix = {arXiv},
       eprint = {1110.1034},
 primaryClass = {astro-ph.SR},
       adsurl = {https://ui.adsabs.harvard.edu/abs/2012MNRAS.419.3147A}
}

@ARTICLE{Oshagh_2017,
       author = {{Oshagh}, M. and {Santos}, N.~C. and {Figueira}, P. and {Barros}, S.~C.~C. and {Donati}, J. -F. and {Adibekyan}, V. and {Faria}, J.~P. and {Watson}, C.~A. and {Cegla}, H.~M. and {Dumusque}, X. and {H{\'e}brard}, E. and {Demangeon}, O. and {Dreizler}, S. and {Boisse}, I. and {Deleuil}, M. and {Bonfils}, X. and {Pepe}, F. and {Udry}, S.},
        title = "{Understanding stellar activity-induced radial velocity jitter using simultaneous K2 photometry and HARPS RV measurements}",
      journal = {\aap},
         year = 2017,
        month = oct,
       volume = {606},
          eid = {A107},
        pages = {A107},
          doi = {10.1051/0004-6361/201731139},
archivePrefix = {arXiv},
       eprint = {1707.01827},
 primaryClass = {astro-ph.EP},
       adsurl = {https://ui.adsabs.harvard.edu/abs/2017A&A...606A.107O}
}

@ARTICLE{Robnik_2020,
       author = {{Robnik}, Jakob and {Seljak}, Uro{\v{s}}},
        title = "{Kepler Data Analysis: Non-Gaussian Noise and Fourier Gaussian Process Analysis of Stellar Variability}",
      journal = {\aj},
         year = 2020,
        month = may,
       volume = {159},
       number = {5},
          eid = {224},
        pages = {224},
          doi = {10.3847/1538-3881/ab8460},
archivePrefix = {arXiv},
       eprint = {1910.01167},
 primaryClass = {astro-ph.EP},
       adsurl = {https://ui.adsabs.harvard.edu/abs/2020AJ....159..224R}
}

@ARTICLE{Liyg_2023,
       author = {{Li}, Yaguang and {Bedding}, Timothy R. and {Stello}, Dennis and {Huber}, Daniel and {Hon}, Marc and {Joyce}, Meridith and {Li}, Tanda and {Perkins}, Jean and {White}, Timothy R. and {Zinn}, Joel C. and {Howard}, Andrew W. and {Isaacson}, Howard and {Hey}, Daniel R. and {Kjeldsen}, Hans},
        title = "{A prescription for the asteroseismic surface correction}",
      journal = {\mnras},
         year = 2023,
        month = jul,
       volume = {523},
       number = {1},
        pages = {916-927},
          doi = {10.1093/mnras/stad1445},
archivePrefix = {arXiv},
       eprint = {2208.01176},
 primaryClass = {astro-ph.SR},
       adsurl = {https://ui.adsabs.harvard.edu/abs/2023MNRAS.523..916L}
}

@article{paxton++2019-mesa,
	adsurl = {https://ui.adsabs.harvard.edu/abs/2019ApJS..243...10P},
	archiveprefix = {arXiv},
	author = {{Paxton}, Bill and {Smolec}, R. and {Schwab}, Josiah and {Gautschy}, A. and {Bildsten}, Lars and {Cantiello}, Matteo and {Dotter}, Aaron and {Farmer}, R. and {Goldberg}, Jared A. and {Jermyn}, Adam S. and {Kanbur}, S.~M. and {Marchant}, Pablo and {Thoul}, Anne and {Townsend}, Richard H.~D. and {Wolf}, William M. and {Zhang}, Michael and {Timmes}, F.~X.},
	doi = {10.3847/1538-4365/ab2241},
	eid = {10},
	eprint = {1903.01426},
	journal = {\apjs},
	month = jul,
	number = {1},
	pages = {10},
	primaryclass = {astro-ph.SR},
	title = {{Modules for Experiments in Stellar Astrophysics (MESA): Pulsating Variable Stars, Rotation, Convective Boundaries, and Energy Conservation}},
	volume = {243},
	year = 2019}

@article{paxton++2011-mesa,
	adsurl = {https://ui.adsabs.harvard.edu/abs/2011ApJS..192....3P},
	archiveprefix = {arXiv},
	author = {{Paxton}, Bill and {Bildsten}, Lars and {Dotter}, Aaron and {Herwig}, Falk and {Lesaffre}, Pierre and {Timmes}, Frank},
	doi = {10.1088/0067-0049/192/1/3},
	eid = {3},
	eprint = {1009.1622},
	journal = {\apjs},
	month = {Jan},
	number = {1},
	pages = {3},
	primaryclass = {astro-ph.SR},
	title = {{Modules for Experiments in Stellar Astrophysics (MESA)}},
	volume = {192},
	year = {2011}}

@article{paxton++2013-mesa,
	adsurl = {https://ui.adsabs.harvard.edu/abs/2013ApJS..208....4P},
	archiveprefix = {arXiv},
	author = {{Paxton}, Bill and {Cantiello}, Matteo and {Arras}, Phil and {Bildsten}, Lars and {Brown}, Edward F. and {Dotter}, Aaron and {Mankovich}, Christopher and {Montgomery}, M.~H. and {Stello}, Dennis and {Timmes}, F.~X. and {Townsend}, Richard},
	doi = {10.1088/0067-0049/208/1/4},
	eid = {4},
	eprint = {1301.0319},
	journal = {\apjs},
	month = {Sep},
	number = {1},
	pages = {4},
	primaryclass = {astro-ph.SR},
	title = {{Modules for Experiments in Stellar Astrophysics (MESA): Planets, Oscillations, Rotation, and Massive Stars}},
	volume = {208},
	year = {2013}}

@article{paxton++2018-mesa,
	adsurl = {https://ui.adsabs.harvard.edu/abs/2018ApJS..234...34P},
	archiveprefix = {arXiv},
	author = {{Paxton}, Bill and {Schwab}, Josiah and {Bauer}, Evan B. and {Bildsten}, Lars and {Blinnikov}, Sergei and {Duffell}, Paul and {Farmer}, R. and {Goldberg}, Jared A. and {Marchant}, Pablo and {Sorokina}, Elena and {Thoul}, Anne and {Townsend}, Richard H.~D. and {Timmes}, F.~X.},
	doi = {10.3847/1538-4365/aaa5a8},
	eid = {34},
	eprint = {1710.08424},
	journal = {\apjs},
	month = {Feb},
	number = {2},
	pages = {34},
	primaryclass = {astro-ph.SR},
	title = {{Modules for Experiments in Stellar Astrophysics (MESA): Convective Boundaries, Element Diffusion, and Massive Star Explosions}},
	volume = {234},
	year = {2018}}

@article{paxton++2015-mesa,
	adsurl = {https://ui.adsabs.harvard.edu/abs/2015ApJS..220...15P},
	archiveprefix = {arXiv},
	author = {{Paxton}, Bill and {Marchant}, Pablo and {Schwab}, Josiah and {Bauer}, Evan B. and {Bildsten}, Lars and {Cantiello}, Matteo and {Dessart}, Luc and {Farmer}, R. and {Hu}, H. and {Langer}, N. and {Townsend}, R.~H.~D. and {Townsley}, Dean M. and {Timmes}, F.~X.},
	doi = {10.1088/0067-0049/220/1/15},
	eid = {15},
	eprint = {1506.03146},
	journal = {\apjs},
	month = {Sep},
	number = {1},
	pages = {15},
	primaryclass = {astro-ph.SR},
	title = {{Modules for Experiments in Stellar Astrophysics (MESA): Binaries, Pulsations, and Explosions}},
	volume = {220},
	year = {2015}}

@ARTICLE{jermyn++2023-mesa,
       author = {{Jermyn}, Adam S. and {Bauer}, Evan B. and {Schwab}, Josiah and {Farmer}, R. and {Ball}, Warrick H. and {Bellinger}, Earl P. and {Dotter}, Aaron and {Joyce}, Meridith and {Marchant}, Pablo and {Mombarg}, Joey S.~G. and {Wolf}, William M. and {Sunny Wong}, Tin Long and {Cinquegrana}, Giulia C. and {Farrell}, Eoin and {Smolec}, R. and {Thoul}, Anne and {Cantiello}, Matteo and {Herwig}, Falk and {Toloza}, Odette and {Bildsten}, Lars and {Townsend}, Richard H.~D. and {Timmes}, F.~X.},
        title = "{Modules for Experiments in Stellar Astrophysics (MESA): Time-dependent Convection, Energy Conservation, Automatic Differentiation, and Infrastructure}",
      journal = {\apjs},
         year = 2023,
        month = mar,
       volume = {265},
       number = {1},
          eid = {15},
        pages = {15},
          doi = {10.3847/1538-4365/acae8d},
archivePrefix = {arXiv},
       eprint = {2208.03651},
 primaryClass = {astro-ph.SR},
       adsurl = {https://ui.adsabs.harvard.edu/abs/2023ApJS..265...15J}
}

@article{townsend+2013-gyre,
	adsurl = {https://ui.adsabs.harvard.edu/abs/2013MNRAS.435.3406T},
	archiveprefix = {arXiv},
	author = {{Townsend}, R.~H.~D. and {Teitler}, S.~A.},
	doi = {10.1093/mnras/stt1533},
	eprint = {1308.2965},
	journal = {\mnras},
	month = {Nov},
	number = {4},
	pages = {3406-3418},
	primaryclass = {astro-ph.SR},
	title = {{GYRE: an open-source stellar oscillation code based on a new Magnus Multiple Shooting scheme}},
	volume = {435},
	year = {2013}}

@ARTICLE{Ball_2014,
       author = {{Ball}, W.~H. and {Gizon}, L.},
        title = "{A new correction of stellar oscillation frequencies for near-surface effects}",
      journal = {\aap},
         year = 2014,
        month = aug,
       volume = {568},
          eid = {A123},
        pages = {A123},
          doi = {10.1051/0004-6361/201424325},
archivePrefix = {arXiv},
       eprint = {1408.0986},
 primaryClass = {astro-ph.SR},
       adsurl = {https://ui.adsabs.harvard.edu/abs/2014A&A...568A.123B}
}

@ARTICLE{Ong_2021,
       author = {{Ong}, J.~M. Joel and {Basu}, Sarbani and {Roxburgh}, Ian W.},
        title = "{Mixed Modes and Asteroseismic Surface Effects. I. Analytic Treatment}",
      journal = {\apj},
         year = 2021,
        month = oct,
       volume = {920},
       number = {1},
          eid = {8},
        pages = {8},
          doi = {10.3847/1538-4357/ac12ca},
archivePrefix = {arXiv},
       eprint = {2107.03405},
 primaryClass = {astro-ph.SR},
       adsurl = {https://ui.adsabs.harvard.edu/abs/2021ApJ...920....8O}
}

@ARTICLE{Bedding_2007,
       author = {{Bedding}, Timothy R. and {Kjeldsen}, Hans and {Arentoft}, Torben and {Bouchy}, Francois and {Brandbyge}, Jacob and {Brewer}, Brendon J. and {Butler}, R. Paul and {Christensen-Dalsgaard}, J{\o}rgen and {Dall}, Thomas and {Frandsen}, S{\o}ren and {Karoff}, Christoffer and {Kiss}, L{\'a}szl{\'o} L. and {Monteiro}, Mario J.~P.~F.~G. and {Pijpers}, Frank P. and {Teixeira}, Teresa C. and {Tinney}, C.~G. and {Baldry}, Ivan K. and {Carrier}, Fabien and {O'Toole}, Simon J.},
        title = "{Solar-like Oscillations in the G2 Subgiant {\ensuremath{\beta}} Hydri from Dual-Site Observations}",
      journal = {\apj},
         year = 2007,
        month = jul,
       volume = {663},
       number = {2},
        pages = {1315-1324},
          doi = {10.1086/518593},
archivePrefix = {arXiv},
       eprint = {astro-ph/0703747},
 primaryClass = {astro-ph},
       adsurl = {https://ui.adsabs.harvard.edu/abs/2007ApJ...663.1315B}
}

@ARTICLE{Davies_2016,
       author = {{Davies}, G.~R. and {Silva Aguirre}, V. and {Bedding}, T.~R. and {Handberg}, R. and {Lund}, M.~N. and {Chaplin}, W.~J. and {Huber}, D. and {White}, T.~R. and {Benomar}, O. and {Hekker}, S. and {Basu}, S. and {Campante}, T.~L. and {Christensen-Dalsgaard}, J. and {Elsworth}, Y. and {Karoff}, C. and {Kjeldsen}, H. and {Lundkvist}, M.~S. and {Metcalfe}, T.~S. and {Stello}, D.},
        title = "{Oscillation frequencies for 35 Kepler solar-type planet-hosting stars using Bayesian techniques and machine learning}",
      journal = {\mnras},
         year = 2016,
        month = feb,
       volume = {456},
       number = {2},
        pages = {2183-2195},
          doi = {10.1093/mnras/stv2593},
archivePrefix = {arXiv},
       eprint = {1511.02105},
 primaryClass = {astro-ph.SR},
       adsurl = {https://ui.adsabs.harvard.edu/abs/2016MNRAS.456.2183D}
}

@ARTICLE{Handberg_2011,
       author = {{Handberg}, R. and {Campante}, T.~L.},
        title = "{Bayesian peak-bagging of solar-like oscillators using MCMC: a comprehensive guide}",
      journal = {\aap},
         year = 2011,
        month = mar,
       volume = {527},
          eid = {A56},
        pages = {A56},
          doi = {10.1051/0004-6361/201015451},
archivePrefix = {arXiv},
       eprint = {1101.0084},
 primaryClass = {astro-ph.SR},
       adsurl = {https://ui.adsabs.harvard.edu/abs/2011A&A...527A..56H}
}

@ARTICLE{Liyg_2020,
       author = {{Li}, Yaguang and {Bedding}, Timothy R. and {Li}, Tanda and {Bi}, Shaolan and {Stello}, Dennis and {Zhou}, Yixiao and {White}, Timothy R.},
        title = "{Asteroseismology of 36 Kepler subgiants - I. Oscillation frequencies, linewidths, and amplitudes}",
      journal = {\mnras},
         year = 2020,
        month = jun,
       volume = {495},
       number = {2},
        pages = {2363-2386},
          doi = {10.1093/mnras/staa1335},
archivePrefix = {arXiv},
       eprint = {2005.06460},
 primaryClass = {astro-ph.SR},
       adsurl = {https://ui.adsabs.harvard.edu/abs/2020MNRAS.495.2363L}
}

@ARTICLE{Kallinger_2014,
       author = {{Kallinger}, T. and {De Ridder}, J. and {Hekker}, S. and {Mathur}, S. and {Mosser}, B. and {Gruberbauer}, M. and {Garc{\'\i}a}, R.~A. and {Karoff}, C. and {Ballot}, J.},
        title = "{The connection between stellar granulation and oscillation as seen by the Kepler mission}",
      journal = {\aap},
         year = 2014,
        month = oct,
       volume = {570},
          eid = {A41},
        pages = {A41},
          doi = {10.1051/0004-6361/201424313},
archivePrefix = {arXiv},
       eprint = {1408.0817},
 primaryClass = {astro-ph.SR},
       adsurl = {https://ui.adsabs.harvard.edu/abs/2014A&A...570A..41K}
}

@article{Campante_2016,
	doi = {10.3847/0004-637x/830/2/138},
	url = {https://doi.org/10.3847%2F0004-637x%2F830%2F2%2F138},
	year = 2016,
	month = {oct},
	publisher = {American Astronomical Society},
	volume = {830},
	number = {2},
	pages = {138},
	author = {T. L. Campante and M. Schofield and J. S. Kuszlewicz and L. Bouma and W. J. Chaplin and D. Huber and J. Christensen-Dalsgaard and H. Kjeldsen and D. Bossini and T. S. H. North and T. Appourchaux and D. W. Latham and J. Pepper and G. R. Ricker and K. G. Stassun and R. Vanderspek and J. N. Winn},
	title = {{THE} {ASTEROSEISMIC} {POTENTIAL} {OF}$\less$i$\greater${TESS}$\less$/i$\greater$: {EXOPLANET}-{HOST} {STARS}
},
	journal = {The Astrophysical Journal}
}

@article{Foreman_Mackey_2017,
	doi = {10.3847/1538-3881/aa9332},
	url = {https://doi.org/10.3847%2F1538-3881%2Faa9332},
	year = 2017,
	month = {nov},
	publisher = {American Astronomical Society},
	volume = {154},
	number = {6},
	pages = {220},
	author = {Daniel Foreman-Mackey and Eric Agol and Sivaram Ambikasaran and Ruth Angus},
	title = {Fast and Scalable Gaussian Process Modeling with Applications to Astronomical Time Series},
	journal = {The Astronomical Journal}
}

@article{Xu_2019,
doi = {10.3847/1538-3881/ab1b47},
url = {https://dx.doi.org/10.3847/1538-3881/ab1b47},
year = {2019},
month = {may},
publisher = {The American Astronomical Society},
volume = {157},
number = {6},
pages = {243},
author = {Xin Xu and Jessi Cisewski-Kehe and Allen B. Davis and Debra A. Fischer and John M. Brewer},
title = {Modeling the Echelle Spectra Continuum with Alpha Shapes and Local Regression Fitting},
journal = {The Astronomical Journal}
}

@ARTICLE{Rosenthal_2021,
       author = {{Rosenthal}, Lee J. and {Fulton}, Benjamin J. and {Hirsch}, Lea A. and {Isaacson}, Howard T. and {Howard}, Andrew W. and {Dedrick}, Cayla M. and {Sherstyuk}, Ilya A. and {Blunt}, Sarah C. and {Petigura}, Erik A. and {Knutson}, Heather A. and {Behmard}, Aida and {Chontos}, Ashley and {Crepp}, Justin R. and {Crossfield}, Ian J.~M. and {Dalba}, Paul A. and {Fischer}, Debra A. and {Henry}, Gregory W. and {Kane}, Stephen R. and {Kosiarek}, Molly and {Marcy}, Geoffrey W. and {Rubenzahl}, Ryan A. and {Weiss}, Lauren M. and {Wright}, Jason T.},
        title = "{The California Legacy Survey. I. A Catalog of 178 Planets from Precision Radial Velocity Monitoring of 719 Nearby Stars over Three Decades}",
      journal = {\apjs},
         year = 2021,
        month = jul,
       volume = {255},
       number = {1},
          eid = {8},
        pages = {8},
          doi = {10.3847/1538-4365/abe23c},
archivePrefix = {arXiv},
       eprint = {2105.11583},
 primaryClass = {astro-ph.EP},
       adsurl = {https://ui.adsabs.harvard.edu/abs/2021ApJS..255....8R}
}

@ARTICLE{Yee_2017,
       author = {{Yee}, Samuel W. and {Petigura}, Erik A. and {von Braun}, Kaspar},
        title = "{Precision Stellar Characterization of FGKM Stars using an Empirical Spectral Library}",
      journal = {\apj},
         year = 2017,
        month = feb,
       volume = {836},
       number = {1},
          eid = {77},
        pages = {77},
          doi = {10.3847/1538-4357/836/1/77},
archivePrefix = {arXiv},
       eprint = {1701.00922},
 primaryClass = {astro-ph.SR},
       adsurl = {https://ui.adsabs.harvard.edu/abs/2017ApJ...836...77Y}
}

@ARTICLE{Gaia_2021,
       author = {{Gaia Collaboration} and {Brown}, A.~G.~A. and {Vallenari}, A. and {Prusti}, T. and {de Bruijne}, J.~H.~J. and {Babusiaux}, C. and {Biermann}, M. and {Creevey}, O.~L. and {Evans}, D.~W. and {Eyer}, L. and {Hutton}, A. and {Jansen}, F. and {Jordi}, C. and {Klioner}, S.~A. and {Lammers}, U. and {Lindegren}, L. and {Luri}, X. and {Mignard}, F. and {Panem}, C. and {Pourbaix}, D. and {Randich}, S. and {Sartoretti}, P. and {Soubiran}, C. and {Walton}, N.~A. and {Arenou}, F. and {Bailer-Jones}, C.~A.~L. and {Bastian}, U. and {Cropper}, M. and {Drimmel}, R. and {Katz}, D. and {Lattanzi}, M.~G. and {van Leeuwen}, F. and {Bakker}, J. and {Cacciari}, C. and {Casta{\~n}eda}, J. and {De Angeli}, F. and {Ducourant}, C. and {Fabricius}, C. and {Fouesneau}, M. and {Fr{\'e}mat}, Y. and {Guerra}, R. and {Guerrier}, A. and {Guiraud}, J. and {Jean-Antoine Piccolo}, A. and {Masana}, E. and {Messineo}, R. and {Mowlavi}, N. and {Nicolas}, C. and {Nienartowicz}, K. and {Pailler}, F. and {Panuzzo}, P. and {Riclet}, F. and {Roux}, W. and {Seabroke}, G.~M. and {Sordo}, R. and {Tanga}, P. and {Th{\'e}venin}, F. and {Gracia-Abril}, G. and {Portell}, J. and {Teyssier}, D. and {Altmann}, M. and {Andrae}, R. and {Bellas-Velidis}, I. and {Benson}, K. and {Berthier}, J. and {Blomme}, R. and {Brugaletta}, E. and {Burgess}, P.~W. and {Busso}, G. and {Carry}, B. and {Cellino}, A. and {Cheek}, N. and {Clementini}, G. and {Damerdji}, Y. and {Davidson}, M. and {Delchambre}, L. and {Dell'Oro}, A. and {Fern{\'a}ndez-Hern{\'a}ndez}, J. and {Galluccio}, L. and {Garc{\'\i}a-Lario}, P. and {Garcia-Reinaldos}, M. and {Gonz{\'a}lez-N{\'u}{\~n}ez}, J. and {Gosset}, E. and {Haigron}, R. and {Halbwachs}, J. -L. and {Hambly}, N.~C. and {Harrison}, D.~L. and {Hatzidimitriou}, D. and {Heiter}, U. and {Hern{\'a}ndez}, J. and {Hestroffer}, D. and {Hodgkin}, S.~T. and {Holl}, B. and {Jan{\ss}en}, K. and {Jevardat de Fombelle}, G. and {Jordan}, S. and {Krone-Martins}, A. and {Lanzafame}, A.~C. and {L{\"o}ffler}, W. and {Lorca}, A. and {Manteiga}, M. and {Marchal}, O. and {Marrese}, P.~M. and {Moitinho}, A. and {Mora}, A. and {Muinonen}, K. and {Osborne}, P. and {Pancino}, E. and {Pauwels}, T. and {Petit}, J. -M. and {Recio-Blanco}, A. and {Richards}, P.~J. and {Riello}, M. and {Rimoldini}, L. and {Robin}, A.~C. and {Roegiers}, T. and {Rybizki}, J. and {Sarro}, L.~M. and {Siopis}, C. and {Smith}, M. and {Sozzetti}, A. and {Ulla}, A. and {Utrilla}, E. and {van Leeuwen}, M. and {van Reeven}, W. and {Abbas}, U. and {Abreu Aramburu}, A. and {Accart}, S. and {Aerts}, C. and {Aguado}, J.~J. and {Ajaj}, M. and {Altavilla}, G. and {{\'A}lvarez}, M.~A. and {{\'A}lvarez Cid-Fuentes}, J. and {Alves}, J. and {Anderson}, R.~I. and {Anglada Varela}, E. and {Antoja}, T. and {Audard}, M. and {Baines}, D. and {Baker}, S.~G. and {Balaguer-N{\'u}{\~n}ez}, L. and {Balbinot}, E. and {Balog}, Z. and {Barache}, C. and {Barbato}, D. and {Barros}, M. and {Barstow}, M.~A. and {Bartolom{\'e}}, S. and {Bassilana}, J. -L. and {Bauchet}, N. and {Baudesson-Stella}, A. and {Becciani}, U. and {Bellazzini}, M. and {Bernet}, M. and {Bertone}, S. and {Bianchi}, L. and {Blanco-Cuaresma}, S. and {Boch}, T. and {Bombrun}, A. and {Bossini}, D. and {Bouquillon}, S. and {Bragaglia}, A. and {Bramante}, L. and {Breedt}, E. and {Bressan}, A. and {Brouillet}, N. and {Bucciarelli}, B. and {Burlacu}, A. and {Busonero}, D. and {Butkevich}, A.~G. and {Buzzi}, R. and {Caffau}, E. and {Cancelliere}, R. and {C{\'a}novas}, H. and {Cantat-Gaudin}, T. and {Carballo}, R. and {Carlucci}, T. and {Carnerero}, M.~I. and {Carrasco}, J.~M. and {Casamiquela}, L. and {Castellani}, M. and {Castro-Ginard}, A. and {Castro Sampol}, P. and {Chaoul}, L. and {Charlot}, P. and {Chemin}, L. and {Chiavassa}, A. and {Cioni}, M. -R.~L. and {Comoretto}, G. and {Cooper}, W.~J. and {Cornez}, T. and {Cowell}, S. and {Crifo}, F. and {Crosta}, M. and {Crowley}, C. and {Dafonte}, C. and {Dapergolas}, A. and {David}, M. and {David}, P. and {de Laverny}, P. and {De Luise}, F. and {De March}, R. and {De Ridder}, J. and {de Souza}, R. and {de Teodoro}, P. and {de Torres}, A. and {del Peloso}, E.~F. and {del Pozo}, E. and {Delbo}, M. and {Delgado}, A. and {Delgado}, H.~E. and {Delisle}, J. -B. and {Di Matteo}, P. and {Diakite}, S. and {Diener}, C. and {Distefano}, E. and {Dolding}, C. and {Eappachen}, D. and {Edvardsson}, B. and {Enke}, H. and {Esquej}, P. and {Fabre}, C. and {Fabrizio}, M. and {Faigler}, S. and {Fedorets}, G. and {Fernique}, P. and {Fienga}, A. and {Figueras}, F. and {Fouron}, C. and {Fragkoudi}, F. and {Fraile}, E. and {Franke}, F. and {Gai}, M. and {Garabato}, D. and {Garcia-Gutierrez}, A. and {Garc{\'\i}a-Torres}, M. and {Garofalo}, A. and {Gavras}, P. and {Gerlach}, E. and {Geyer}, R. and {Giacobbe}, P. and {Gilmore}, G. and {Girona}, S. and {Giuffrida}, G. and {Gomel}, R. and {Gomez}, A. and {Gonzalez-Santamaria}, I. and {Gonz{\'a}lez-Vidal}, J.~J. and {Granvik}, M. and {Guti{\'e}rrez-S{\'a}nchez}, R. and {Guy}, L.~P. and {Hauser}, M. and {Haywood}, M. and {Helmi}, A. and {Hidalgo}, S.~L. and {Hilger}, T. and {H{\l}adczuk}, N. and {Hobbs}, D. and {Holland}, G. and {Huckle}, H.~E. and {Jasniewicz}, G. and {Jonker}, P.~G. and {Juaristi Campillo}, J. and {Julbe}, F. and {Karbevska}, L. and {Kervella}, P. and {Khanna}, S. and {Kochoska}, A. and {Kontizas}, M. and {Kordopatis}, G. and {Korn}, A.~J. and {Kostrzewa-Rutkowska}, Z. and {Kruszy{\'n}ska}, K. and {Lambert}, S. and {Lanza}, A.~F. and {Lasne}, Y. and {Le Campion}, J. -F. and {Le Fustec}, Y. and {Lebreton}, Y. and {Lebzelter}, T. and {Leccia}, S. and {Leclerc}, N. and {Lecoeur-Taibi}, I. and {Liao}, S. and {Licata}, E. and {Lindstr{\o}m}, E.~P. and {Lister}, T.~A. and {Livanou}, E. and {Lobel}, A. and {Madrero Pardo}, P. and {Managau}, S. and {Mann}, R.~G. and {Marchant}, J.~M. and {Marconi}, M. and {Marcos Santos}, M.~M.~S. and {Marinoni}, S. and {Marocco}, F. and {Marshall}, D.~J. and {Martin Polo}, L. and {Mart{\'\i}n-Fleitas}, J.~M. and {Masip}, A. and {Massari}, D. and {Mastrobuono-Battisti}, A. and {Mazeh}, T. and {McMillan}, P.~J. and {Messina}, S. and {Michalik}, D. and {Millar}, N.~R. and {Mints}, A. and {Molina}, D. and {Molinaro}, R. and {Moln{\'a}r}, L. and {Montegriffo}, P. and {Mor}, R. and {Morbidelli}, R. and {Morel}, T. and {Morris}, D. and {Mulone}, A.~F. and {Munoz}, D. and {Muraveva}, T. and {Murphy}, C.~P. and {Musella}, I. and {Noval}, L. and {Ord{\'e}novic}, C. and {Orr{\`u}}, G. and {Osinde}, J. and {Pagani}, C. and {Pagano}, I. and {Palaversa}, L. and {Palicio}, P.~A. and {Panahi}, A. and {Pawlak}, M. and {Pe{\~n}alosa Esteller}, X. and {Penttil{\"a}}, A. and {Piersimoni}, A.~M. and {Pineau}, F. -X. and {Plachy}, E. and {Plum}, G. and {Poggio}, E. and {Poretti}, E. and {Poujoulet}, E. and {Pr{\v{s}}a}, A. and {Pulone}, L. and {Racero}, E. and {Ragaini}, S. and {Rainer}, M. and {Raiteri}, C.~M. and {Rambaux}, N. and {Ramos}, P. and {Ramos-Lerate}, M. and {Re Fiorentin}, P. and {Regibo}, S. and {Reyl{\'e}}, C. and {Ripepi}, V. and {Riva}, A. and {Rixon}, G. and {Robichon}, N. and {Robin}, C. and {Roelens}, M. and {Rohrbasser}, L. and {Romero-G{\'o}mez}, M. and {Rowell}, N. and {Royer}, F. and {Rybicki}, K.~A. and {Sadowski}, G. and {Sagrist{\`a} Sell{\'e}s}, A. and {Sahlmann}, J. and {Salgado}, J. and {Salguero}, E. and {Samaras}, N. and {Sanchez Gimenez}, V. and {Sanna}, N. and {Santove{\~n}a}, R. and {Sarasso}, M. and {Schultheis}, M. and {Sciacca}, E. and {Segol}, M. and {Segovia}, J.~C. and {S{\'e}gransan}, D. and {Semeux}, D. and {Shahaf}, S. and {Siddiqui}, H.~I. and {Siebert}, A. and {Siltala}, L. and {Slezak}, E. and {Smart}, R.~L. and {Solano}, E. and {Solitro}, F. and {Souami}, D. and {Souchay}, J. and {Spagna}, A. and {Spoto}, F. and {Steele}, I.~A. and {Steidelm{\"u}ller}, H. and {Stephenson}, C.~A. and {S{\"u}veges}, M. and {Szabados}, L. and {Szegedi-Elek}, E. and {Taris}, F. and {Tauran}, G. and {Taylor}, M.~B. and {Teixeira}, R. and {Thuillot}, W. and {Tonello}, N. and {Torra}, F. and {Torra}, J. and {Turon}, C. and {Unger}, N. and {Vaillant}, M. and {van Dillen}, E. and {Vanel}, O. and {Vecchiato}, A. and {Viala}, Y. and {Vicente}, D. and {Voutsinas}, S. and {Weiler}, M. and {Wevers}, T. and {Wyrzykowski}, {\L}. and {Yoldas}, A. and {Yvard}, P. and {Zhao}, H. and {Zorec}, J. and {Zucker}, S. and {Zurbach}, C. and {Zwitter}, T.},
        title = "{Gaia Early Data Release 3. Summary of the contents and survey properties}",
      journal = {\aap},
         year = 2021,
        month = may,
       volume = {649},
          eid = {A1},
        pages = {A1},
          doi = {10.1051/0004-6361/202039657},
archivePrefix = {arXiv},
       eprint = {2012.01533},
 primaryClass = {astro-ph.GA},
       adsurl = {https://ui.adsabs.harvard.edu/abs/2021A&A...649A...1G}
}

@ARTICLE{Smith_2012,
       author = {{Smith}, Jeffrey C. and {Stumpe}, Martin C. and {Van Cleve}, Jeffrey E. and {Jenkins}, Jon M. and {Barclay}, Thomas S. and {Fanelli}, Michael N. and {Girouard}, Forrest R. and {Kolodziejczak}, Jeffery J. and {McCauliff}, Sean D. and {Morris}, Robert L. and {Twicken}, Joseph D.},
        title = "{Kepler Presearch Data Conditioning II - A Bayesian Approach to Systematic Error Correction}",
      journal = {\pasp},
         year = 2012,
        month = sep,
       volume = {124},
       number = {919},
        pages = {1000},
          doi = {10.1086/667697},
archivePrefix = {arXiv},
       eprint = {1203.1383},
 primaryClass = {astro-ph.IM},
       adsurl = {https://ui.adsabs.harvard.edu/abs/2012PASP..124.1000S}
}

@ARTICLE{Stumpe_2012,
       author = {{Stumpe}, Martin C. and {Smith}, Jeffrey C. and {Van Cleve}, Jeffrey E. and {Twicken}, Joseph D. and {Barclay}, Thomas S. and {Fanelli}, Michael N. and {Girouard}, Forrest R. and {Jenkins}, Jon M. and {Kolodziejczak}, Jeffery J. and {McCauliff}, Sean D. and {Morris}, Robert L.},
        title = "{Kepler Presearch Data Conditioning I{\textemdash}Architecture and Algorithms for Error Correction in Kepler Light Curves}",
      journal = {\pasp},
         year = 2012,
        month = sep,
       volume = {124},
       number = {919},
        pages = {985},
          doi = {10.1086/667698},
archivePrefix = {arXiv},
       eprint = {1203.1382},
 primaryClass = {astro-ph.IM},
       adsurl = {https://ui.adsabs.harvard.edu/abs/2012PASP..124..985S}
}

@ARTICLE{Stumpe_2014,
       author = {{Stumpe}, Martin C. and {Smith}, Jeffrey C. and {Catanzarite}, Joseph H. and {Van Cleve}, Jeffrey E. and {Jenkins}, Jon M. and {Twicken}, Joseph D. and {Girouard}, Forrest R.},
        title = "{Multiscale Systematic Error Correction via Wavelet-Based Bandsplitting in Kepler Data}",
      journal = {\pasp},
         year = 2014,
        month = jan,
       volume = {126},
       number = {935},
        pages = {100},
          doi = {10.1086/674989},
       adsurl = {https://ui.adsabs.harvard.edu/abs/2014PASP..126..100S}
}

@MISC{lightkurve,
   author = {{Lightkurve Collaboration} and {Cardoso}, J.~V.~d.~M. and
             {Hedges}, C. and {Gully-Santiago}, M. and {Saunders}, N. and
             {Cody}, A.~M. and {Barclay}, T. and {Hall}, O. and
             {Sagear}, S. and {Turtelboom}, E. and {Zhang}, J. and
             {Tzanidakis}, A. and {Mighell}, K. and {Coughlin}, J. and
             {Bell}, K. and {Berta-Thompson}, Z. and {Williams}, P. and
             {Dotson}, J. and {Barentsen}, G.},
    title = "{Lightkurve: Kepler and TESS time series analysis in Python}",
howpublished = {Astrophysics Source Code Library},
     year = 2018,
    month = dec,
archivePrefix = "ascl",
   eprint = {1812.013},
   adsurl = {http://adsabs.harvard.edu/abs/2018ascl.soft12013L},
}

@inproceedings{Crane_2006,
author = {Jeffrey D. Crane and Stephen A. Shectman and R. Paul Butler},
title = {{The Carnegie Planet Finder Spectrograph}},
volume = {6269},
booktitle = {Ground-based and Airborne Instrumentation for Astronomy},
editor = {Ian S. McLean and Masanori Iye},
organization = {International Society for Optics and Photonics},
publisher = {SPIE},
pages = {626931},
year = {2006},
doi = {10.1117/12.672339},
URL = {https://doi.org/10.1117/12.672339}
}

@inproceedings{Crane_2008,
author = {Jeffrey D. Crane and Stephen A. Shectman and R. Paul Butler and Ian B. Thompson and Gregory S. Burley},
title = {{The Carnegie Planet Finder Spectrograph: a status report}},
volume = {7014},
booktitle = {Ground-based and Airborne Instrumentation for Astronomy II},
editor = {Ian S. McLean and Mark M. Casali},
organization = {International Society for Optics and Photonics},
publisher = {SPIE},
pages = {701479},
year = {2008},
doi = {10.1117/12.789637},
URL = {https://doi.org/10.1117/12.789637}
}

@inproceedings{Crane_2010,
author = {Jeffrey D. Crane and Stephen A. Shectman and R. Paul Butler and Ian B. Thompson and Christoph Birk and Patricio Jones and Gregory S. Burley},
title = {{The Carnegie Planet Finder Spectrograph: integration and commissioning}},
volume = {7735},
booktitle = {Ground-based and Airborne Instrumentation for Astronomy III},
editor = {Ian S. McLean and Suzanne K. Ramsay and Hideki Takami},
organization = {International Society for Optics and Photonics},
publisher = {SPIE},
pages = {773553},
year = {2010},
doi = {10.1117/12.857792},
URL = {https://doi.org/10.1117/12.857792}
}

@article{Butler_1996,
doi = {10.1086/133755},
url = {https://dx.doi.org/10.1086/133755},
year = {1996},
month = {jun},
publisher = {The Astronomical Society of the Pacific},
volume = {108},
number = {724},
pages = {500},
author = {R. P. Butler and G. W. Marcy and E. Williams and C. McCarthy and P. Dosanjh and S. S. Vogt},
title = {ATTAINING DOPPLER PRECISION OF 3 M S-1},
journal = {Publications of the Astronomical Society of the Pacific}
}

@article{Guerrero_2021,
doi = {10.3847/1538-4365/abefe1},
url = {https://dx.doi.org/10.3847/1538-4365/abefe1},
year = {2021},
month = {jun},
publisher = {The American Astronomical Society},
volume = {254},
number = {2},
pages = {39},
author = {Natalia M. Guerrero and S. Seager and Chelsea X. Huang and Andrew Vanderburg and Aylin Garcia Soto and Ismael Mireles and Katharine Hesse and William Fong and Ana Glidden and Avi Shporer and David W. Latham and Karen A. Collins and Samuel N. Quinn and Jennifer Burt and Diana Dragomir and Ian Crossfield and Roland Vanderspek and Michael Fausnaugh and Christopher J. Burke and George Ricker and Tansu Daylan and Zahra Essack and Maximilian N. Günther and Hugh P. Osborn and Joshua Pepper and Pamela Rowden and Lizhou Sha and Steven Villanueva Jr. and Daniel A. Yahalomi and Liang Yu and Sarah Ballard and Natalie M. Batalha and David Berardo and Ashley Chontos and Jason A. Dittmann and Gilbert A. Esquerdo and Thomas Mikal-Evans and Rahul Jayaraman and Akshata Krishnamurthy and Dana R. Louie and Nicholas Mehrle and Prajwal Niraula and Benjamin V. Rackham and Joseph E. Rodriguez and Stephen J. L. Rowden and Clara Sousa-Silva and David Watanabe and Ian Wong and Zhuchang Zhan and Goran Zivanovic and Jessie L. Christiansen and David R. Ciardi and Melanie A. Swain and Michael B. Lund and Susan E. Mullally and Scott W. Fleming and David R. Rodriguez and Patricia T. Boyd and Elisa V. Quintana and Thomas Barclay and Knicole D. Colón and S. A. Rinehart and Joshua E. Schlieder and Mark Clampin and Jon M. Jenkins and Joseph D. Twicken and Douglas A. Caldwell and Jeffrey L. Coughlin and Chris Henze and Jack J. Lissauer and Robert L. Morris and Mark E. Rose and Jeffrey C. Smith and Peter Tenenbaum and Eric B. Ting and Bill Wohler and G. Á. Bakos and Jacob L. Bean and Zachory K. Berta-Thompson and Allyson Bieryla and Luke G. Bouma and Lars A. Buchhave and Nathaniel Butler and David Charbonneau and John P. Doty and Jian Ge and Matthew J. Holman and Andrew W. Howard and Lisa Kaltenegger and Stephen R. Kane and Hans Kjeldsen and Laura Kreidberg and Douglas N. C. Lin and Charlotte Minsky and Norio Narita and Martin Paegert and András Pál and Enric Palle and Dimitar D. Sasselov and Alton Spencer and Alessandro Sozzetti and Keivan G. Stassun and Guillermo Torres and Stephane Udry and Joshua N. Winn},
title = {The TESS Objects of Interest Catalog from the TESS Prime Mission},
journal = {The Astrophysical Journal Supplement Series},
}

@article{Chaplin_2019,
doi = {10.3847/1538-3881/ab0c01},
url = {https://dx.doi.org/10.3847/1538-3881/ab0c01},
year = {2019},
month = {apr},
publisher = {The American Astronomical Society},
volume = {157},
number = {4},
pages = {163},
author = {W. J. Chaplin and H. M. Cegla and C. A. Watson and G. R. Davies and W. H. Ball},
title = {Filtering Solar-Like Oscillations for Exoplanet Detection in Radial Velocity Observations},
journal = {The Astronomical Journal},
}

@ARTICLE{Vines_2022,
       author = {{Vines}, Jose I. and {Jenkins}, James S.},
        title = "{ARIADNE: Measuring accurate and precise stellar parameters through SED fitting}",
      journal = {\mnras},
         year = 2022,
        month = apr,
          doi = {10.1093/mnras/stac956},
archivePrefix = {arXiv},
       eprint = {2204.03769},
 primaryClass = {astro-ph.SR},
       adsurl = {https://ui.adsabs.harvard.edu/abs/2022MNRAS.tmp..920V}
}

@inproceedings{Diego_1990,
author = {Francisco Diego and Andrew Charalambous and Adrian C. Fish and David D. Walker},
title = {{Final tests and commissioning of the UCL echelle spectrograph}},
volume = {1235},
booktitle = {Instrumentation in Astronomy VII},
editor = {David L. Crawford},
organization = {International Society for Optics and Photonics},
publisher = {SPIE},
pages = {562 -- 576},
year = {1990},
doi = {10.1117/12.19119},
URL = {https://doi.org/10.1117/12.19119}
}

@ARTICLE{Valenti_1995,
       author = {{Valenti}, Jeff A. and {Butler}, R. Paul and {Marcy}, Geoffrey W.},
        title = "{Determining Spectrometer Instrumental Profiles Using FTS Reference Spectra}",
      journal = {\pasp},
         year = 1995,
        month = oct,
       volume = {107},
        pages = {966},
          doi = {10.1086/133645},
       adsurl = {https://ui.adsabs.harvard.edu/abs/1995PASP..107..966V}
}

@book{rasmussen2006gaussian,
  title={Gaussian Processes for Machine Learning},
  author={Rasmussen, C. E. and Williams, C. K. I.},
  year={2006},
  publisher={MIT Press},
  address={Massachusetts},
  isbn={026218253X},
  note={\textcopyright~2006 Massachusetts Institute of Technology. \url{http://www.GaussianProcess.org/gpml}}
}

@ARTICLE{Michel_2009,
       author = {{Michel}, E. and {Samadi}, R. and {Baudin}, F. and {Barban}, C. and {Appourchaux}, T. and {Auvergne}, M.},
        title = "{Intrinsic photometric characterisation of stellar oscillations and granulation. Solar reference values and CoRoT response functions}",
      journal = {\aap},
         year = 2009,
        month = mar,
       volume = {495},
       number = {3},
        pages = {979-987},
          doi = {10.1051/0004-6361:200810353},
archivePrefix = {arXiv},
       eprint = {0809.1078},
 primaryClass = {astro-ph},
       adsurl = {https://ui.adsabs.harvard.edu/abs/2009A&A...495..979M}
}

@ARTICLE{Guo_2022,
       author = {{Guo}, Zhao and {Ford}, Eric B. and {Stello}, Dennis and {Luhn}, Jacob K. and {Mahadevan}, Suvrath and {Gupta}, Arvind F. and {Yu}, Jie},
        title = "{Modeling Stellar Oscillations and Granulation in Radial Velocity Time Series: A Fourier-based Method}",
      journal = {arXiv e-prints},
         year = 2022,
        month = feb,
          eid = {arXiv:2202.06094},
        pages = {arXiv:2202.06094},
          doi = {10.48550/arXiv.2202.06094},
archivePrefix = {arXiv},
       eprint = {2202.06094},
 primaryClass = {astro-ph.SR},
       adsurl = {https://ui.adsabs.harvard.edu/abs/2022arXiv220206094G}
}

@ARTICLE{Queloz_2001,
       author = {{Queloz}, D. and {Henry}, G.~W. and {Sivan}, J.~P. and {Baliunas}, S.~L. and {Beuzit}, J.~L. and {Donahue}, R.~A. and {Mayor}, M. and {Naef}, D. and {Perrier}, C. and {Udry}, S.},
        title = "{No planet for HD 166435}",
      journal = {\aap},
         year = 2001,
        month = nov,
       volume = {379},
        pages = {279-287},
          doi = {10.1051/0004-6361:20011308},
archivePrefix = {arXiv},
       eprint = {astro-ph/0109491},
 primaryClass = {astro-ph},
       adsurl = {https://ui.adsabs.harvard.edu/abs/2001A&A...379..279Q}
}

@ARTICLE{Santos_2002,
       author = {{Santos}, N.~C. and {Mayor}, M. and {Naef}, D. and {Pepe}, F. and {Queloz}, D. and {Udry}, S. and {Burnet}, M. and {Clausen}, J.~V. and {Helt}, B.~E. and {Olsen}, E.~H. and {Pritchard}, J.~D.},
        title = "{The CORALIE survey for southern extra-solar planets. IX. A 1.3-day period brown dwarf disguised as a planet}",
      journal = {\aap},
         year = 2002,
        month = sep,
       volume = {392},
        pages = {215-229},
          doi = {10.1051/0004-6361:20020876},
archivePrefix = {arXiv},
       eprint = {astro-ph/0206213},
 primaryClass = {astro-ph},
       adsurl = {https://ui.adsabs.harvard.edu/abs/2002A&A...392..215S}
}

@ARTICLE{Feng2023MNRAS,
       author = {{Feng}, Fabo and {Butler}, R. Paul and {Vogt}, Steven S. and {Holden}, Bradford and {Rui}, Yicheng},
        title = "{Revised orbits of the two nearest Jupiters}",
      journal = {\mnras},
         year = 2023,
        month = oct,
       volume = {525},
       number = {1},
        pages = {607-619},
          doi = {10.1093/mnras/stad2297},
archivePrefix = {arXiv},
       eprint = {2307.13622},
 primaryClass = {astro-ph.EP},
       adsurl = {https://ui.adsabs.harvard.edu/abs/2023MNRAS.525..607F}
}

@ARTICLE{Xiao2024MNRAS,
       author = {{Xiao}, Guang-Yao and {Feng}, Fabo and {Shectman}, Stephen A. and {Tinney}, C.~G. and {Teske}, Johanna K. and {Carter}, B.~D. and {Jones}, H.~R.~A. and {Wittenmyer}, Robert A. and {D{\'\i}az}, Mat{\'\i}as R. and {Crane}, Jeffrey D. and {Wang}, Sharon X. and {Bailey}, J. and {O'Toole}, S.~J. and {Feinstein}, Adina D. and {Rice}, Malena and {Essack}, Zahra and {Montet}, Benjamin T. and {Shporer}, Avi and {Butler}, R. Paul},
        title = "{HD 222237 b: a long period super-Jupiter around a nearby star revealed by radial-velocity and Hipparcos-Gaia astrometry}",
      journal = {\mnras},
         year = 2024,
        month = sep,
          doi = {10.1093/mnras/stae2151},
archivePrefix = {arXiv},
       eprint = {2409.08067},
 primaryClass = {astro-ph.EP},
       adsurl = {https://ui.adsabs.harvard.edu/abs/2024MNRAS.tmp.2178X}
}

@ARTICLE{vanLeeuwen2007,
       author = {{van Leeuwen}, F.},
        title = "{Validation of the new Hipparcos reduction}",
      journal = {\aap},
         year = 2007,
        month = nov,
       volume = {474},
       number = {2},
        pages = {653-664},
          doi = {10.1051/0004-6361:20078357},
archivePrefix = {arXiv},
       eprint = {0708.1752},
 primaryClass = {astro-ph},
       adsurl = {https://ui.adsabs.harvard.edu/abs/2007A&A...474..653V}
}

@ARTICLE{GaiaCollaboration2018,
       author = {{Gaia Collaboration} and {Brown}, A.~G.~A. and {Vallenari}, A. and {Prusti}, T. and {de Bruijne}, J.~H.~J. and {Babusiaux}, C. and {Bailer-Jones}, C.~A.~L. and {Biermann}, M. and {Evans}, D.~W. and {Eyer}, L. and {Jansen}, F. and {Jordi}, C. and {Klioner}, S.~A. and {Lammers}, U. and {Lindegren}, L. and {Luri}, X. and {Mignard}, F. and {Panem}, C. and {Pourbaix}, D. and {Randich}, S. and {Sartoretti}, P. and {Siddiqui}, H.~I. and {Soubiran}, C. and {van Leeuwen}, F. and {Walton}, N.~A. and {Arenou}, F. and {Bastian}, U. and {Cropper}, M. and {Drimmel}, R. and {Katz}, D. and {Lattanzi}, M.~G. and {Bakker}, J. and {Cacciari}, C. and {Casta{\~n}eda}, J. and {Chaoul}, L. and {Cheek}, N. and {De Angeli}, F. and {Fabricius}, C. and {Guerra}, R. and {Holl}, B. and {Masana}, E. and {Messineo}, R. and {Mowlavi}, N. and {Nienartowicz}, K. and {Panuzzo}, P. and {Portell}, J. and {Riello}, M. and {Seabroke}, G.~M. and {Tanga}, P. and {Th{\'e}venin}, F. and {Gracia-Abril}, G. and {Comoretto}, G. and {Garcia-Reinaldos}, M. and {Teyssier}, D. and {Altmann}, M. and {Andrae}, R. and {Audard}, M. and {Bellas-Velidis}, I. and {Benson}, K. and {Berthier}, J. and {Blomme}, R. and {Burgess}, P. and {Busso}, G. and {Carry}, B. and {Cellino}, A. and {Clementini}, G. and {Clotet}, M. and {Creevey}, O. and {Davidson}, M. and {De Ridder}, J. and {Delchambre}, L. and {Dell'Oro}, A. and {Ducourant}, C. and {Fern{\'a}ndez-Hern{\'a}ndez}, J. and {Fouesneau}, M. and {Fr{\'e}mat}, Y. and {Galluccio}, L. and {Garc{\'\i}a-Torres}, M. and {Gonz{\'a}lez-N{\'u}{\~n}ez}, J. and {Gonz{\'a}lez-Vidal}, J.~J. and {Gosset}, E. and {Guy}, L.~P. and {Halbwachs}, J. -L. and {Hambly}, N.~C. and {Harrison}, D.~L. and {Hern{\'a}ndez}, J. and {Hestroffer}, D. and {Hodgkin}, S.~T. and {Hutton}, A. and {Jasniewicz}, G. and {Jean-Antoine-Piccolo}, A. and {Jordan}, S. and {Korn}, A.~J. and {Krone-Martins}, A. and {Lanzafame}, A.~C. and {Lebzelter}, T. and {L{\"o}ffler}, W. and {Manteiga}, M. and {Marrese}, P.~M. and {Mart{\'\i}n-Fleitas}, J.~M. and {Moitinho}, A. and {Mora}, A. and {Muinonen}, K. and {Osinde}, J. and {Pancino}, E. and {Pauwels}, T. and {Petit}, J. -M. and {Recio-Blanco}, A. and {Richards}, P.~J. and {Rimoldini}, L. and {Robin}, A.~C. and {Sarro}, L.~M. and {Siopis}, C. and {Smith}, M. and {Sozzetti}, A. and {S{\"u}veges}, M. and {Torra}, J. and {van Reeven}, W. and {Abbas}, U. and {Abreu Aramburu}, A. and {Accart}, S. and {Aerts}, C. and {Altavilla}, G. and {{\'A}lvarez}, M.~A. and {Alvarez}, R. and {Alves}, J. and {Anderson}, R.~I. and {Andrei}, A.~H. and {Anglada Varela}, E. and {Antiche}, E. and {Antoja}, T. and {Arcay}, B. and {Astraatmadja}, T.~L. and {Bach}, N. and {Baker}, S.~G. and {Balaguer-N{\'u}{\~n}ez}, L. and {Balm}, P. and {Barache}, C. and {Barata}, C. and {Barbato}, D. and {Barblan}, F. and {Barklem}, P.~S. and {Barrado}, D. and {Barros}, M. and {Barstow}, M.~A. and {Bartholom{\'e} Mu{\~n}oz}, S. and {Bassilana}, J. -L. and {Becciani}, U. and {Bellazzini}, M. and {Berihuete}, A. and {Bertone}, S. and {Bianchi}, L. and {Bienaym{\'e}}, O. and {Blanco-Cuaresma}, S. and {Boch}, T. and {Boeche}, C. and {Bombrun}, A. and {Borrachero}, R. and {Bossini}, D. and {Bouquillon}, S. and {Bourda}, G. and {Bragaglia}, A. and {Bramante}, L. and {Breddels}, M.~A. and {Bressan}, A. and {Brouillet}, N. and {Br{\"u}semeister}, T. and {Brugaletta}, E. and {Bucciarelli}, B. and {Burlacu}, A. and {Busonero}, D. and {Butkevich}, A.~G. and {Buzzi}, R. and {Caffau}, E. and {Cancelliere}, R. and {Cannizzaro}, G. and {Cantat-Gaudin}, T. and {Carballo}, R. and {Carlucci}, T. and {Carrasco}, J.~M. and {Casamiquela}, L. and {Castellani}, M. and {Castro-Ginard}, A. and {Charlot}, P. and {Chemin}, L. and {Chiavassa}, A. and {Cocozza}, G. and {Costigan}, G. and {Cowell}, S. and {Crifo}, F. and {Crosta}, M. and {Crowley}, C. and {Cuypers}, J. and {Dafonte}, C. and {Damerdji}, Y. and {Dapergolas}, A. and {David}, P. and {David}, M. and {de Laverny}, P. and {De Luise}, F. and {De March}, R. and {de Martino}, D. and {de Souza}, R. and {de Torres}, A. and {Debosscher}, J. and {del Pozo}, E. and {Delbo}, M. and {Delgado}, A. and {Delgado}, H.~E. and {Di Matteo}, P. and {Diakite}, S. and {Diener}, C. and {Distefano}, E. and {Dolding}, C. and {Drazinos}, P. and {Dur{\'a}n}, J. and {Edvardsson}, B. and {Enke}, H. and {Eriksson}, K. and {Esquej}, P. and {Eynard Bontemps}, G. and {Fabre}, C. and {Fabrizio}, M. and {Faigler}, S. and {Falc{\~a}o}, A.~J. and {Farr{\`a}s Casas}, M. and {Federici}, L. and {Fedorets}, G. and {Fernique}, P. and {Figueras}, F. and {Filippi}, F. and {Findeisen}, K. and {Fonti}, A. and {Fraile}, E. and {Fraser}, M. and {Fr{\'e}zouls}, B. and {Gai}, M. and {Galleti}, S. and {Garabato}, D. and {Garc{\'\i}a-Sedano}, F. and {Garofalo}, A. and {Garralda}, N. and {Gavel}, A. and {Gavras}, P. and {Gerssen}, J. and {Geyer}, R. and {Giacobbe}, P. and {Gilmore}, G. and {Girona}, S. and {Giuffrida}, G. and {Glass}, F. and {Gomes}, M. and {Granvik}, M. and {Gueguen}, A. and {Guerrier}, A. and {Guiraud}, J. and {Guti{\'e}rrez-S{\'a}nchez}, R. and {Haigron}, R. and {Hatzidimitriou}, D. and {Hauser}, M. and {Haywood}, M. and {Heiter}, U. and {Helmi}, A. and {Heu}, J. and {Hilger}, T. and {Hobbs}, D. and {Hofmann}, W. and {Holland}, G. and {Huckle}, H.~E. and {Hypki}, A. and {Icardi}, V. and {Jan{\ss}en}, K. and {Jevardat de Fombelle}, G. and {Jonker}, P.~G. and {Juh{\'a}sz}, {\'A}. L. and {Julbe}, F. and {Karampelas}, A. and {Kewley}, A. and {Klar}, J. and {Kochoska}, A. and {Kohley}, R. and {Kolenberg}, K. and {Kontizas}, M. and {Kontizas}, E. and {Koposov}, S.~E. and {Kordopatis}, G. and {Kostrzewa-Rutkowska}, Z. and {Koubsky}, P. and {Lambert}, S. and {Lanza}, A.~F. and {Lasne}, Y. and {Lavigne}, J. -B. and {Le Fustec}, Y. and {Le Poncin-Lafitte}, C. and {Lebreton}, Y. and {Leccia}, S. and {Leclerc}, N. and {Lecoeur-Taibi}, I. and {Lenhardt}, H. and {Leroux}, F. and {Liao}, S. and {Licata}, E. and {Lindstr{\o}m}, H.~E.~P. and {Lister}, T.~A. and {Livanou}, E. and {Lobel}, A. and {L{\'o}pez}, M. and {Managau}, S. and {Mann}, R.~G. and {Mantelet}, G. and {Marchal}, O. and {Marchant}, J.~M. and {Marconi}, M. and {Marinoni}, S. and {Marschalk{\'o}}, G. and {Marshall}, D.~J. and {Martino}, M. and {Marton}, G. and {Mary}, N. and {Massari}, D. and {Matijevi{\v{c}}}, G. and {Mazeh}, T. and {McMillan}, P.~J. and {Messina}, S. and {Michalik}, D. and {Millar}, N.~R. and {Molina}, D. and {Molinaro}, R. and {Moln{\'a}r}, L. and {Montegriffo}, P. and {Mor}, R. and {Morbidelli}, R. and {Morel}, T. and {Morris}, D. and {Mulone}, A.~F. and {Muraveva}, T. and {Musella}, I. and {Nelemans}, G. and {Nicastro}, L. and {Noval}, L. and {O'Mullane}, W. and {Ord{\'e}novic}, C. and {Ord{\'o}{\~n}ez-Blanco}, D. and {Osborne}, P. and {Pagani}, C. and {Pagano}, I. and {Pailler}, F. and {Palacin}, H. and {Palaversa}, L. and {Panahi}, A. and {Pawlak}, M. and {Piersimoni}, A.~M. and {Pineau}, F. -X. and {Plachy}, E. and {Plum}, G. and {Poggio}, E. and {Poujoulet}, E. and {Pr{\v{s}}a}, A. and {Pulone}, L. and {Racero}, E. and {Ragaini}, S. and {Rambaux}, N. and {Ramos-Lerate}, M. and {Regibo}, S. and {Reyl{\'e}}, C. and {Riclet}, F. and {Ripepi}, V. and {Riva}, A. and {Rivard}, A. and {Rixon}, G. and {Roegiers}, T. and {Roelens}, M. and {Romero-G{\'o}mez}, M. and {Rowell}, N. and {Royer}, F. and {Ruiz-Dern}, L. and {Sadowski}, G. and {Sagrist{\`a} Sell{\'e}s}, T. and {Sahlmann}, J. and {Salgado}, J. and {Salguero}, E. and {Sanna}, N. and {Santana-Ros}, T. and {Sarasso}, M. and {Savietto}, H. and {Schultheis}, M. and {Sciacca}, E. and {Segol}, M. and {Segovia}, J.~C. and {S{\'e}gransan}, D. and {Shih}, I. -C. and {Siltala}, L. and {Silva}, A.~F. and {Smart}, R.~L. and {Smith}, K.~W. and {Solano}, E. and {Solitro}, F. and {Sordo}, R. and {Soria Nieto}, S. and {Souchay}, J. and {Spagna}, A. and {Spoto}, F. and {Stampa}, U. and {Steele}, I.~A. and {Steidelm{\"u}ller}, H. and {Stephenson}, C.~A. and {Stoev}, H. and {Suess}, F.~F. and {Surdej}, J. and {Szabados}, L. and {Szegedi-Elek}, E. and {Tapiador}, D. and {Taris}, F. and {Tauran}, G. and {Taylor}, M.~B. and {Teixeira}, R. and {Terrett}, D. and {Teyssandier}, P. and {Thuillot}, W. and {Titarenko}, A. and {Torra Clotet}, F. and {Turon}, C. and {Ulla}, A. and {Utrilla}, E. and {Uzzi}, S. and {Vaillant}, M. and {Valentini}, G. and {Valette}, V. and {van Elteren}, A. and {Van Hemelryck}, E. and {van Leeuwen}, M. and {Vaschetto}, M. and {Vecchiato}, A. and {Veljanoski}, J. and {Viala}, Y. and {Vicente}, D. and {Vogt}, S. and {von Essen}, C. and {Voss}, H. and {Votruba}, V. and {Voutsinas}, S. and {Walmsley}, G. and {Weiler}, M. and {Wertz}, O. and {Wevers}, T. and {Wyrzykowski}, {\L}. and {Yoldas}, A. and {{\v{Z}}erjal}, M. and {Ziaeepour}, H. and {Zorec}, J. and {Zschocke}, S. and {Zucker}, S. and {Zurbach}, C. and {Zwitter}, T.},
        title = "{Gaia Data Release 2. Summary of the contents and survey properties}",
      journal = {\aap},
         year = 2018,
        month = aug,
       volume = {616},
          eid = {A1},
        pages = {A1},
          doi = {10.1051/0004-6361/201833051},
archivePrefix = {arXiv},
       eprint = {1804.09365},
 primaryClass = {astro-ph.GA},
       adsurl = {https://ui.adsabs.harvard.edu/abs/2018A&A...616A...1G}
}

@ARTICLE{GaiaCollaboration2023,
       author = {{Gaia Collaboration} and {Vallenari}, A. and {Brown}, A.~G.~A. and {Prusti}, T. and {de Bruijne}, J.~H.~J. and {Arenou}, F. and {Babusiaux}, C. and {Biermann}, M. and {Creevey}, O.~L. and {Ducourant}, C. and {Evans}, D.~W. and {Eyer}, L. and {Guerra}, R. and {Hutton}, A. and {Jordi}, C. and {Klioner}, S.~A. and {Lammers}, U.~L. and {Lindegren}, L. and {Luri}, X. and {Mignard}, F. and {Panem}, C. and {Pourbaix}, D. and {Randich}, S. and {Sartoretti}, P. and {Soubiran}, C. and {Tanga}, P. and {Walton}, N.~A. and {Bailer-Jones}, C.~A.~L. and {Bastian}, U. and {Drimmel}, R. and {Jansen}, F. and {Katz}, D. and {Lattanzi}, M.~G. and {van Leeuwen}, F. and {Bakker}, J. and {Cacciari}, C. and {Casta{\~n}eda}, J. and {De Angeli}, F. and {Fabricius}, C. and {Fouesneau}, M. and {Fr{\'e}mat}, Y. and {Galluccio}, L. and {Guerrier}, A. and {Heiter}, U. and {Masana}, E. and {Messineo}, R. and {Mowlavi}, N. and {Nicolas}, C. and {Nienartowicz}, K. and {Pailler}, F. and {Panuzzo}, P. and {Riclet}, F. and {Roux}, W. and {Seabroke}, G.~M. and {Sordo}, R. and {Th{\'e}venin}, F. and {Gracia-Abril}, G. and {Portell}, J. and {Teyssier}, D. and {Altmann}, M. and {Andrae}, R. and {Audard}, M. and {Bellas-Velidis}, I. and {Benson}, K. and {Berthier}, J. and {Blomme}, R. and {Burgess}, P.~W. and {Busonero}, D. and {Busso}, G. and {C{\'a}novas}, H. and {Carry}, B. and {Cellino}, A. and {Cheek}, N. and {Clementini}, G. and {Damerdji}, Y. and {Davidson}, M. and {de Teodoro}, P. and {Nu{\~n}ez Campos}, M. and {Delchambre}, L. and {Dell'Oro}, A. and {Esquej}, P. and {Fern{\'a}ndez-Hern{\'a}ndez}, J. and {Fraile}, E. and {Garabato}, D. and {Garc{\'\i}a-Lario}, P. and {Gosset}, E. and {Haigron}, R. and {Halbwachs}, J. -L. and {Hambly}, N.~C. and {Harrison}, D.~L. and {Hern{\'a}ndez}, J. and {Hestroffer}, D. and {Hodgkin}, S.~T. and {Holl}, B. and {Jan{\ss}en}, K. and {Jevardat de Fombelle}, G. and {Jordan}, S. and {Krone-Martins}, A. and {Lanzafame}, A.~C. and {L{\"o}ffler}, W. and {Marchal}, O. and {Marrese}, P.~M. and {Moitinho}, A. and {Muinonen}, K. and {Osborne}, P. and {Pancino}, E. and {Pauwels}, T. and {Recio-Blanco}, A. and {Reyl{\'e}}, C. and {Riello}, M. and {Rimoldini}, L. and {Roegiers}, T. and {Rybizki}, J. and {Sarro}, L.~M. and {Siopis}, C. and {Smith}, M. and {Sozzetti}, A. and {Utrilla}, E. and {van Leeuwen}, M. and {Abbas}, U. and {{\'A}brah{\'a}m}, P. and {Abreu Aramburu}, A. and {Aerts}, C. and {Aguado}, J.~J. and {Ajaj}, M. and {Aldea-Montero}, F. and {Altavilla}, G. and {{\'A}lvarez}, M.~A. and {Alves}, J. and {Anders}, F. and {Anderson}, R.~I. and {Anglada Varela}, E. and {Antoja}, T. and {Baines}, D. and {Baker}, S.~G. and {Balaguer-N{\'u}{\~n}ez}, L. and {Balbinot}, E. and {Balog}, Z. and {Barache}, C. and {Barbato}, D. and {Barros}, M. and {Barstow}, M.~A. and {Bartolom{\'e}}, S. and {Bassilana}, J. -L. and {Bauchet}, N. and {Becciani}, U. and {Bellazzini}, M. and {Berihuete}, A. and {Bernet}, M. and {Bertone}, S. and {Bianchi}, L. and {Binnenfeld}, A. and {Blanco-Cuaresma}, S. and {Blazere}, A. and {Boch}, T. and {Bombrun}, A. and {Bossini}, D. and {Bouquillon}, S. and {Bragaglia}, A. and {Bramante}, L. and {Breedt}, E. and {Bressan}, A. and {Brouillet}, N. and {Brugaletta}, E. and {Bucciarelli}, B. and {Burlacu}, A. and {Butkevich}, A.~G. and {Buzzi}, R. and {Caffau}, E. and {Cancelliere}, R. and {Cantat-Gaudin}, T. and {Carballo}, R. and {Carlucci}, T. and {Carnerero}, M.~I. and {Carrasco}, J.~M. and {Casamiquela}, L. and {Castellani}, M. and {Castro-Ginard}, A. and {Chaoul}, L. and {Charlot}, P. and {Chemin}, L. and {Chiaramida}, V. and {Chiavassa}, A. and {Chornay}, N. and {Comoretto}, G. and {Contursi}, G. and {Cooper}, W.~J. and {Cornez}, T. and {Cowell}, S. and {Crifo}, F. and {Cropper}, M. and {Crosta}, M. and {Crowley}, C. and {Dafonte}, C. and {Dapergolas}, A. and {David}, M. and {David}, P. and {de Laverny}, P. and {De Luise}, F. and {De March}, R. and {De Ridder}, J. and {de Souza}, R. and {de Torres}, A. and {del Peloso}, E.~F. and {del Pozo}, E. and {Delbo}, M. and {Delgado}, A. and {Delisle}, J. -B. and {Demouchy}, C. and {Dharmawardena}, T.~E. and {Di Matteo}, P. and {Diakite}, S. and {Diener}, C. and {Distefano}, E. and {Dolding}, C. and {Edvardsson}, B. and {Enke}, H. and {Fabre}, C. and {Fabrizio}, M. and {Faigler}, S. and {Fedorets}, G. and {Fernique}, P. and {Fienga}, A. and {Figueras}, F. and {Fournier}, Y. and {Fouron}, C. and {Fragkoudi}, F. and {Gai}, M. and {Garcia-Gutierrez}, A. and {Garcia-Reinaldos}, M. and {Garc{\'\i}a-Torres}, M. and {Garofalo}, A. and {Gavel}, A. and {Gavras}, P. and {Gerlach}, E. and {Geyer}, R. and {Giacobbe}, P. and {Gilmore}, G. and {Girona}, S. and {Giuffrida}, G. and {Gomel}, R. and {Gomez}, A. and {Gonz{\'a}lez-N{\'u}{\~n}ez}, J. and {Gonz{\'a}lez-Santamar{\'\i}a}, I. and {Gonz{\'a}lez-Vidal}, J.~J. and {Granvik}, M. and {Guillout}, P. and {Guiraud}, J. and {Guti{\'e}rrez-S{\'a}nchez}, R. and {Guy}, L.~P. and {Hatzidimitriou}, D. and {Hauser}, M. and {Haywood}, M. and {Helmer}, A. and {Helmi}, A. and {Sarmiento}, M.~H. and {Hidalgo}, S.~L. and {Hilger}, T. and {H{\l}adczuk}, N. and {Hobbs}, D. and {Holland}, G. and {Huckle}, H.~E. and {Jardine}, K. and {Jasniewicz}, G. and {Jean-Antoine Piccolo}, A. and {Jim{\'e}nez-Arranz}, {\'O}. and {Jorissen}, A. and {Juaristi Campillo}, J. and {Julbe}, F. and {Karbevska}, L. and {Kervella}, P. and {Khanna}, S. and {Kontizas}, M. and {Kordopatis}, G. and {Korn}, A.~J. and {K{\'o}sp{\'a}l}, {\'A}. and {Kostrzewa-Rutkowska}, Z. and {Kruszy{\'n}ska}, K. and {Kun}, M. and {Laizeau}, P. and {Lambert}, S. and {Lanza}, A.~F. and {Lasne}, Y. and {Le Campion}, J. -F. and {Lebreton}, Y. and {Lebzelter}, T. and {Leccia}, S. and {Leclerc}, N. and {Lecoeur-Taibi}, I. and {Liao}, S. and {Licata}, E.~L. and {Lindstr{\o}m}, H.~E.~P. and {Lister}, T.~A. and {Livanou}, E. and {Lobel}, A. and {Lorca}, A. and {Loup}, C. and {Madrero Pardo}, P. and {Magdaleno Romeo}, A. and {Managau}, S. and {Mann}, R.~G. and {Manteiga}, M. and {Marchant}, J.~M. and {Marconi}, M. and {Marcos}, J. and {Marcos Santos}, M.~M.~S. and {Mar{\'\i}n Pina}, D. and {Marinoni}, S. and {Marocco}, F. and {Marshall}, D.~J. and {Martin Polo}, L. and {Mart{\'\i}n-Fleitas}, J.~M. and {Marton}, G. and {Mary}, N. and {Masip}, A. and {Massari}, D. and {Mastrobuono-Battisti}, A. and {Mazeh}, T. and {McMillan}, P.~J. and {Messina}, S. and {Michalik}, D. and {Millar}, N.~R. and {Mints}, A. and {Molina}, D. and {Molinaro}, R. and {Moln{\'a}r}, L. and {Monari}, G. and {Mongui{\'o}}, M. and {Montegriffo}, P. and {Montero}, A. and {Mor}, R. and {Mora}, A. and {Morbidelli}, R. and {Morel}, T. and {Morris}, D. and {Muraveva}, T. and {Murphy}, C.~P. and {Musella}, I. and {Nagy}, Z. and {Noval}, L. and {Oca{\~n}a}, F. and {Ogden}, A. and {Ordenovic}, C. and {Osinde}, J.~O. and {Pagani}, C. and {Pagano}, I. and {Palaversa}, L. and {Palicio}, P.~A. and {Pallas-Quintela}, L. and {Panahi}, A. and {Payne-Wardenaar}, S. and {Pe{\~n}alosa Esteller}, X. and {Penttil{\"a}}, A. and {Pichon}, B. and {Piersimoni}, A.~M. and {Pineau}, F. -X. and {Plachy}, E. and {Plum}, G. and {Poggio}, E. and {Pr{\v{s}}a}, A. and {Pulone}, L. and {Racero}, E. and {Ragaini}, S. and {Rainer}, M. and {Raiteri}, C.~M. and {Rambaux}, N. and {Ramos}, P. and {Ramos-Lerate}, M. and {Re Fiorentin}, P. and {Regibo}, S. and {Richards}, P.~J. and {Rios Diaz}, C. and {Ripepi}, V. and {Riva}, A. and {Rix}, H. -W. and {Rixon}, G. and {Robichon}, N. and {Robin}, A.~C. and {Robin}, C. and {Roelens}, M. and {Rogues}, H.~R.~O. and {Rohrbasser}, L. and {Romero-G{\'o}mez}, M. and {Rowell}, N. and {Royer}, F. and {Ruz Mieres}, D. and {Rybicki}, K.~A. and {Sadowski}, G. and {S{\'a}ez N{\'u}{\~n}ez}, A. and {Sagrist{\`a} Sell{\'e}s}, A. and {Sahlmann}, J. and {Salguero}, E. and {Samaras}, N. and {Sanchez Gimenez}, V. and {Sanna}, N. and {Santove{\~n}a}, R. and {Sarasso}, M. and {Schultheis}, M. and {Sciacca}, E. and {Segol}, M. and {Segovia}, J.~C. and {S{\'e}gransan}, D. and {Semeux}, D. and {Shahaf}, S. and {Siddiqui}, H.~I. and {Siebert}, A. and {Siltala}, L. and {Silvelo}, A. and {Slezak}, E. and {Slezak}, I. and {Smart}, R.~L. and {Snaith}, O.~N. and {Solano}, E. and {Solitro}, F. and {Souami}, D. and {Souchay}, J. and {Spagna}, A. and {Spina}, L. and {Spoto}, F. and {Steele}, I.~A. and {Steidelm{\"u}ller}, H. and {Stephenson}, C.~A. and {S{\"u}veges}, M. and {Surdej}, J. and {Szabados}, L. and {Szegedi-Elek}, E. and {Taris}, F. and {Taylor}, M.~B. and {Teixeira}, R. and {Tolomei}, L. and {Tonello}, N. and {Torra}, F. and {Torra}, J. and {Torralba Elipe}, G. and {Trabucchi}, M. and {Tsounis}, A.~T. and {Turon}, C. and {Ulla}, A. and {Unger}, N. and {Vaillant}, M.~V. and {van Dillen}, E. and {van Reeven}, W. and {Vanel}, O. and {Vecchiato}, A. and {Viala}, Y. and {Vicente}, D. and {Voutsinas}, S. and {Weiler}, M. and {Wevers}, T. and {Wyrzykowski}, {\L}. and {Yoldas}, A. and {Yvard}, P. and {Zhao}, H. and {Zorec}, J. and {Zucker}, S. and {Zwitter}, T.},
        title = "{Gaia Data Release 3. Summary of the content and survey properties}",
      journal = {\aap},
         year = 2023,
        month = jun,
       volume = {674},
          eid = {A1},
        pages = {A1},
          doi = {10.1051/0004-6361/202243940},
archivePrefix = {arXiv},
       eprint = {2208.00211},
 primaryClass = {astro-ph.GA},
       adsurl = {https://ui.adsabs.harvard.edu/abs/2023A&A...674A...1G}
}
\bibliographystyle{aasjournal}



\end{document}